\documentclass[preprint,12pt]{aastex}
\usepackage{emulateapj5}

\newcommand{\ee}[1]{\mbox{${} \times 10^{#1}$}}
\newcommand{\eten}[1]{\mbox{$10^{#1}$}}

\newcommand{\degree}{\mbox{$^{\circ}$}}

\newcommand{\as}{\mbox{\arcsec}}

\newcommand\cmv{\mbox{cm$^{-3}$}}

\def\lsim {$\rlap{\raise.4ex\hbox{$<$}}\lower.55ex\hbox{$\sim$}\,$}

\newcommand{\iras}{\mbox{\it IRAS}}

\newcommand\submm{submillimeter}

\newcommand\fir{far-infrared}

\newcommand{\lsun}{\mbox{L$_\odot$}}
\newcommand{\msun}{\mbox{M$_\odot$}}
\newcommand{\ta}{{$T_A^*$}}

\newcommand{\lbol}{\mbox{$L_{bol}$}} 
\newcommand{\tbol}{\mbox{$T_{bol}$}} 

\newcommand{\mn}{\mbox{$M_n$}} 
\newcommand{\mean}[1]{\mbox{$\langle#1\rangle$}} 
\newcommand{\lsmm}{\mbox{$L_{smm}$}} 

\input{epsf}

\newcommand{\Snu}{\mbox{$S_{\nu}$}}

\newcommand{\rn}{\mbox{$r_n$}} 
\newcommand{\lobs}{\mbox{$L_{bol}$}} 
\newcommand{\lint}{\mbox{$L_{bol}$}} 
\newcommand{\md}{\mbox{$M(< r_{dec})$}} 
\newcommand{\miso}{\mbox{$M_{iso}$}} 
\newcommand{\mden}{\mbox{$M(< r_n)$}} 
\newcommand{\HII}{\ion{H}{2}}
\newcommand{\uchii}{\mbox{{\rm UC}\HII}}
\newcommand{\rdust}{\mbox{$r_{dec}$}} 
\newcommand{\chisq}{\mbox{$\chi_r^2$}}
\newcommand{\chisqsed}{\mbox{$\chi_{SED}^2$}}
\newcommand{\chisqsharc}{\mbox{$\chi_{350}^2$}}

\slugcomment{\footnotesize {\LaTeX}ed at \number\time\ min., \today}
\shorttitle{Physical Conditions for Massive Star Formation }
\shortauthors{Mueller et al.}

\begin{document}


\title {\bf The Physical Conditions for Massive Star Formation: 
Dust Continuum Maps and Modeling}
\author {Kaisa E. Mueller, Yancy L. Shirley, Neal J. Evans II, 
and Heather R. Jacobson}
\affil{Department of Astronomy, The University of Texas at Austin,
       Austin, Texas 78712--1083}
\email{mueller, yshirley, nje, hrj@astro.as.utexas.edu}

 
\begin{abstract}

Fifty-one dense cores associated with water masers were mapped at 350
\micron.  These cores are very luminous, $10^3 < \lobs/\lsun < 10^6$,
indicative of the formation of massive stars. Dust continuum contour
maps, radial intensity profiles, and photometry are presented for
these sources. The submillimeter dust emission peak is, on average,
nearly coincident with the water maser position. The spectral energy
distributions and normalized radial profiles of dust continuum
emission were modeled for 31 sources using a one-dimensional dust
radiative transfer code, assuming a power law density distribution in
the envelope, $n = n_f (r/r_f)^{-p}$. The best fit density power law
exponent, $p$, ranged from 0.75 to 2.5 with $\langle p \rangle = 1.8
\pm 0.4$, similar to the mean value found by Beuther et al.\ (2002) in
a large sample of massive star forming regions.  The mean value of $p$
is also comparable to that found in regions forming only low mass
stars, but $\langle n_f \rangle$ is over two orders of magnitude
greater for the massive cores. The mean $p$ is incompatible with a
logatropic sphere ($p$ = 1), but other star formation models cannot be
ruled out.  Different mass estimates are compared and mean masses of
gas and dust are reported within a half-power radius determined from
the dust emission, $\langle {\rm log} (\md) \rangle$ = 2.0 $\pm$ 0.6,
and within a radius where the total density exceeds $10^4$ \cmv,
$\mean {{\rm log} (\mden)} = 2.5 \pm 0.6$.  Evolutionary indicators
commonly used for low mass star formation, such as \tbol\ and $\lbol
/\lsmm$, may have some utility for regions forming massive stars.
Additionally, for comparison with extragalactic star formation
studies, the luminosity to dust mass ratio is calculated for these
sources, $\langle \lobs/M_D \rangle = 1.4\ee4$ \lsun/\msun, with a
method most parallel to that used in studies of distant galaxies. This
ratio is similar to that seen in high redshift starburst galaxies.

\end{abstract}
 


\section{Introduction}

The study of regions forming massive stars is essential to our
understanding of how stars are born. Since most stars form in clusters
associated with high mass stars (e.g., Carpenter 2000), many recent
studies have focused on better understanding the physical conditions
in these regions (e.g., van der Tak et al.\ 2000, Sridharan et al.\ 2002, 
Beuther et al.\ 2002, Hatchell et al.\ 2000,  
Walsh et al.\ 2001, Osorio, Lizano, \& D'Alessio 1999, Garay \& Lizano 
1999). 
The density distribution in the envelopes of regions
forming massive stars is an important observational constraint for
star formation models.  The density distribution is usually 
a power law ($n \propto r^{-p}$). For example, McLaughlin \& Pudriz's (1997) 
logatropic sphere model predicts a shallow power law with $p = 1$ in 
the static envelope, whereas Shu's
(1977) inside-out collapse model for isolated star formation has an initial
density distribution with $p = 2$. 

A power law density distribution has been fitted to observations of 
low mass star forming regions. Shirley et
al.\ (2002a) and Young et al.\ (2002) find a combined $\langle p \rangle
= 1.6 \pm 0.4$ for Class 0 (Andr\'{e},
Ward-Thompson, \& Barsony 1993) and Class I sources (Lada \& Wilking 1984, 
Lada 1987, Myers \& Ladd 1993, Chen et al.\ 1995).
 All ``errors" on mean values in this paper refer to the standard
deviation of the distribution of values about the mean. Shirley et
al.\ (2002a) and Young et al.\ (2002) also report that aspherical cores
have shallower power laws. If the aspherical cores are left out of the
average for the low mass cores, $\langle p \rangle$ goes to 1.8. 
While low mass cores are
well-studied, it is only recently that the density structure of
high mass cores has been investigated for large samples. For example, 
van der Tak
et al.\ (2000) found a shallow density structure, $\langle p \rangle =
1$ to 1.5, for a sample of 14 regions forming massive stars, while
Beuther et al. (2002) reported $\langle p \rangle$ = 1.6 $\pm$ 0.5 for a
larger sample of 69 massive star forming regions.

Studies of massive star forming cores also have important implications
for understanding extragalactic star formation, including starburst
galaxies. The \fir\ luminosity to dust mass ratio, $L/M$, is a tool 
often used in
extragalactic studies to characterize star formation
since it is proportional to the star formation rate per unit mass
(Kennicutt 1998). To learn if starburst galaxies are forming stars by
mechanisms similar to those in the Milky Way, but on grander scales,
it is important to investigate the star formation efficiency and $L/M$
for dense gas in more accessible Galactic star formation regions to
provide a point of comparison between these modes of star formation.

\subsection{The Sample}

The objects in this study were selected from the sample of Plume et
al.\ (1992, 1997) of massive star forming cores associated with water
masers.  Table \ref{obstab} lists the sources and their observed
properties. Water masers are associated with regions of very dense gas
($n \geq 10^{10}$ \cmv ; Elitzur et al.\ 1989). Each of the cores had
been mapped in the CS $J=5\rightarrow 4$ transition (Shirley et
al.\ 2002b) and detected in the CS $J=7\rightarrow 6$ transition (\ta\
$> 1.0$ K; Plume et al.\ 1992). The critical density of CS
$J=5\rightarrow 4$ is $n_c = 8.9\ee6$ \cmv; however, a
density, $n_{eff}$, of 2.2\ee6 \cmv\ will produce an observable
line of 1 K for a gas temperature of 10 K (Evans 1999). For a gas temperature of
100 K, which may better describe massive cores, the effective critical
density is even lower, $n_{eff} = 6.0\ee4$ \cmv\ (Evans 1999).  Consequently,
models constrained by multiple transitions are needed to determine
density. Plume
et al.\ (1997) reported $\langle {\rm log} (n) \rangle = 5.9$ from LVG models
of multiple CS transitions for the regions from which our sample was
taken.  Therefore, these objects were known to contain a significant
amount of dense gas; however, their mass and density structures were
not well known. Many (43\%) of the regions in our study were also known to
be associated with \uchii\ regions.

\begin{figure}
\plotone{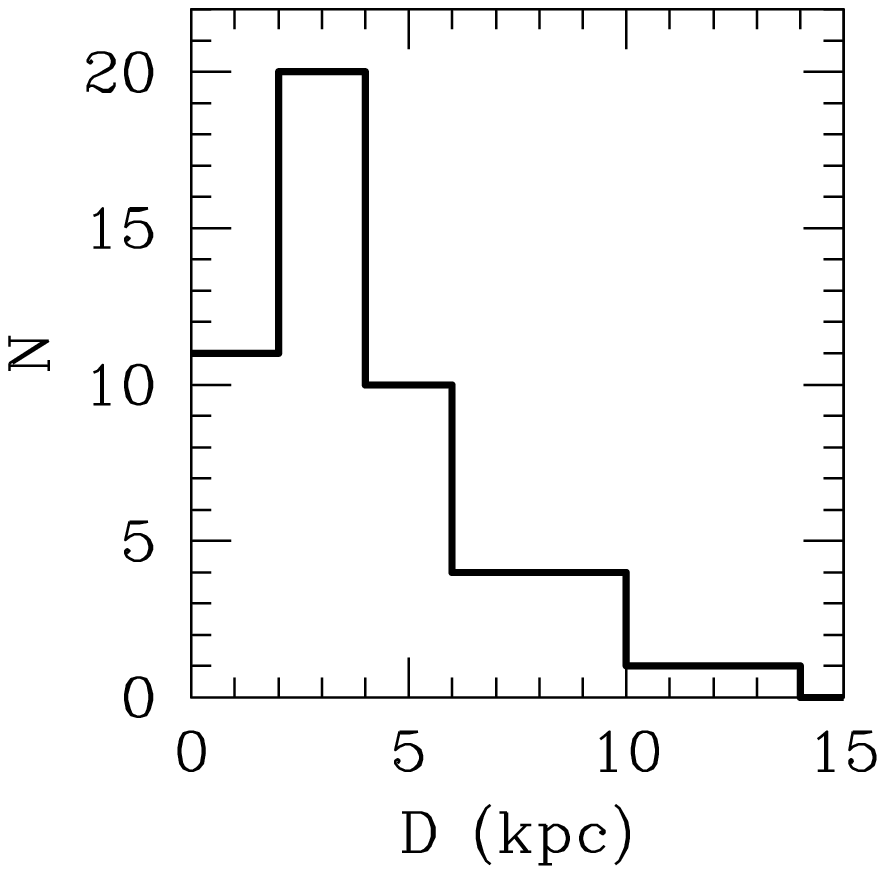}
\figcaption{\label{dist}
The distribution of distances in our sample. $N$ is the number of sources.}
\end{figure}

The sample covers a large range of distances, from Ori-IRC2 at 
a distance of 450 pc to G12.21$-$0.10 at 13.7 kpc. The distances were
found in the literature (see Table 1), 
and spectrophotometric distances were used when available.  
The mean distance in the sample is 3.9 kpc, while the median distance
is 2.8 kpc. Figure \ref{dist} shows the distribution of distances
in the sample.

Our sample includes nine objects described in van der Tak et
al.\ (2000). The van der 
Tak et al.\ (2000) sources were selected
to be luminous, visible from the Northern hemisphere, and, in most cases, 
bright in the mid-infrared. For
comparison, the sample of high mass protostellar objects studied by
Beuther et al.\ (2002) were selected from objects north of $-20$\degree\
declination detected in CS $J=2\rightarrow 1$ (Bronfman et al.\ 1996),
with far-infrared colors characteristic of \uchii\ regions.
Their sources are also bright at far-infrared wavelengths. However, 
the sources in the Beuther et al.\ (2002; also Sridharan et al.\ 2002) 
sample were also chosen to have low radio continuum flux ($<$ 25 mJy; 
Sridharan et al.\ 2002) to ensure that their sources were isolated and 
{\it not} typical \uchii\ regions. Surprisingly, 
our sample has only two objects in common with that of Beuther et al.\ 
(S231 and G12.89$+$0.49).

\section{Observations}

\subsection{Observation Techniques}

\begin{figure}
\epsscale{0.8}
\plotone{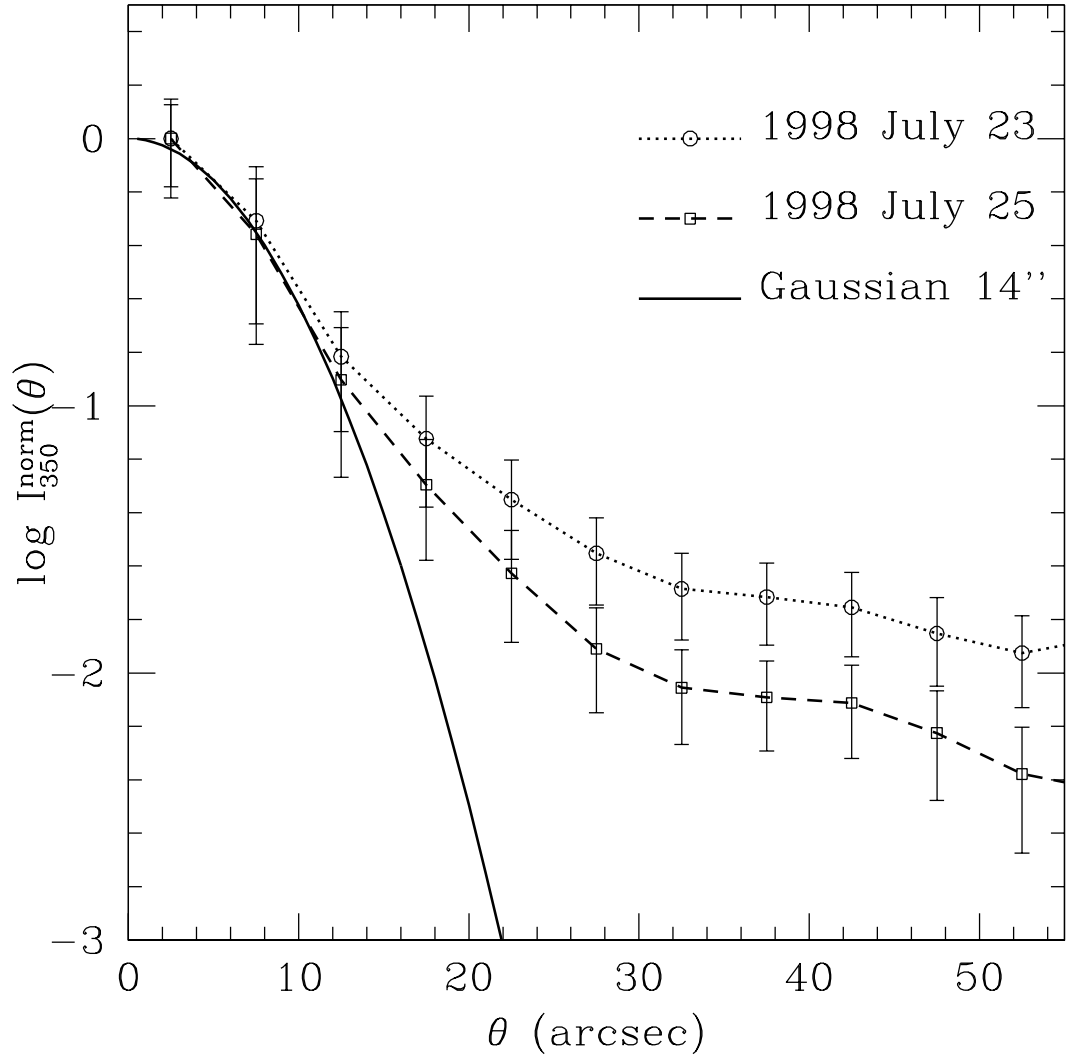}
\figcaption{\label{beamfig}
Radial profiles of Uranus on 1998 July 23 and 1998 July 25 used to measure the
beam plotted with a 14\as\ Gaussian. The error bars represent the weighted mean
variation in the annulus for which each point of the profile is calculated.
We adopt a FWHM beam size, $\theta_{mb}$, of 14\arcsec.}
\end{figure}

	Fifty-one regions forming massive stars were observed with
SHARC (the Submillimeter High Angular Resolution Camera), described by
Hunter,
Benford, \& Serabyn (1996), during 5 nights in 1997 (December 21 and 22)
and 1998 (July 15, 23, and 25) on the 10.4 m Caltech Submillimeter
Telescope.  SHARC is a one-dimensional bolometer array with a FWHM
beam size, $\theta_{mb}$, of 14\as\ (see Figure \ref{beamfig} and 
$\S$2.3).

	SHARC's linear array consists of 24 detectors (Hunter et
al.\ 1996); therefore, the telescope must be scanned in azimuth 
at constant elevation to map the source.  Each 350 \micron\ map 
consists of approximately 11 scans
extending 240\as\ in azimuth scanning at a rate of 4\as\ per second.
The individual scans are shifted by 4\as\ in elevation to extend
the mapped region and to eliminate gaps in the map due to bad pixels
(pixels 1, 5, 15, and 16 of the 24 pixels in the SHARC array).  The
scanning rate and elevation shifts were selected to be slightly
smaller than the size of the pixel, 5\as\, in the focal plane of the
array, to obtain better sampling (Hunter et al.\ 1996). 
The secondary was chopped at 1.123 Hz with a chop throw of
90\as\ to 100\as\ in the azimuth direction.  SHARC observations were
conducted only during very dry conditions with $\tau _{cso} <$ 0.06
(see \S 2.3 below).

\subsection{Image Reduction}

All of the data were reduced and restored with the standard 
program CAMERA. The restoration algorithm is based on a technique
described by Emerson, Klein, \& Haslam (1979). The despiking routine
was used on those maps that had pixels with spikes above 10 $\sigma$,
which we identified by visual inspection during reduction. The routine
replaces the spiked pixel with the average value of the adjacent
pixels. In some cases the source was highly peaked, so a higher sigma
was used to ensure the central pixel was not removed by the despiking
routine. The night of 1998 July 15 was unusually windy.  Five maps
made on this night (W28A2, G12.89$+$0.49, G12.21$-$0.10, G24.49$-$0.04,
W43S) showed signs of being affected by the wind and were corrected
with linear destriping. The destriping affected the maximum pixel
value, in most cases decreasing it by less than 10\%.

After the data were reduced with CAMERA, gray scale images of the
restored and combined bolometer maps were created. The Image Reduction
and Analysis Facility (IRAF) was used to find the value of the maximum
pixel, the average background, and $\sigma$ for each map. The voltages in
each map were multiplied by the extinction correction,

\begin{equation}
\label{extcorr}
V_{corr} = V_{obs} e^{\tau sec(z)}
\end{equation}

\noindent where $\sec(z)$ is the average airmass at the time of
observation. The determination of $\tau$ is described below in \S 2.3.

\begin{figure} 
\centering
 \vspace*{7.5cm}
   \leavevmode
   \includegraphics{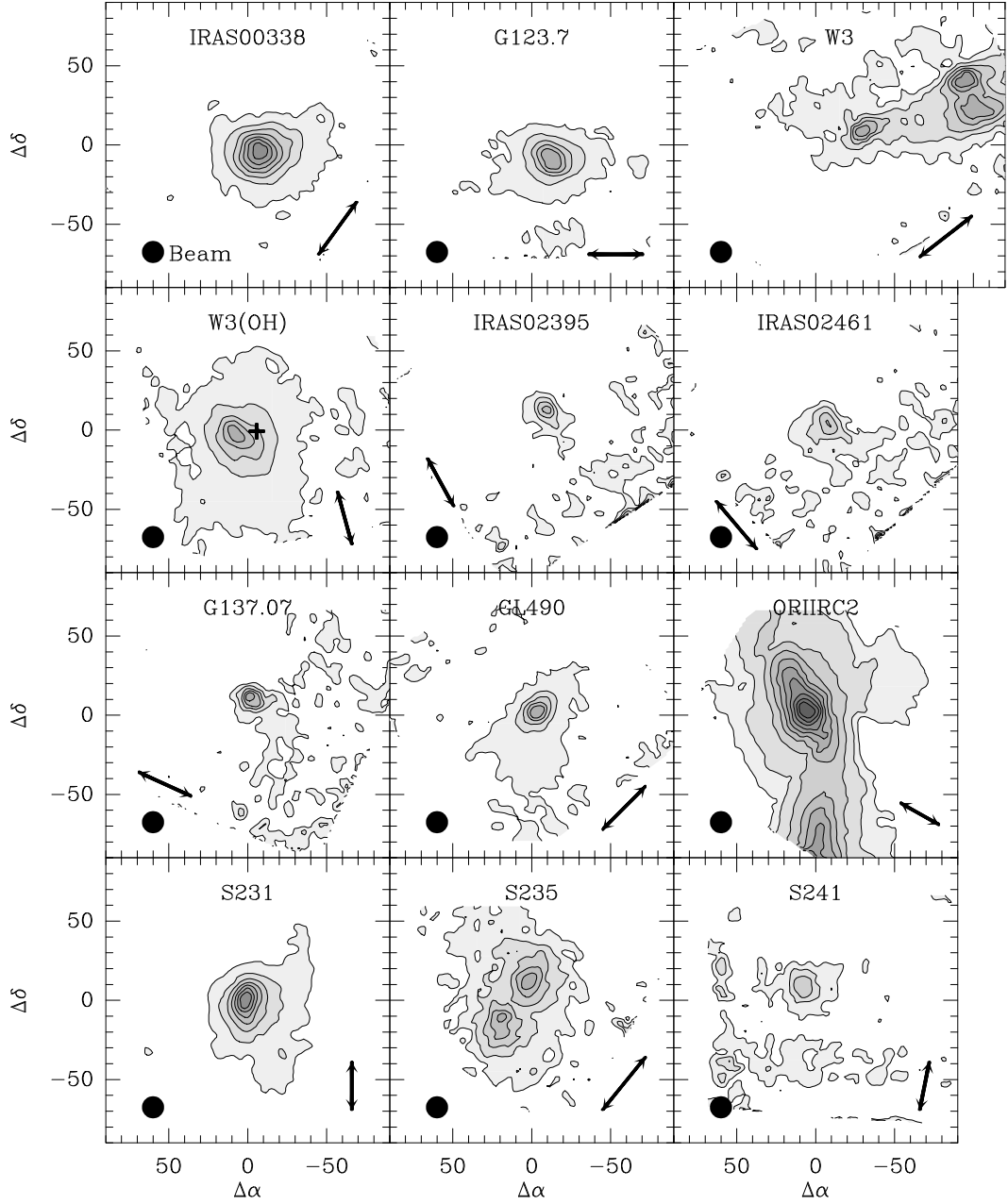}
\vskip 3.5in
\figcaption{\label{contfig}
350 \micron\ dust emission contour maps. The arrows indicate the 
chop direction but are not indicative of the chop length ($\sim$ 100\as). The 
plus signs indicate the positions of known \uchii\ regions (W3(OH): Wilner et al.\
1995).  
The contour levels are as follows:
IRAS 00338$+$6312 (4$\sigma$), 
G123.07$-$6.31 (3$\sigma$, then in increments of 20\% (10$\sigma$) of the peak),
W3 (4$\sigma$),
W3(OH) (3$\sigma$, 20\% (10$\sigma$)),
IRAS 02395$+$6244 (2$\sigma$),
IRAS 02461$+$6147 (2$\sigma$),
G137.07$-$3.00 (2$\sigma$),
GL490 (3$\sigma$, 20\% (6$\sigma$)),
Ori-IRC2 (3$\sigma$, 10\% (7$\sigma$)),
S231 (4$\sigma$),
S235 (2$\sigma$),
S241 (2$\sigma$).}
\end{figure}

\begin{figure}
\centering
 \vspace*{7.5cm}
   \leavevmode
   \includegraphics{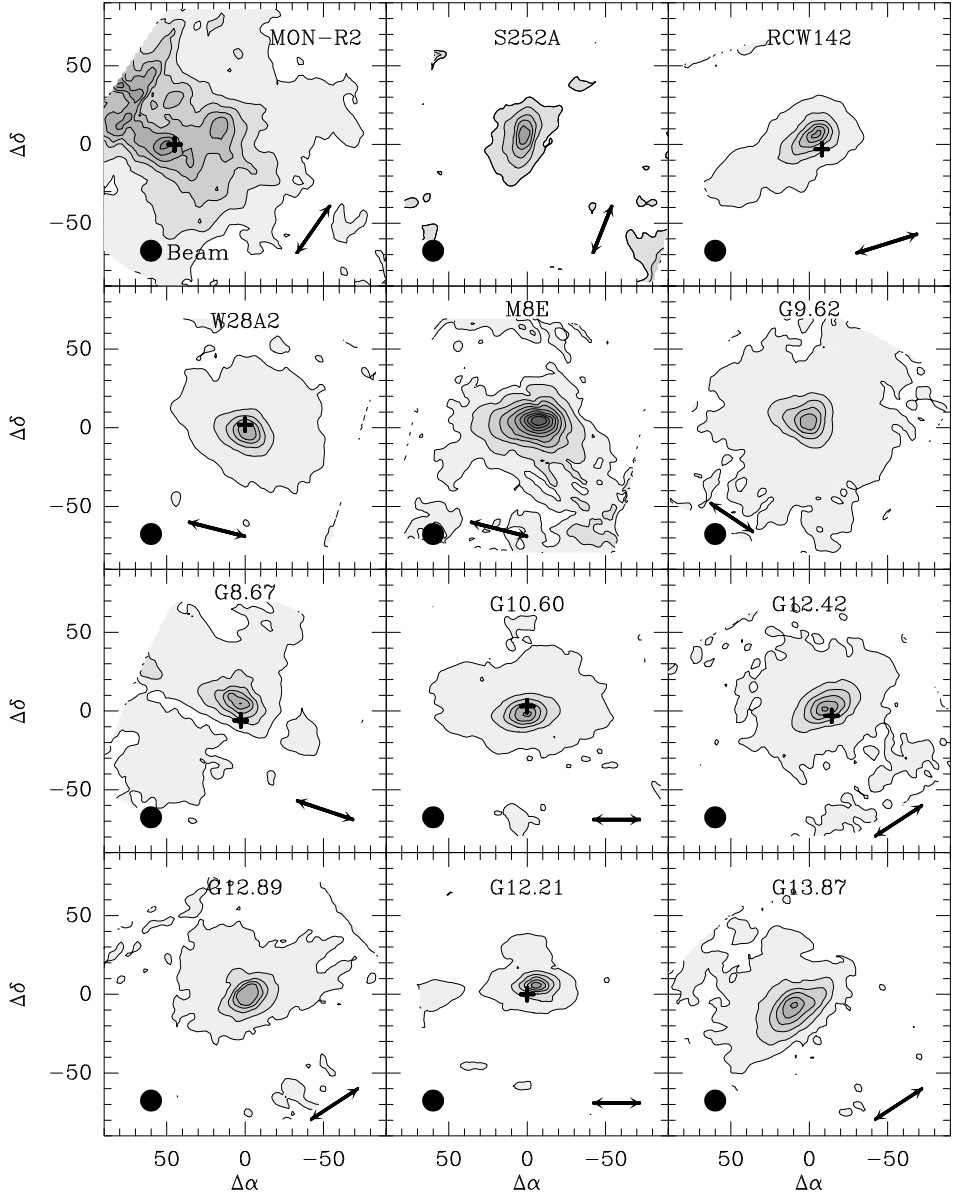}
\vskip 3.2in
\figcaption{\label{cont2fig} 
350 \micron\ dust emission contour maps. The arrows indicate the 
chop direction but are not indicative of the chop length ($\sim$ 100\as). The 
plus signs indicate the positions of known \uchii\ regions (MonR2, W28A2, G8.67, 
G10.60, G12.21: Wood \& Churchwell 1989; RCW142: Walsh et al.\ 1998;
 G12.42: Jaffe et al.\ 1984).   
The contour levels are as follows:
MonR2 (4$\sigma$), 
S252A (4$\sigma$),
RCW142 (3$\sigma$, then in increments of 20\% (26$\sigma$) of the peak),
W28A2 (3$\sigma$, 20\% (29$\sigma$)),
M8E (3$\sigma$, 10\% (5$\sigma$)),
G9.62$+$0.10 (3$\sigma$, 20\% (28$\sigma$)),
G8.67$-$0.36 (3$\sigma$, 20\% (25$\sigma$)),
G10.60$-$0.40 (3$\sigma$, 20\% (30$\sigma$)),
G12.42$+$0.50 (3$\sigma$, 20\% (17$\sigma$)),
G12.89$+$0.49 (3$\sigma$, 20\% (12$\sigma$)),
G12.21$-$0.10 (3$\sigma$, 10\% (20$\sigma$)),
G13.87$+$0.28 (3$\sigma$, 10\% (9$\sigma$)).
} 
\end{figure}

Figures \ref{contfig}--\ref{contend} show the dust continuum contour 
maps. The
contour levels are even multiples of $\sigma$ or 
10\% or 20\% of the peak signal with an additional lowest contour at
3 $\sigma$. The (0,0) position is the location of the water maser from
Plume et al.\ (1992, 1997). The positions of known \uchii\ regions are 
indicated by plus signs on the contour maps. Three of the 22 marked
\uchii\ regions (G12.21$-$0.10, G23.95+0.16, and W43S) were listed as 
\uchii\ regions by Wood \& Churchwell (1989) but, in the same study, were
 reported to have diameters greater than 0.1 pc. Therefore, these 
sources may also be classified at {\it compact} \HII\ regions.

Normalized radial
intensity profiles were created as in Shirley et al.\ (2000). The
intensity was azimuthally averaged and normalized to the peak emission.
The normalized intensities, $I(b)/I(0)$, were plotted versus the 
impact parameter, $b = \theta D$, a line of sight offset from
the center by an angle $\theta$, for a source at distance $D$. The
radial profiles are truncated at a radius, $r_{prof}$, when the signal
fell to 1 $\sigma$ or at 60\arcsec, whichever is smaller. We do not 
considervdata beyond a 60\as\ radius where simulation in
our models of the effects of chopping becomes problematic. 
Photometry was also provided by the radial profile
program. The sky-subtracted fluxes for 30\arcsec\ and 120\arcsec\
diameter apertures, $\theta_{ap}$, are listed in Table \ref{obstab}.

\subsection{Calibration}

The extinction coefficients at 350 \micron, $\tau_{350}$, were
determined using skydips from the CSO tipper ($\tau_{cso}$) at 225
GHz.  A scaling between $\tau_{cso}$ and $\tau_{350}$ has been
determined by comparing skydips between the 225 GHz and 350 \micron\
tippers,

\begin{equation}
\label{tau}
\tau_{350} = (23.5 \pm 0.2)\tau_{cso}
\end{equation}
 
\noindent(R. Chamberlin 2000, private communication). To check the relationship
between $\tau_{350}$ and $\tau_{cso}$, Uranus was observed as it set
during two nights in 1998 July with exceptionally stable sky opacity
($\sigma _{\tau }/\langle \tau \rangle \leq 0.06$).
 The variation of peak voltage on
Uranus with airmass determined $\tau_{350}$.  The resulting
$\tau_{350}$ was consistent with Equation (\ref{tau}).  Since we were
unable to observe a source as it set during each night, and since
$\tau_{350}$ may vary throughout the night, we determined the
extinction correction, Equation (\ref{extcorr}), for each image by
scaling from $\tau_{cso}$ measurements using Equation (\ref{tau}).

Full maps of Uranus and secondary calibrator sources, NGC 2071IR and
W3(OH), were used to determine the calibration factors, $C^{\theta}$,
for each observing run.  Sky subtractions were made for each image by
measuring the voltage (V$_{sky}$) through multiple 20\as\ apertures
($\theta_{sky}$) away from the source and averaging the measured sky
voltage.

Total voltages measured in apertures of diameter $\theta_{ap}$ were then
corrected for sky emission by
 
\begin{equation}
\label{skycorr}
V(\theta _{ap})_{corr} = V(\theta _{ap})_{obs} - V_{sky}
\frac{\pi\theta^{2}_{ap}}{\pi\theta_{sky}^2} .
\end{equation}

\noindent The calibration factors $C^{\theta}$ were calculated for
images taken in the 1998 July run by measuring the flux at 350
\micron\ in 30\as\ and 120\as\ diameter apertures of two maps of
Uranus.  The total flux of Uranus in 1998 July was 266.5 Jy.
Calibration factors for sources observed in 1998 July were, with
statistical errors, $C^{30} = (9.6\pm1.5)$ mJy V$^{-1}$ and
 $C^{120} = (8.5
\pm 2.0)$ mJy V$^{-1}$ for 30\as\ and 120\as\ apertures, respectively.
Because no planets were visible to serve as pointlike sources in 1997
December, maps of NGC 2071IR and W3(OH) were used for calibration.  The
assumed fluxes for NGC 2071IR and W3(OH) were 177 Jy beam$^{-1}$ and 498
Jy beam$^{-1}$, respectively (Sandell 1994).  Voltages measured in 30\as\ and
120\as\ apertures in three maps of each secondary calibrator were used
to determine calibration factors.  The 1997 December calibration
factors averaged over the six maps were, with statistical errors,
$C^{30} = (9.6\pm1.3)$ mJy V$^{-1}$ and $C^{120} = (4.7 \pm 0.7)$ 
mJy V$^{-1}$.
While some of the statistical uncertainties are smaller than 20\%, we
expect systematic errors to not allow fluxes to be determined at 350
\micron\ to better than 20\% (Hunter et al.\ 2000).  Furthermore,
calibrations in 1997 December determined from NGC 2071IR and W3(OH) are
expected to be worse since neither calibrator is a point source (see Fig. 
\ref{contfig} for a map of W3(OH)).
This effect is obvious in $C^{120}$, which is nearly a
factor of two lower when using the secondary calibrators. $C^{30}$ was
not affected by the extended emission and is stable from 1997 December
to 1998 July. For this reason, $C^{120}$ from 1998 July is used to
calibrate all of the 120\arcsec\ aperture fluxes.

Radial beam profiles were obtained from two maps of Uranus in 1998 July
(Figure \ref{beamfig}). The data were binned in 5\as\ bins.
Broad sidelobes are seen beyond 15\as.
At 40\as, the sidelobe power is as high as $-17$ dB.  The general
shape of the two profiles is consistent, but there are variations
in the strength of the sidelobes. Fitting the beam profiles with a 
Gaussian, we find that the FWHM is at least 14\as\ which is significantly
different from the beam size reported in previous studies using SHARC
at 350 \micron\ (10\as, van der Tak et al. 2000; 11\as, Hunter et al.\ 2000). 
The difference could be the result of poor focus or temporal changes in the beam. 
Young et al.\ (2002) report a 3\as\ change in the FWHM of the
James Clerk Maxwell Telescope beam at 450 \micron\ in the course of a 
single night. The beam measurements were not frequent enough to
characterize the changes of the CSO beam over the course of 
the observations in this study; therefore, we adopt $\theta_{mb}$ = 14\as.

\section{Results}

Most of the cores appear slightly elongated in the contour maps
(Figures \ref{contfig}--\ref{contend}). This is likely an effect of 
chopping during
observations as the maps are usually elongated along the chop
direction. The chop direction is indicated on the maps with an arrow.
However, some of the sources have extended asymmetrical
emission at 350 \micron\ distinct from the chopping asymmetry with an
intensity of a few $\sigma$ (e.g., ON2S, G40.50$+$2.54). Some (14\%) of the
maps have double or multiple peaks, indicating the presence of more
than one luminosity source.  These sources are noted
in Table \ref{obstab}. A few have close embedded double peaks
(e.g., G23.95$+$0.16, S235), while others show more spatially-distinct
peaks. For example, the NGC 7538 region shows 3 distinct peaks in 
Figure \ref{contend}; these were mapped separately, but 
are plotted in their relative locations.

The contour maps also show that the 350 \micron\ dust peak is often
nearly coincident with the water maser position at the center of the map. 
Thirty cores (59\%) have the dust centroid within $\theta_{mb}/2$ 
of the maser position (see Table \ref{obstab}). The mean absolute 
distance of the dust centroid to the maser position is 8\as.
About half of the 22 known \uchii\ regions in the sample 
are also close ($< \theta_{mb}/2$) to the maser, 
however, the mean absolute distance of the 
\uchii\ position from the water maser is 10\as. This separation is similar to 
that of the \uchii\ region from the dust peak with mean absolute 
distance of 11\as. Samples based on water masers favor an earlier phase
of star formation than \uchii\ region samples (Cesaroni et al.\ 1988, 
Shirley et al.\ 2002b). The high coincidence of the dust centroid and maser
position in these regions implies that the dust emission may be 
primarily tracing the earlier stages as well.

The FWHM size of each source, $\theta_{dec}$, was determined by
deconvolving the telescope main beam from the observed FWHM of the
core by subtracting $\theta_{mb}$ from the observed FWHM in quadrature.  
The observed FWHM was determined from the radial profile of each source.
The deconvolved half-power radius is defined to be $\rdust =
(D/2)\theta_{dec}$, where $D$ is the distance to the source. For the entire
sample, $\langle \rdust \rangle = 0.16 \pm 0.10$ pc, with a median of 
0.14 pc. Figure \ref{rfig} shows the distribution of \rdust. 

Photometry from the literature is collected in Table \ref{sedtab}.
The observed spectral energy distribution (SED) was used to 
calculate the bolometric luminosity, \lobs, for each source, 
using only data taken with beam
sizes $\ge$ 20\as, with
the exception of the endpoints to the SED. 
Luminosities range over
three orders of magnitude, about 10$^3$ to 10$^6$ \lsun\ with
$\langle \lobs \rangle = 2.5\ee5$ \lsun\ and a median $\lobs = 5.1\ee{4}$
\lsun. If all the luminosity is attributed to a single star, the 
range of spectral types would be B3 to O4, with the median being an O9
star. Considering multiple sources and   
accretion luminosity would lower these spectral types.

Many recent studies have used power laws to fit directly the radial intensity
profiles and infer the density structure of star forming regions
(e.g., Shirley et al.\ 2000, Beuther et al.\ 2002). We did not use this
technique; however, for comparison purposes, we
describe some general trends in the radial profiles of our
sample. Generally, the radial profiles follow a power law from about
12\arcsec\ (approximately $\theta_{mb}$) to 40\arcsec. The profiles are
flattened toward the interior, and beyond 40\arcsec, they sometimes
deviate from the power law. The flattening at small angles is likely 
due to beam effects, but could also result from fragmentation of the 
core. The change
in the slope beyond 40\arcsec\ is not consistently steeper or
shallower. Some radial profiles are 
distorted by the presence of multiple peaks.

\section{Models}

One of the major motivations for this study was to learn what
density distributions fit the data. A power law density
distribution was assumed of the form $n(r) = n_f (r/r_f)^{-p}$, for $p$
in a range of 0.5 to 2.5. We chose $r_f = 1000$ AU for convenient
comparison to other studies, but this $r_f$ is well inside our beam.
The values of $n_f$ should be taken only as indicative of likely mean
densities as substructure is very likely on those scales.
The observed radial intensity profiles and SEDs were modeled using 
a modified version of the 
one-dimensional dust
continuum radiative transfer code by Egan, Leung, and Spagna (1988)
and an observation simulation code described by Evans et al.\ (2001). 
The radiative transfer code calculates the radial temperature distribution,
$T_D(r)$ self-consistently for each input model of $n(r)$. 

We also included contributions from the interstellar
radiation field (ISRF) to the temperature distribution at the edge of
the cloud. Figure \ref{tempfig} shows that $T_D(r)$ is approximately a
power law.  At small radii, $T_D(r)$ deviates by rising more steeply
towards the center than a strict power law.  Additionally, the ISRF
causes an upturn in $T_D(r)$ at large radii (about 1 pc for a
10$^4$ \lsun\ source). The ISRF makes a substantial contribution to
the temperature profile in regions forming low mass stars where the
internal luminosity is low (Shirley et al.\ 2002a, Young et al.\ 2002)
but has little effect in regions forming massive stars because the
temperature profile is dominated by the embedded source. While these
sources may exist in regions of enhanced ISRF, the effects on the
models are negligible even in our least luminous sources ($L_{bol} 
= 10^3$ \lsun) unless the ISRF is a factor of 10
stronger than the standard value (see Evans et al.\ 2001 for a plot of
the ISRF). For a more typical source with $L_{bol} = 10^4$ \lsun, the
ISRF field must be at least 200 times stronger to change the best fit
$p$ by 0.25.

For each of our models, the input density,
$n_f$, was normalized so that the model flux at 350 \micron\ matched
our observations, given a dust opacity at 350 \micron.
Dust opacities were adopted from column 5 of the table in 
Ossenkopf and Henning (1994; hereafter OH5), which were calculated for coagulated 
dust grains with ice mantles. These dust opacities (OH5) have been previously
shown to match observations of massive star formation regions by van
der Tak et al.\ (1999, 2000) as well as low mass star forming regions
(Evans et al.\ 2001, Shirley et al.\ 2002a, Young et al.\ 2002). 
We also considered opacities for grains without ice
mantles taken from column 2 of Ossenkopf \& Henning (1994; OH2) 
because temperatures in regions with young stars may be high
enough to destroy the mantles. The OH2 opacities are higher at $\lambda > 350$
\micron\ ($\kappa_{OH2}/\kappa_{OH5} = 1.8$ at 700 \micron), but lower
at shorter wavelengths. The crossover point is near 350 \micron, with
$\kappa_{OH2}/\kappa_{OH5} = 1.1$ at 350 \micron\ (see Fig. 2 of Evans
et al.\ 2001). Using OH2 opacities thus
results in best fit $n_f$ lower by about 10\%.
Using OH2 opacities also resulted in higher model fluxes at long wavelengths,
but lower fluxes at short wavelengths.
No difference was found in the best fit $p$ between the OH2 and OH5 
opacities. 

For each value of $p$ and $n_f$, other
model input parameters were adjusted to fit the observed values. 
The temperature of the star
was taken to be the value corresponding to a star with $L = \lint$
(Thompson 1984). However, the results are very insensitive to the
stellar temperature (van der Tak et al.\ 2000). 
The outer, $r_o$, and inner, $r_i$, radii for each source are based on the
angular extent of the observed radial profile and
the chopper throw.  We set $r_o$ to be the sum of 
the chopper throw and twice the extent of the observed profile.
In particular, $r_o$ is large
enough to allow simulation of chopping.
The inner model radius was taken to be $r_o$ divided by
1000, so that it is small enough to be unresolved. The model's
sensitivity to the radii was tested in M8E (see \S 4.1 below).
 The best fit values were found to be insensitive to both
inner and outer radii (see \S 4.1, Shirley et al.\ 2002a, Young et
al.\ 2002).

The observation simulation program uses the
temperature distribution output from the radiative transfer code for
the density model being tested. The code calculates observed fluxes and
luminosities and generates a radial profile of normalized
intensity. The model is convolved with the observed
beam and chopping is simulated to produce a more realistic
radial profile for comparison with observations. For our models, an average
one-dimensional representation of the actual beam was used (Fig.\ \ref{beamfig}).  
The fit of the models with the
observations was quantified by calculation of the reduced chi squared
($\chisq$).
The $\chisq$ value for the radial profile follows the definition in 
Evans et al.\ (2001) and is denoted \chisqsharc.
We also compute a \chisq\ value for the fit to the SED, denoted
\chisqsed.
The shortest wavelength point was generally left out of
the calculation of \chisqsed, because the model
underestimates the flux at $\lambda \le$ 30 \micron. This effect is
well known in spherical models that do not account for holes and
inhomogeneities in the cloud that allow these wavelengths to escape
(e.g., van der Tak et al.\ 2000).
  
The modeling scheme and the dependence of the derived quantities on the 
model parameters are discussed by Evans et al.\ (2002) and the sensitivity
of the best fit $p$ to uncertainties in other parameters is quantified
for low-luminosity regions by Shirley et al.\ (2002a). They found that
the largest source of uncertainty in $p$ is the strength of the interstellar
radiation field. For the luminous sources studied here, this is a minor
source of uncertainty, as already noted.

\subsection{M8E: A Model in Detail}

M8E was used as a test case for checking the effects of the input
parameters on the models. The 350 \micron\ contour map of M8E is shown
in Figure \ref{cont2fig}. A spectro-photometric distance of 1.8 kpc
(Blitz, Fich \& Stark 1982) was used. The observed bolometric
luminosity, \lobs, was 1.47\ee4 \lsun. The best fit model
yielded $\lbol = 1.45\ee4\ \lsun$, calculated from convolving the
model emission with the beams used for the observations. The
sensitivity of the model to the input internal luminosity was tested
by decreasing and increasing the parameter by a factor of two. While
changing the luminosity affected the model SED by increasing or
decreasing the flux at certain wavelengths, it had no effect on the
radial profile fit.

The fiducial density, $n_f$, was fixed by
matching the model flux at 350 \micron\ to the observed value for each
modeled density distribution.
The best fit $n_f = 1.2$ $\times$ 10$^7$ cm$^{-3}$ for M8E. 
For M8E, $r_o$ = 4.3 $\times$ 10$^5$ AU (2.1 pc) and $r_i$ = 430 AU.
 To test the sensitivity of the
models to these parameters, $r_i$ was doubled and, independently,
$r_o$ was halved. Both tests yielded the same best fit $p$ as the
model with $r_i$ and $r_o$ calculated in the standard way. Other
resulting model values, such as $\lbol$ and $T_{bol}$, also
did not vary significantly. Therefore, we concluded that the models
are insensitive to changes of a factor of two in $r_i$ and $r_o$.

Figure \ref{m8efig} shows the best fit model for this source, a
density power law with $p = 1.75$ ($\chisqsed = 13.5$, $\chisqsharc
 = 0.52$).
As with a few of our sample, the best fit density
distribution to the radial profile did not agree with the best fit to
the SED for M8E. The \chisqsed\ was lower 
(8.3) with  $p = 1.5$ and a lower fiducial density and central
temperature. However, the slope of the model radial profile is clearly
too shallow ($\chisqsharc = 4.3$, shown as a dashed line in
Fig.\ \ref{m8efig}). With M8E and the other sources that showed a
disagreement in the best fit of the SED and radial profile, we report
the best fit $p$ to the radial profile because the shape of the opacity
law and density inhomogeneities strongly affect the SED at wavelengths
where the dust is optically thick. The radial profiles are much more
effective in constraining the value of $p$. The $p = 2.0$ model (dotted 
line in Figure \ref{m8efig}) produced a slope very similar to what was
observed, but was not a best fit because the radial profile steepens
at a smaller radius than observed. 

These models illustrate
(Fig.\ \ref{m8efig}) a conservative uncertainty in $p$ of $\pm$0.25; while $p =
1.5$ or 2 clearly do not fit the data as well as $p = 1.75$, they do
lie at the edges of the error bars in the data. 
The $\chisqsharc$ for models with $p$ = 1.6--1.9 were near
or less than one. The models were quantitatively distinguishable for
$\delta p = \pm 0.2$ ($\chisqsharc = 2.4$ for $p = 1.55$).
The limiting uncertainty in $p$ appears to be the signal-to-noise of the
profiles, rather than any systematic effect. However, this uncertainty
assumes that a power law density profile is a reasonable model; many sources
are known to be more complex on smaller scales.

Another source of uncertainty is the knowledge of the beam profile. 
Shirley et al.\ (2002a) tested the effects of the James Clerk Maxwell 
Telescope (JCMT) beam uncertainties and found $\delta p = 0.1$.
As were all sources, M8E was modeled with the 
average of the two observed beam profiles shown in Figure \ref{beamfig}. 
For M8E, we also tried each of the two beam profiles, which were 
observed on different nights. 
Using the individual beam profiles did not change the result
($\delta p < 0.05$) for M8E.
For these data, the beam uncertainties are negligible compared with 
those in the data, as long as the measured beam is used. If the beam
is represented by a Gaussian, the effects are much larger (\S 6.1).

\subsection{Model Results}

Of our sample of 51 regions forming massive stars, 31 could be modeled.
We required that the source was not confused with
multiple sources, that the map had high signal-to-noise, 
and that a range of flux density measurements were
available in the literature. Most of our modeled sources are well resolved,
 i.e., the {\it deconvolved} source size is at least the beam size, 
$\theta_{dec}/\theta_{mb} \ge 1$. We also consider some smaller sources to 
be resolved and require $\theta_{dec}/\theta_{mb} \ge 0.8$ (or a source 
with an observed FWHM of 1.3 times the FWHM of the beam) for modeling. 
Table \ref{modtab}
lists $\theta_{dec}/\theta_{mb}$ for the sample. Three modeled 
sources have $0.8 < \theta_{dec}/\theta_{mb} < 1$ (W28A2, W43S, G31.41$+$0.31). 
The model results for these sources show steep density profiles ($p$ = 
2.25--2.5) and should be considered less robust than
the rest of the sample, because they may not be resolved due to uncertainties 
in the beam. However, the steeper profiles may also be real;
 in $\S$6.3 we discuss the correlation between
$p$ and $\theta_{dec}/\theta_{mb}$. Because of this correlation, our models 
are biased against sources with $p > 2.5$ which would not be considered 
resolved. The requirement on $\theta_{dec}$ also produces a bias
against small sources. For a source at the median distance of 2.8 kpc,
sources with $\rdust < 0.14$ pc would not qualify for modeling. If 
the minimum 350 \micron\ flux density of our modeled sample and the median 
distance are assumed, the minimum required mass for our models is 61 \msun.
The sources with small angular extents were also in some cases
the most luminous and distant sources (e.g., G12.21$-$0.10 with $L_{bol}$ = 
5.5\ee5 \lsun\ and $D = 13.7$ kpc).  

Figures \ref{modfig} and \ref{modend} show the best fit model with the observed 
SED and radial
profile for a sub-sample of sources that illustrate the full range of 
best fitting $p$ values. Figure \ref{phist}
shows a histogram of the best fit $p$ values. The mean and standard
deviation of the histogram are $\langle p \rangle = 1.8 \pm 0.4$.
The standard deviation is about twice our estimated uncertainties on a
single fit, providing marginal evidence for a range of actual values of
$p$. Figure \ref{phist} also shows the distribution of densities
at 1000 AU, $n_f$. For this sample, $\langle n_f \rangle$ = 1.2\ee8 \cmv\
 and the median value is 1.4\ee7 \cmv. In Section 6.3, we compare the distributions 
of $p$ and $n_f$ to those found in studies of low mass star formation.  

The models generally fit the observed radial profiles very
well. The value of \chisqsharc\ for the best fit models was less
than one for nearly all the models. The majority of the \chisqsed\ values
were under 10 over the range of wavelengths where they were
computed. For reasons described in \S 4, the model typically underestimates
the emission at shorter wavelengths, where the dust is becoming 
optically thick.
Models with density inhomogeneities (clumps, cavities, etc.) or flattened
structure might match the emission at shorter wavelengths, but such models
introduce many free parameters.  

Recent studies of low mass star forming cores using the same modeling
techniques as presented here have found a correlation between $p$ and
the aspect ratio of the core (Shirley et al.\ 2002a and Young et
al.\ 2002). Because our source shapes were affected by chopping, we
cannot measure reliable aspect ratios in enough sources to make this
comparison.

The modeled density power law exponent can also be affected by the 
inclusion of a compact central source, such as an \uchii\ region. For 
low mass star forming cores, the inclusion of a compact source, in that
case a disk, decreased $p$ for the envelope by 0.5 (Young et al.\ 2002). 

The \uchii\ region at the position of the water maser in W3(OH) has a 
3 mm flux density of 3.5 Jy (Wilner et al.\ 1995). 
For a conservative
upper limit on the amount of flux the \uchii\ region contributes at 350 \micron,
we have assumed 
that all of the 3 mm flux is due to ionized gas rather than dust and that the 
\uchii\ region is optically thick at 350 \micron\  
so that $S_{\nu} \propto \nu^2$.  These assumptions give  
the maximum contribution of an \uchii\ region 
at 350 \micron\ as 23\% of the total observed flux. Including a compact source
with this maximum contribution 
in our dust model, assuming that all of the emission from the \uchii\ region 
is included only in the central beam of our observations (as in Young et al.\
2002), steepens the radial profile and, therefore, steepens the best fit $p$
to the radial profile. For W3(OH), $p$ increased by 0.3. 
The result is that the possible presence of a compact source introduces uncertainty 
into the models, because the modeled density power law of the envelope is steeper 
than the density structure if there were no compact source. 
We conclude that a \uchii\ region introduces 
uncertainty in our model results for the density structure of the envelope, 
$\delta p = -0.3$, if it contributes more than 20\% of the total 350 \micron\ 
flux.

Rick Forster at the Berkely-Illinois-Maryland Array (BIMA) generously
provided recent 3 mm flux densities for several \uchii\ regions near the 
center of cores in this study. Three of the four sources have negligible
contribution from the \uchii\ region at 350 \micron.  Therefore, 
the simple presence of a \uchii\ region does not 
necessarily imply an uncertainty in $p$. 

Our results for the density structure of massive star forming regions can be 
compared with theoretical predictions.
The mean value of $p$ (1.8) is incompatible with a logatropic sphere 
($p$ = 1; McLaughlin \& Pudritz 1997) at about a 2 $\sigma$ level. 
However, both $p$ = 1.5 and $p$ = 2 are possible, so the inside-out collapse
model of star formation (Shu 1977) cannot be ruled out. While a changing opacity 
as a function of radius might affect the values of $p$ and this conclusion, the 
opacity of a grain with (OH5) and without (OH2) an ice mantle 
is nearly the same at 350 \micron, so evaporation of mantles will 
not have a large effect (see \S 4).  

\section{Masses}

\subsection{Integrated Mass}

In a power-law mass distribution, the mass can only be defined within
some specified radius. For sources with models, we can compute the 
integrated mass
within a particular radius, $R$, from

\begin{eqnarray}
\label{mint}
M_{int} & = & 4 \pi \mu m n_f r_f^{p} \int^R_0 r^{2-p} dr \nonumber\\
& = & 4 \pi \mu m n_f r_f^{p} \frac{R^{3-p} - r_i^{3-p}}{(3 - p)} \;\; p < 3 \;\;,
\end{eqnarray}

\noindent where $\mu = 2.3$, $m$ is the hydrogen atom mass, 
$n_f$ is the gas density at $r_f$ (1000 AU), and $p$ is the best fit density
power law exponent. These masses refer to the total mass of gas and dust 
within $R$. The mass is proportional to $R^{3-p}$; for the mean value of $p$, 
$M_{int} \propto R^{1.2}$.

One choice for $R$ is \rdust, the radius corresponding to the deconvolved
source size. While this size has no intrinsic meaning if the density 
distribution is truly a power law, it is a fiducial size that is model
independent.  For the modeled sample, the integrated masses (Table
\ref{masstab}) within $\rdust$ range from 6 to 1500 \msun\ and  
 $\langle r_{dec}^{mod} \rangle = 0.16 \pm 0.09$ pc, the same as 
for the complete sample; 
 $\langle M(< \rdust) \rangle = 250 \pm 380 $ \msun\ with a median value
of 120 \msun, and 
$\langle {\rm log} (M(< \rdust)) \rangle = 2.0 \pm 0.6$ for this subsample.

A second, more physical choice for $R$ is the radius inside which the
density is actually enhanced over the surrounding cloud. This choice would
define the total mass of the actual core, but we generally lack information for
particular sources on the ambient density. 
In a study of extended cloud conditions in regions
of massive star formation, including some studied here, Allers et al.\
(2002) find a typical ambient gas density of $n \sim 10^4$ \cmv. 
The models for cores allow determination of \rn, defined to be the
radius at which $n = 10^4$ \cmv. 
For the modeled sample of 31 sources, $\langle \rn \rangle = 0.42$ pc, about
2.5 times \mean{\rdust}.
Setting $R = \rn$ yields a core mass (\mden) given in Table \ref{masstab} for
each core. Averaged over the cores with models,  $\langle \mden
 \rangle$ = 720 $\pm$ 860 \msun\ and $\langle {\rm log} (\mden)
\rangle$ =  2.5 $\pm$ 0.6. 
Figure \ref{masshist} shows the distribution of sizes (\rdust\ and $r_n$) and 
integrated masses ($M(< \rdust)$ and \mden) for the modeled sample.

\subsection{Isothermal Mass}

In order to estimate the mass for those sources without models, and
hence a best fit $p$, the measured flux density was used to calculate
a mass by assuming a single representative temperature. The
isothermal total mass, $M_{iso}$, was calculated according to the equation:

\begin{eqnarray}
\label{misoeq}
M_{iso} & = & \frac{S_{\nu} D^2}{B_{\nu} \kappa_{\nu}} \nonumber\\
& = & 5.09 \ee{-8} \msun\ S_{\nu} ({\rm Jy}) D^2 ({\rm pc}) 
(e^{41 K / T_{iso}} - 1),
\end{eqnarray}

\noindent which assumes a single dust temperature,
$T_{iso}$. S$_{\nu}$ is the observed flux at 350 \micron\ in a 120\as\
aperture and $D$ is the distance. We assumed the OH5 opacity,
$\kappa_{\nu} = 0.10$ cm$^2$ g$^{-1}$ of gas and dust at 350 \micron. 
Using the modeled sources with integrated
masses, $M_{int}$, within a 120\as\ aperture, 
we can calculate what assumption about dust temperature
would give the best agreement with the masses from the models. The
``isothermal temperature", T$_{iso}$, is given by the equation:

\begin{equation}
\label{t}
T_{iso} = \frac{h\nu/k}{{\rm ln}(1 + (2h\nu^3 \kappa_{\nu} M_{int}
(\msun)) / (c^2 S_{\nu}({\rm Jy}) D^2 ({\rm pc})))}
\end{equation}

\noindent (Shirley et al.\ 2002a). For all the other cores, we assumed
the mean temperature derived from the modeled sources, $\langle
T_{iso} \rangle$ = 29 $\pm$ 9 K. 
This method allowed us to estimate the
masses for the complete sample (see Table \ref{masstab} and Figure
\ref{rfig}). We find $\langle M_{iso} \rangle = 2020 \pm 4410$ \msun,
a median $\miso = 400$ \msun, 
and $\langle {\rm log} (M_{iso}) \rangle = 2.8 \pm 0.7$. The distribution
of $M_{iso}$ is skewed to lower masses, but it has a tail of very massive
cores.

\subsection{Mean Column Density}

The mean column density, $\Sigma$, has been used 
by McKee \& Tan (2002) (and Tan \& McKee 2002) to derive fundamental 
physical quantities in regions forming massive stars. The observed
$\Sigma$ is a key parameter in the determination of the mean pressure,
accretion rate, and star formation time (McKee \& Tan 2002). Tan \& McKee
(2002) use the virial masses and radii from Plume et al.\ (1997) and 
report $\Sigma_{vir} \approx$ 1 g cm$^{-2}$. For comparison, we calculate 
$\Sigma$ using the masses derived from the dust continuum for our sub-sample
of Plume et al.\ (1997). For the modeled sample, using $M(<r_n)$
 and $r_n$, $\Sigma_{mod} = 0.19 \pm 0.12 $ g cm$^{-2}$ which is significantly
 lower than $\Sigma_{vir}$ (Tan \& McKee 2002). However, expanding the 
calculation to the complete sample results in a more consistent value.
Using the isothermal mass, $M_{iso}$ and 
$r_n$ (or $\langle r_n \rangle$ for the sources that were not modeled), 
$\Sigma$ = 0.73
$\pm$ 1.7 g cm$^{-2}$. The mass accretion rate, $\dot{m_*}$, varies as 
$\Sigma^{3/4}$ 
and the star formation time, $t_{*f}$, as $\Sigma^{-3/4}$; therefore, 
a lower $\Sigma$ 
decreases the $\dot{m_*}$ and increases the $t_{*f}$ given by McKee \& Tan (2002).

It is important to note the inverse dependence of the masses and the 
mean column density on the opacity. Shirley et al.\ (2002b) compare the
masses determined from the density distributions reported here with 
virial masses and find that the virial masses are larger on average by 3.4. 
While this is quite good agreement considering that the opacities from 
different dust models can vary up to a factor of 10 (Ossenkopf \& Henning 
1994), the masses and mean column density may be a factor of three larger.

\section{Discussion}

\subsection{Comparison with Previous Models}

The average density distribution exponent, 1.8, is significantly
higher than the 1.0 to 1.5 reported by van der Tak et
al.\ (2000). Steeper best fit power laws were also found
for all of the sources in common. The van der Tak et al.\ (2000) best fit
models were based on molecular line emission observations, and were
not necessarily the best fits to their dust continuum emission (see their
Figure 8). However, there is still a discrepancy between their modeled
350 \micron\ radial profiles and our models even though the same method was
used.  The difference stems from
convolving the model with an observed beam rather than a 10\as\ Gaussian, as
was done with the van der Tak et al.\ (2000) models. Figure \ref{gaussfig}
compares a modeled radial profile for GL2591 using a 10\as\ Gaussian and the
observed beam with $p = 2.0$ and all other parameters the same. The
Gaussian beam requires a much shallower density power law to fit the
observations. In the case of GL2591, van der Tak et al.\ (2000)
reported a best fit based on molecular line emission of $p = 1.0$ and show
that the dust emission radial profile would be fit by $p \sim 1.25$ (see 
their Figure 8). We find $p = 2.0$; this was the largest discrepancy caused
by the beam for the overlapping sample. On average, van der Tak et al.\ (2000)
report a shallower power law by 0.4.  If a 14\as\ Gaussian, which we 
find better describes the beam at the time of our observations, is used, 
the difference is also $\delta p = -0.75$ for GL2591; the main effect is 
in the sidelobes.

\subsection{Comparison with Other Studies}

\subsubsection{Hatchell et al.\ 2000}

Hatchell et al.\ (2000) modeled the 450 and 850 \micron\ dust emission
for five massive star forming regions selected by emission from dense
molecular gas. Three of their sources are 
included in this study (G12.21$-$0.10, G31.41$+$0.31 and G13.87$+$0.28). 
Hatchell  
et al.\ (2000) fit the SED and radial profile at 450 and 850 \micron\ using
the dust radiative transfer code DUSTY (Ivezi$\acute{c}$ \& Elitzur 1997).
For G13.87$+$0.28, the only modeled source in common, Hatchell et al.\ (2000) 
report $p$ = 1.5, which is shallower than our best fit $p$ = 1.75. However, 
although $p$ = 1.0 and 2.0 were ruled out by Hatchell et al.\ (2000), 
intermediate $p$ values were not investigated in their study. 

Hatchell et al.\ (2000) find substantially higher masses (by a few times
ten) than in our study. This discrepancy is due to the use of 
different dust properties. Hatchell et al.\ (2000) used Draine \& Lee (1984) 
ice-free graphite and silicon dust grains as opposed to OH5 coagulated 
grains with ice mantles. The Draine \& Lee (1984) opacities are much smaller
and result in higher mass estimates.

Hatchell et al.\ (2000) included compact central cores in some of their models
to fit the radial profiles. They found that $p$ = 1.5 density distribution
was too shallow, but a central core with a $p$ = 1.5 envelope fit the data.
Hatchell et al.\ (2000) used the core model for G12.21$-$0.10 and G31.41$+$0.31, 
because the 450 and 850 \micron\ radial profiles fell off steeply within 20\as. 
The 350 \micron\ radial intensity profiles for these sources are also 
steeper than average. We report $p$ = 2.25 for G31.41$+$0.31 although it
was not well resolved with $\theta_{dec}/\theta_{mb} = 0.8$. 
 G12.21$-$0.10 did not fit our modeling criteria
because $\theta_{dec}/\theta_{mb} < 0.5$, so we considered it unresolved.

\subsubsection{Beuther et al.\ 2002}

Beuther et al.\ (2002) studied the density structure of 69 
regions forming massive stars (Sridharan at al.\ 2002)
 with 1.2 mm continuum and CS
emission. Their resolution at 1.2 mm was very similar to ours at 350 \micron,
providing an interesting comparison.
To determine the density structure, they fit the radial
intensity profiles with a broken power law, steeper in the outer
region, and assumed a power law temperature distribution ($T \propto
r^{-0.4}$) to determine the density power law. 
We tested the effect of the power law temperature distribution by 
modeling a source with $T
\propto r^{-0.4}$ rather than the temperature distribution calculated
from the dust code (see \S 4). Assuming a power law resulted in less
flux at shorter wavelengths, up to 60\% at 25 \micron, and a 20\%
higher $n_f$. However, the radial profile was unaffected and the
best fit $p$ did not change. Therefore, the resultant density
distributions from these studies can be reasonably compared.

Beuther et al.\ (2002) fit the radial intensity profiles on average
with $I \propto r^{-1.2}$ within 32\as\ and $r^{-1.8}$ in the outer
regions. They report a mean density power law index derived from the
mean inner radial index of $\langle p \rangle = 1.6 \pm 0.5$. These
results are consistent with our findings, although we have modeled the
core envelopes well beyond 32\as. As noted in \S 3, we found no systematic
tendency for intensity profiles to steepen beyond 40\as. This difference
between the two studies could reflect differences in the sample or 
differences in the observational details. For example, our chop throw
was about twice that used by Beuther et al.\ (2002).
More fundamentally, the agreement in the
inner regions is very reassuring, indicating that observations from
350 to 1200 \micron\ are tracing the same distribution in these sources.

Their sample has a mean mass, based on the integrated flux density and using
either the unambiguous or near distance, of $\langle M \rangle = 1550$ \msun.
 This should
be compared to our $\mean{\miso} = 2020$ \msun, which was derived in a 
similar way.  However, they used different assumptions about
opacities. Beuther et al.\ (2002) explain that the use of OH5
opacities, as used in our study, ``would result in masses and column
densities about a factor 4 lower''. The factor of 4 makes their average
masses about one fifth of our mean value or about equal to the median \miso\ 
(397 \msun). The Beuther et al.\ (2002) 
``OH5 scale'' mass distribution is shifted to 
lower masses than the distribution of isothermal masses 
in this study (see Fig.\ \ref{rfig}) and 
peaks near log($M$) = 2.2 compared to log($\miso$) = 2.8.
The discrepancy in the masses can further be explained by the use of 
different apertures in 
the measurement of flux densities between the two samples. Beuther et al.\ 
(2002) measure flux densities within
the 5\% level of the peak emission. In our study, the aperture was 
generally larger, 120\as, resulting in larger flux densities. 
Since the isothermal mass is proportional to the
flux density, we expect higher masses than Beuther et al.\ (2002).

Different methods were also used to calculate the
luminosities. Sridharan et al.\ (2002) calculate the luminosities of the
same sample studied by Beuther et al.\ (2002) by
integrating a two-component greybody curve fit to {\it Infrared
Astronomical Satellite} ({\it IRAS}) data. 
They find an average luminosity ($\mean{L} = 10^4$ \lsun) a 
factor of 25 lower than the
mean luminosity reported here from integrating the observed SEDs, which
in many cases also included {\it IRAS} fluxes. However, only a negligible 
fraction of
this discrepancy is the result of different methods. For the two sources 
in common,
 Sridharan et al.\ (2002) find about a 20\% lower luminosity than 
$L_{obs}$ when the sources are placed at the same distances. 
Because of their selection against \uchii\ regions, the Sridharan et al.\
 (2002) sample may contain younger and, therefore,
less luminous sources than our sample. Sridharan et al.\ (2002) report
a lower luminosity to mass ratio in their sample than that found in 
\uchii\ regions, implying that the ratio increases as a core evolves.
 Section 6.4 discusses the luminosity to mass ratio of our sample.

\subsection{Comparison to Low Mass Star Formation, Evolution, and Correlations}

 Figure \ref{phist} shows a histogram of $p$ values for
low mass star forming regions (Young et al.\ 2002 and Shirley et al.\ 2002a) 
alongside the distribution of $p$ for the 
massive stars in this study. Both samples were modeled with the techniques
described here. For the 
low mass cores, $\langle p \rangle = 1.6 \pm 0.4$ and increases to $\langle
p \rangle = 1.8$ if cores with high aspect ratios are left out of the mean (Young
et al.\ 2002). Although the modes
of low and high mass star formation are drastically different, the 
distributions of $p$ are strikingly similar. The mean fiducial density in the 
regions forming massive stars ($\langle n_f \rangle$ = 1.2\ee8 \cmv) 
is over two orders of magnitude greater than
that for the low mass cores ($\langle n_f \rangle _{low mass}$ = 5\ee5 \cmv) 
(Figure \ref{phist}). Because the fiducial density is referred to 1000 AU, 
which we do not resolve,
the absolute value should not be taken too literally; however, the similar values
of $p$ imply that the densities are higher at all radii by about two orders of
magnitude in these sources than in cores forming low mass stars.
The similarity in $p$ suggests that similar processes determine the {\it shape}
 of the density distribution in regions forming low and high mass stars.

In order to address the problem of a coherent evolutionary sequence
for regions forming massive stars, we examine quantities commonly used
as indicators of evolution in regions forming low mass stars. One such
indicator is the bolometric temperature, $T_{bol}$ (Myers \& Ladd
1993), the temperature of a blackbody with the same mean frequency as
the observed SED. The range of bolometric temperatures in this study
was relatively small, 46 -- 173 K, compared to that found by Myers \&
Ladd (1993) for low mass objects, almost two orders of
magnitude. Another indicator is the ratio of total to submillimeter 
luminosity, $\lobs/L_{smm}$. For low mass star forming regions, both
quantities are used as evolutionary indicators to mark the boundary
between Class 0 and Class I sources; $\lobs/L_{smm}$ increases as a source
evolves, and Andr\'{e} et al.\ (1993) described sources with
$\lobs/L_{smm} < 200$ as Class 0. Chen et al.\ (1995) defined Class 0
objects to be sources with $T_{bol} < 70$ K. $T_{bol}$ is plotted against
$\lobs/L_{smm}$ in Figure \ref{tbolfig}, which shows that
while the Class 0 and I definitions may not translate directly to
massive stars, there is some correlation between the two measures. 
The trend implies that high bolometric temperature may be an indicator of
more evolved sources.  Young et al.\ (2002) plot $T_{bol}$
versus $\lobs/L_{smm}$ for both low mass sources and this sample and find
that the Class 0/I boundary is not as clearly defined
in regions forming low mass stars as in high mass regions. 

Since $T_{bol}$ may be an indicator of evolution, we looked for
correlations with other parameters. Figure \ref{tbolfig} shows no
significant correlation between $T_{bol}$ and $p$ or $n_f$.  Another
indicator of evolution is the ratio of envelope to stellar mass ($\miso/
M_*$). As a source evolves and more material is accreted onto the central 
source, the ratio should decrease. To approximate
the stellar mass, $M_*$, we assumed that the luminosity is dominated by 
a single massive star, and that $M_* \propto L^{1/3.5}$. This mass-luminosity 
relationship is for zero age main sequence stars and has been used for 
a range of stellar masses (van der Tak et al.\ 2000, Shirley et al.\ 2002a). 
Observations
of binary stellar masses and luminosities suggest that this relationship
is valid to approximately 60 \msun\ (Scalo 1986). The luminosities in this 
sample suggest very few of these cores have a central source larger than 
60 \msun, and the sources with the highest luminosities are likely forming
more than one massive star. Figure
\ref{tbolfig} also shows that for regions forming massive stars there is 
not a significant correlation between $T_{bol}$ and $\miso/
M_*$. However, the high values for $T_{bol}$ ($>$ 80) occur for
$\miso/M_*$ $<$ 70, indicating a more evolved source.

Figure \ref{ratiofig} plots
$M_{iso}/M_*$ and another potential evolutionary indicator,
the far infrared color
($F_{60}$/$F_{100}$), which is the ratio of
flux densities at 60 \micron\ and 100 \micron, both observed with the
same instrument [either \iras\ or Kuiper Airborne Observatory (D. Jaffe
2001, private communication)]. Our data do not show the same direct 
correlation between
far-infrared color and ratio of envelope to stellar mass as reported
by van der Tak et al.\ (2000). However, their conclusion that 
bluer far-infrared colors only occur for lower mass ratios does apply. 
A more evolved source, as measured by an envelope to stellar mass 
ratio less than 70, may have a $T_{bol} > 100$ and far-infrared
color, $F_{60}$/$F_{100} > 0.7$, while these values are never found for 
$M_{iso}/M_* > 70$. Figure \ref{ratiofig} 
also plots $p$ versus 
$M_{iso} / M_*$ and indicates that less evolved sources (i.e., 
high $M_{iso} / M_*$) may have steeper than average
density distributions.

Possible correlations with the density distribution were also
examined. Van der Tak et al.\ (2000) found no correlations between
envelope mass or the internal luminosity and the density law
exponent, $p$. However, Figure \ref{pfig} shows our larger sample does
suggest some correlation with each of these parameters ($r \sim 0.5$). 
The density distribution steepens with increasing mass and luminosity. 
Figure \ref{pfig} also plots the ratio of the deconvolved source diameter 
to the FWHM beam size, $\theta_{dec}/\theta_{mb}$, and 
the distance versus $p$. There is a strong relationship between $\theta_{dec} /
\theta_{mb}$ and the best fit $p$. Better resolved sources
tend to have shallower density profiles, as do more nearby sources.
 The correlation of $\theta_{dec} / \theta_{mb}$ with $p$ could be the result
of a systematic effect in the models or resolution. Three of the six sources
with $p >$ 2 (G10.60$-$0.40, G31.41$+$0.31, and W43S) are also the most distant, 
suggesting that better resolution is needed to confirm their density 
distributions. The correlation of $\theta_{dec} / \theta_{mb}$ with $p$
could also indicate a physical difference 
in the cores, such as different initial conditions. Young et al.\ (2002) found
the same correlation for low mass cores where distance is not the same limiting
factor as in this study. A source with a steeper density distribution 
(higher $p$) naturally has a smaller size ($\theta_{dec}$) and will be less well
resolved (Figure \ref{pfig}(c)).

\subsection{Luminosity to Mass Ratios}

\subsubsection{Comparison to Galactic Studies}

The luminosity to mass ratio, $L/M$, is often used as a measure of the 
star formation rate per unit mass. Most studies in our Galaxy have used
the mass of the entire molecular cloud, determined from CO emission.
The resulting average ($L/M$ = 0.4 \lsun/\msun) is quite low 
(Bronfman et al.\ 2000). For molecular clouds
that contain \HII\ regions, the luminosity to mass ratio from CO emission
is an order of magnitude higher, $L/M$ = 4 \lsun/\msun\ (Mooney \& 
Solomon 1988), but the dispersion is over 2--3 orders of 
magnitude (Evans 1991; Mead, Kutner, \& Evans 1990). CO emission
traces the less dense gas of the entire molecular cloud, 
while the dust continuum emission at \submm\ wavelengths measures the 
mass of very dense gas actually involved in star formation. We computed values 
of $L_{bol}/\miso$ because $\miso$ could be obtained
for the largest number of sources. Figure 
\ref{mlm} plots log($\lobs/\miso$) versus log(\miso). Both \lobs\ and 
\miso\ are proportional to the square of the distance, so the distance
uncertainties are minimized in the ratio.
The ratio and, hence, the star formation rate per unit mass is constant 
over the entire mass range.
The dispersion of 
log(\lobs/\miso) is about one and a half orders of magnitude, which is 
significantly 
less than the 2--3 orders of magnitude when the mass is traced by CO. 

Figure \ref{mlm} also plots log(\lobs) versus log(\miso) and shows that the 
luminosity to mass ratio traced in massive cores is significantly
higher than the average $L/M$ in molecular clouds with \HII\ regions.
 Figure \ref{lmhist} shows the
distribution of ${\rm log} (\lobs/\miso)$, which peaks near $\langle {\rm log}
(\lobs/\miso) \rangle = 2.0 \pm 0.4$; the median log(\lobs/\miso)
 is also 2.0. The mean
value of the ratio is $\langle \lobs/\miso \rangle = 140 \pm 100$
\lsun/\msun, with a median value of 120 \lsun/\msun. 
This $\langle \lobs/\miso \rangle$ is 30 times that derived
from CO for Galactic molecular clouds containing \HII\ regions.
 Clearly the star formation rate per
unit mass of {\it dense} gas is much higher, indicating that the dense
gas traced by \submm\ emission is the location of massive star formation.

The histogram of $\lobs/\miso$ in Figure \ref{lmhist}
shows that the distribution is strongly skewed. Most sources have low
ratios, but there is a tail of very high ratios, reaching up to 
490 \lsun/\msun. A few sources could be affected by confusion with 
multiple sources in the large \iras\ beam, resulting in a luminosity that 
is too high. This problem is difficult to 
avoid since \iras\ is often the only source of mid- to far-infrared fluxes.
Higher spatial resolution is needed to eliminate multiple source confusion
in flux measurements. The Stratospheric Observatory For Infrared Astronomy 
(SOFIA) that will fly later in the decade will be ideally suited to address
this problem and fix the upper limit to $L/M$ for Galactic sources.

The mean $L/M$ for the Beuther et al.\ (2002) sample 
($\langle L/M \rangle = 20 \pm 18$ \lsun/\msun, Sridharan et al. 2002) is
7 times lower than in this study ($\mean{\lobs/\miso} = 140$ 
\lsun/\msun). 
If we decrease their masses by a factor of four, putting them on the
``OH5 scale", the values are in better agreement ($\mean{L/M} \sim$ 80
 \lsun/\msun).
For sources with a distance ambiguity, we assume their near distance,
which may produce a downward bias in luminosities and masses.
Eliminating this bias by averaging only sources in their sample for 
which the distance is not ambiguous, gives a luminosity to mass
ratio ($\mean{L/M} = 120 \pm 90$ \lsun/\msun) that is consistent with 
the results of our study. However, Sridharan et al.\ 
(2002) report an $L/M$ significantly smaller than that of a sample of 
\uchii\ regions (Hunter 1997, Hunter et al.\ 2000). They suggest that the
cores in their sample are in a younger pre-\uchii\ phase and that  
 $L/M$ increases as the cores evolve and develop \uchii\ regions 
(Sridharan et al. 2002). Hunter et al.\ (2000) 
describe their sample as having \uchii\ region far-infrared colors or 
thermal radio continuum emission. If we also convert 
the opacities used by Hunter et al. to the OH5 scale, 
then the average luminosity to mass ratio for their sample
is $\mean{L/M} = 280 \pm 370$ \lsun/\msun\ with a median of 150.
The mean $L/M$ for the \uchii\ regions studied by Hunter (1997) and 
Hunter et al.\ (2000)
is higher, but the median is not inconsistent with the results of the
work presented here for cores both with and without \uchii\ regions. 
 
In order to test the hypothesis that the luminosity to mass ratio 
increases as a massive star forming region develops an \uchii\ region,
 we divided our sample
 and calculated $\lobs/\miso$ for 
cores with and cores without \uchii\ regions. We do not find the 
same disparity in $L/M$ between our two subsamples as between the results of 
Sridharan et al.\ and Hunter (1997 \& Hunter et al. 2000). For the cores 
associated with \uchii\ regions, $\langle \lobs/\miso \rangle = 170
\pm 130$ (with a median of 120), which is lower than the mean of the sample
of Hunter and Hunter et al., but within the dispersions of the two samples.
The mean ratio of the remaining cores that do not contain an \uchii\ region 
is not significantly lower, $\langle \lobs/\miso \rangle = 130 \pm 90$ with 
a median of 100. However, in a CS study with a larger sample of which ours is 
a subset, Shirley et al.\ (2002b) found that the median 
luminosity to virial mass ratio is more than a factor of two higher
in cores with \uchii\ regions than in cores without \uchii\ regions.

\subsubsection{Comparison with Extragalactic Studies}

Submillimeter wavelengths are becoming increasingly important in the the study 
of star formation in other galaxies, especially at high redshifts (Blain et al.\
2002). A recent submillimeter survey measured the dust masses, $M_D$, and
far-infrared to submillimeter ($\sim$ 1 to 1000 \micron) luminosities, 
$L_{FIR}$, of bright local {\it IRAS} galaxies (Dunne \& Eales 2001). 
Dust is often used as a tracer of mass for more distant sources as well 
(e.g., Omont et al.\ 2001, Calzetti et al.\ 2000, Benford et al.\ 1999). 
These studies assume a single dust temperature, usually 50 K, based on
the far-infrared and submillimeter SEDs of the galaxies (e.g, Calzetti
et al.\ 2000, Benford et al.\ 1999). 
The average $T_{iso}$ of 29 K found in this study is substantially lower.
If we compared the mass within \rdust, we found that $T_{iso}$ was about 50 K,
but when using the full extent, much cooler dust contributed substantially.
The dust mass, $M_D$, of galaxies is usually calculated in the same way as
our $M_{iso}$ (Equation (\ref{misoeq})). However, different opacities are used 
and only
the mass of the dust is calculated, not the mass of gas and dust as with 
$M_{iso}$.  

For high redshift ($z >$ 4) quasars, Omont et al.\ (2002) report a mean 
(and median) dust mass, $\langle M_D \rangle \sim 5\ee8$ \msun\ for a 
range of $2\ee8$ to $1\ee9$ \msun. A typical far-infrared luminosity 
($>$ 50 \micron) for their sample is    
$L_{FIR} \sim 10^{13}$ \lsun, giving $L_{FIR} /
M_D = 2\ee4$ \lsun/\msun. For purposes of comparison, we calculated
an isothermal dust mass, $M_D$, using the same temperature (50 K) and opacities 
as Omont et al.\ (2002). For the regions in this study, $\langle L_{obs} / M_D
\rangle = (1.4 \pm 1.0)\ee4$ \lsun/\msun\ ranging from 2\ee3 to 4.5\ee4 
\lsun/\msun.
The luminosity to mass ratio for high redshift quasars is 
similar to the higher values in our
sample, suggesting that starbursts form stars as if most of their molecular 
material acts like the most extreme regions forming massive stars in the 
Milky Way. However, starbursts might not be the only source of $L_{FIR}$ 
in distant quasars. Omont et al.\ (2002) suggest 
central active galactic nuclei as another source of dust heating. 
In that case, the star-forming $L/M$ would be smaller.

\section{Summary}

We have presented dust continuum maps of 51 regions forming massive
stars with a large range of sizes and masses,  $\langle \rdust \rangle$ =
 0.16 $\pm$ 0.10 pc and $\langle \miso \rangle$
= 2020 $\pm$ 4410 \msun. We find that the peak of the dust emission is more 
often
coincident with the water maser position than the \uchii\ region, implying 
that the dust may better trace the earlier stages of massive
star formation. Modeling
a subset of 31 sources yielded a mean density index for power laws ($n(r) =
(r/r_f)^{-p}$) of $\langle p \rangle$ = 1.8 $\pm$ 0.4. The dispersion is
about twice the expected uncertainty for an individual source, suggesting
some real dispersion among sources. The mean value for $p$ is incompatible with 
a logatropic sphere ($p$ = 1), but $p$ = 1.5 or 2 are possible.
The mean value and variation in $p$
 are similar to those found for low mass regions (Young et al. 2002).

For the modeled sources, integrated masses within two different fiducial radii
were presented in addition to the isothermal masses. The mean mass 
within the 350 \micron\ half-power radius (\rdust) is $\langle \md \rangle
= 250 \pm 380$ \msun. For the larger, more physical radius where the density
falls off to the ambient level ($n = 10^4$ \cmv), $\langle \mden \rangle$
= 720 $\pm$ 860 \msun.  The mean column density for the complete sample 
was found to be $\Sigma = 0.73 \pm 1.7$. However, because of the inverse 
dependency of mass on opacity, comparison with virial masses suggests that
the masses and mean surface density could be about a factor of three larger.

The density structure of massive star-forming cores was found to be
consistent with Beuther et al.\ (2002). The results were also
consistent with van der Tak et al.\ (2000) when the difference in a
Gaussian and the observed beam profile was taken into account. Our
data also confirm some of the conclusions made by van der Tak et
al.\ (2000) with respect to possible evolutionary indicators.  There is
a trend of rising $T_{bol}$ with increasing $L/\lsmm$, suggesting that
either of these quantities could be tracing evolution.

The luminosity to mass ratio, a tracer of star formation rate per
unit mass, has a mean of $\langle \lobs/\miso \rangle$ = 140 $\pm$ 100 
\lsun/\msun\
for dense gas traced by dust emission, with a tail extending up
to about 500 \lsun/\msun. These values are much higher than those
based on masses from CO emission. The mean luminosity to mass ratio
 derived using the dust mass
 is similar to that in extreme starburst galaxies, especially those
seen at substantial redshift.

\section{Acknowledgments}

We thank the staff of the Caltech Submillimeter Observatory for 
assistance in using SHARC. We are very grateful to R. Chamberlain
for providing sky opacity information from the CSO tippers.
We also thank T. Hunter, D. Benford, R. Chamberlin, and D. Lis for 
technical assistance, sometimes via phone calls at inconvenient times.
We are grateful to R. Forster and to D. Jaffe for their generous sharing of
data and useful comments. We thank NSF (Grant AST-9988230) and the 
State of Texas for support.



\clearpage

\clearpage

\clearpage

\clearpage

\clearpage

\begin{figure}
\centering
 \vspace*{7.5cm}
   \leavevmode
   \includegraphics{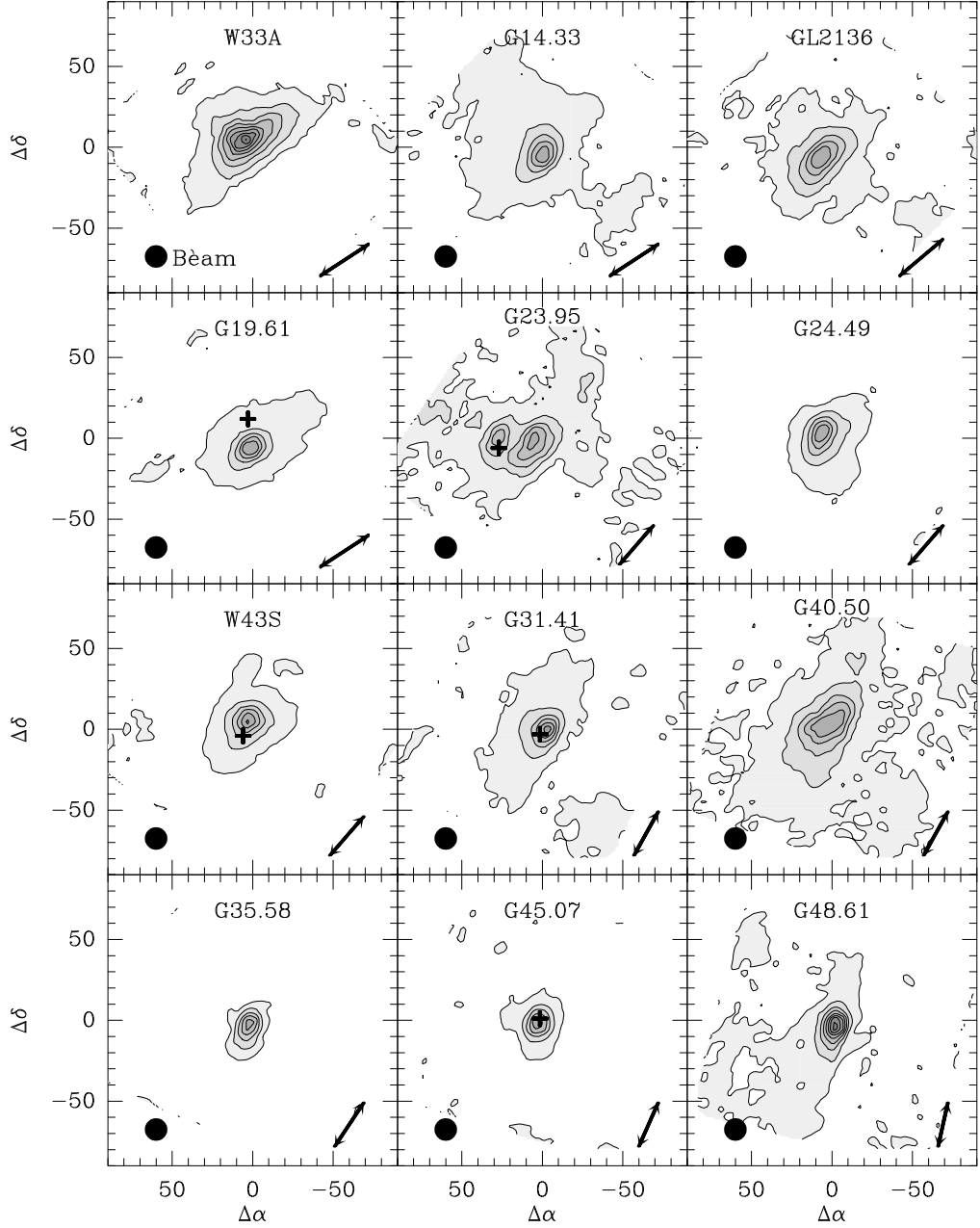}
\vskip 3.5in
\figcaption{
350 \micron\ dust emission contour maps. The arrows indicate the 
chop direction but are not indicative of the chop length ($\sim$ 100\as).
The plus signs indicate the positions of known \uchii\ regions (Wood \& 
Churchwell 1989).  
The contour levels are as follows:
W33A (4$\sigma$), 
G14.33$-$0.64 (3$\sigma$, then in increments of 20\% (17$\sigma$) of the peak),
GL2136 (3$\sigma$, 20\% (8$\sigma$)),
G19.61$-$0.23 (3$\sigma$, 20\% (33$\sigma$)),
G23.95$+$0.16 (3$\sigma$, 20\% (5$\sigma$)),
G24.49$-$0.04 (3$\sigma$, 20\% (15$\sigma$)),
W43S (3$\sigma$, 20\% (29$\sigma$)),
G31.41$+$0.31 (3$\sigma$, 20\% (18$\sigma$)),
G40.50$+$2.54 (3$\sigma$, 20\% (10$\sigma$)),
G35.58$-$0.03 (4$\sigma$),
G45.07$+$0.13 (3$\sigma$, 20\% (6$\sigma$)),
G48.61$+$0.02 (3$\sigma$, 10\% (5$\sigma$)).
}
\end{figure}
 
\clearpage

\begin{figure}
\centering
 \vspace*{7.5cm}
   \leavevmode
   \includegraphics{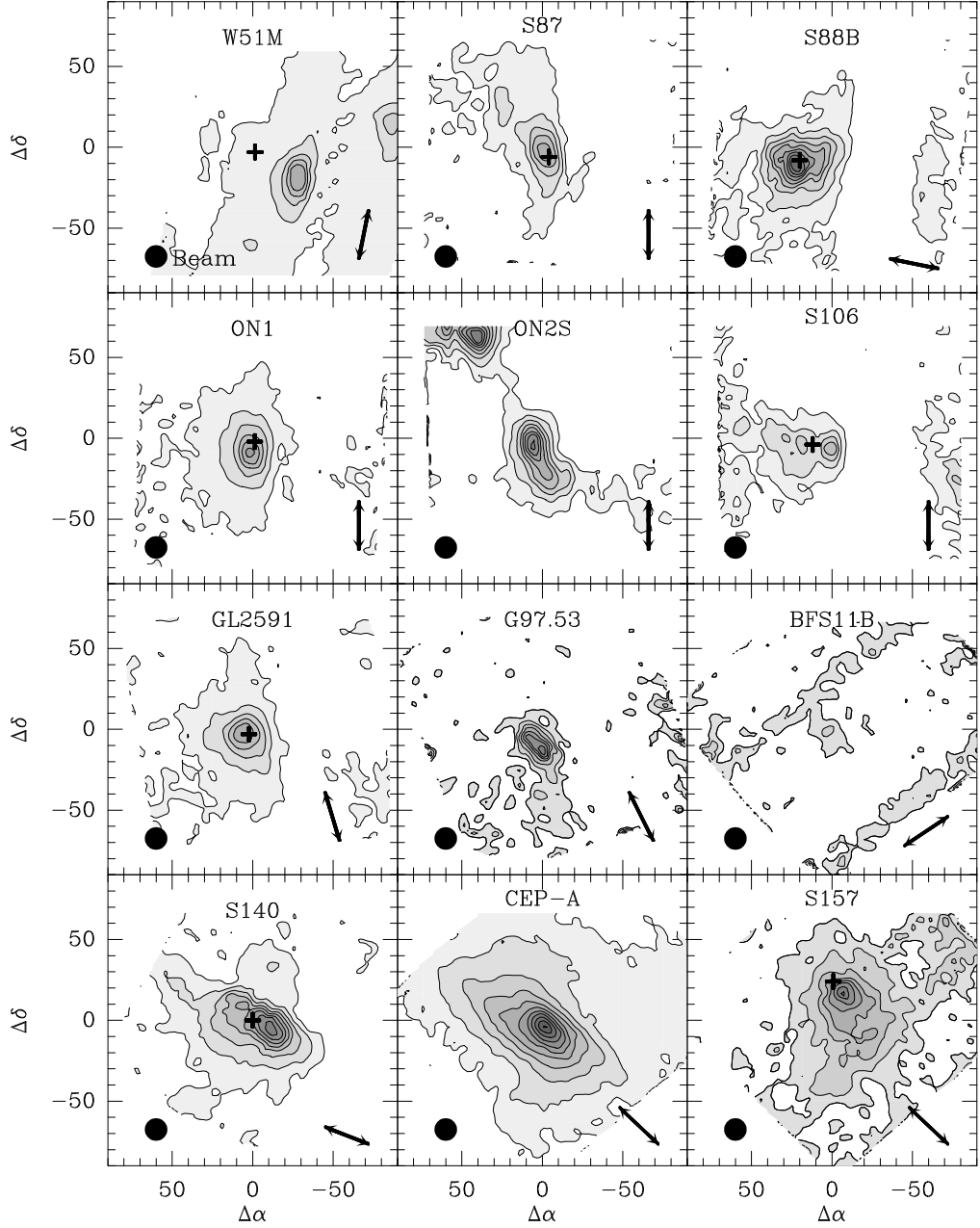}
\vskip 3.5in
\figcaption{
350 \micron\ dust emission contour maps. The arrows indicate the 
chop direction but are not indicative of the chop length ($\sim$ 100\as). 
The plus signs indicate the positions of known \uchii\ regions (W51M: Scott 1978;
S88B: Wood \& Churchwell 1989; S87, ON1, S106, S157: Kurtz et al.\ 1994;
GL2591, S140: Tofani et al.\ 1995).
The contour levels are as follows:
W51M (3$\sigma$, then in increments of 20\% (19$\sigma$) of the peak), 
S87 (4$\sigma$),
S88B (4$\sigma$),
ON1 (3$\sigma$, 20\% (11$\sigma$)),
ON2S (4$\sigma$),
S106 (4$\sigma$),
GL2591 (3$\sigma$, 20\% (11$\sigma$)),
G97.53$+$3.19 (2$\sigma$),
BFS11B (2$\sigma$),
S140 (4$\sigma$),
CEP A (3$\sigma$, 10\% (7$\sigma$)),
S157 (2$\sigma$).
}
\end{figure}

\clearpage

\begin{figure}
\centering
 \vspace*{7.5cm}
   \leavevmode
   \includegraphics{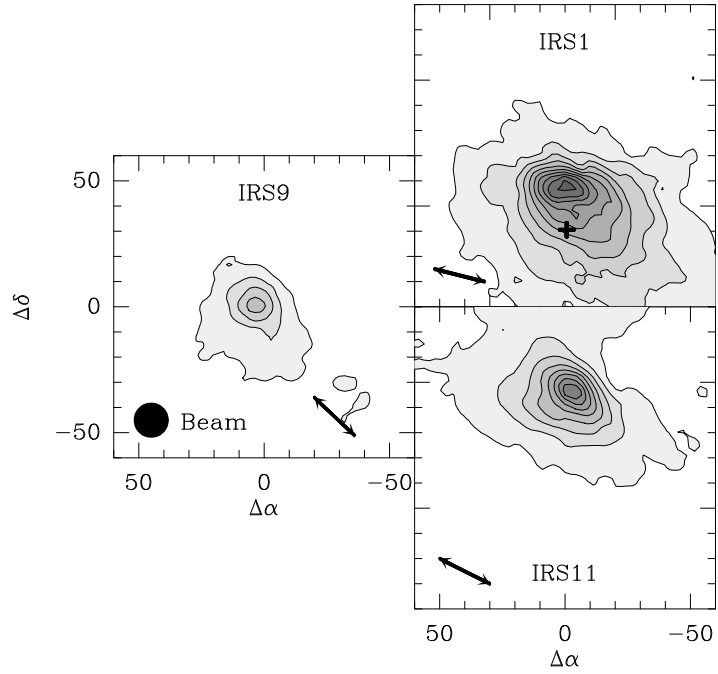}
\vskip 3.5in
\figcaption{\label{contend}
Composite 350 \micron\ contour map of NGC 7538. The arrows indicate the 
chop direction but are not indicative of the chop length ($\sim$ 100\as).
Clockwise from left are the sources IRS9, IRS1, and IRS11. The visible
\HII\ region NCG7538 (S158) lies north of IRS1. The plus sign indicates
the position of the \uchii\ region associated with IRS1 (23$^h$ 11$^m$ 36.6$^s$ 
61\degree\ 11\arcmin\ 36.6\as, Wood \& Churchwell 1989). 
The contour levels are as follows: IRS9 (4$\sigma$), IRS1 (3$\sigma$, 10\% 
(6$\sigma$)), IRS11 (3$\sigma$, 10\% (6$\sigma$)).}
\end{figure}

\clearpage

\begin{figure}
\plotone{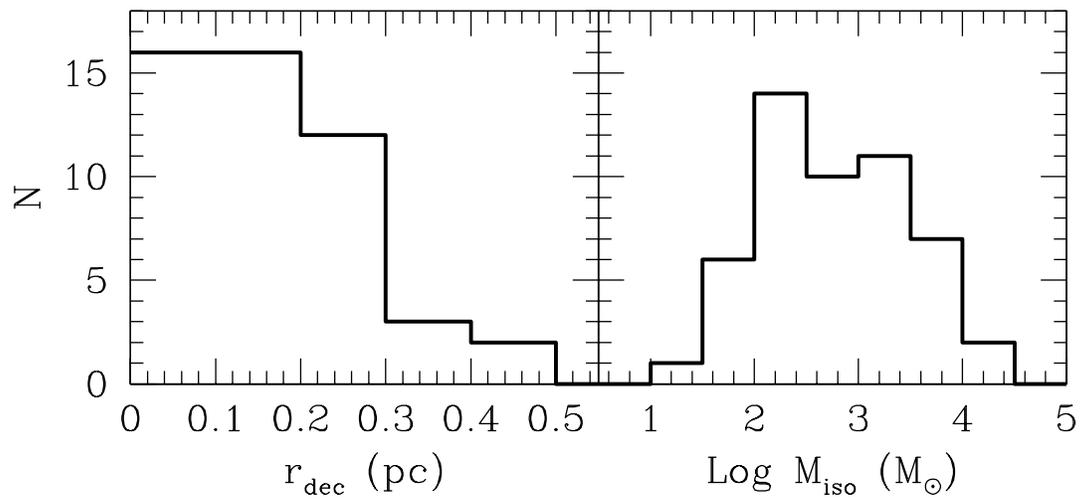}
\figcaption{\label{rfig}
Distribution of radii and masses for the complete sample. \rdust\ is the 
deconvolved half-power size as
determined from the 350 \micron\ dust emission and \miso\ is the isothermal mass.}
\end{figure}

\clearpage

\begin{figure}
\plotone{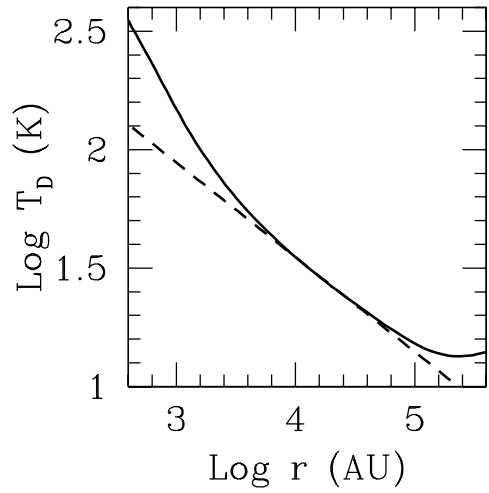}
\figcaption{\label{tempfig}
The solid line is the model dust temperature distribution for internal 
source with $L_{bol} = 10^4$ \lsun. The dashed line is a $T_D(r) \propto r^{-0.4}$
power law. The modeled distribution deviates from a strict power law at large
$r$ due to the ISRF and at small $r$ because of optical depth effects.}
\end{figure}

\clearpage

\begin{figure}
\plotone{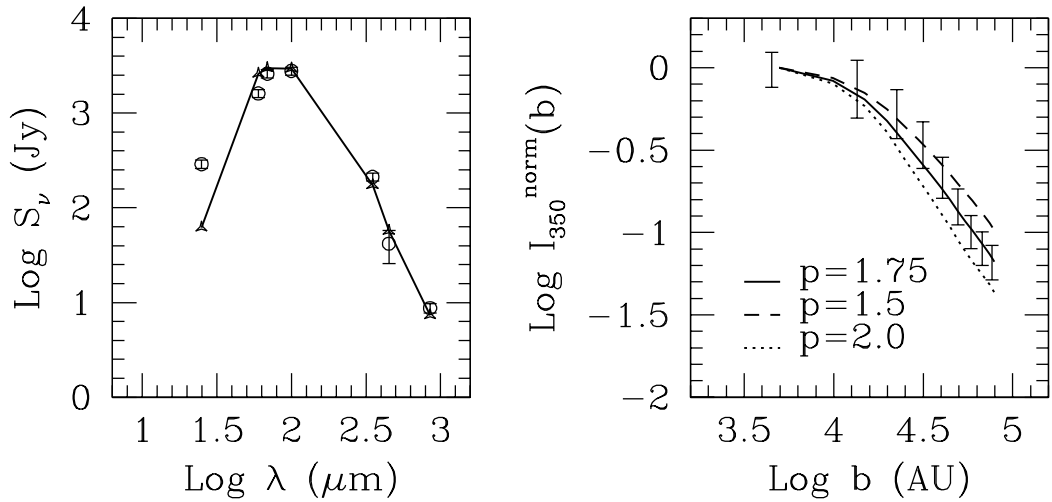}
\figcaption{\label{m8efig}
The left figure shows the observed (circles with error bars) and best fit model 
(solid line) SED ($\chi_R^2 =
14$) for M8E. 
On the right, the solid line is the best fit model for
the observed normalized radial intensity profile (error bars), 
$p = 1.75$ (solid line, 
$\chi_R^2 = 0.52$). The error bars in the radial profile represent the weighted mean
variation in the annulus for which each point of the profile is calculated.
Also plotted are the $p = 1.5$ (dashed line, $\chi_R^2 = 4.3$) and $p = 2.0$ 
(dotted line, $\chi_R^2 = 3.4$) model profiles.}
\end{figure}

\clearpage

\begin{figure}
\centering
 \vspace*{7.8cm}
   \leavevmode
   \includegraphics{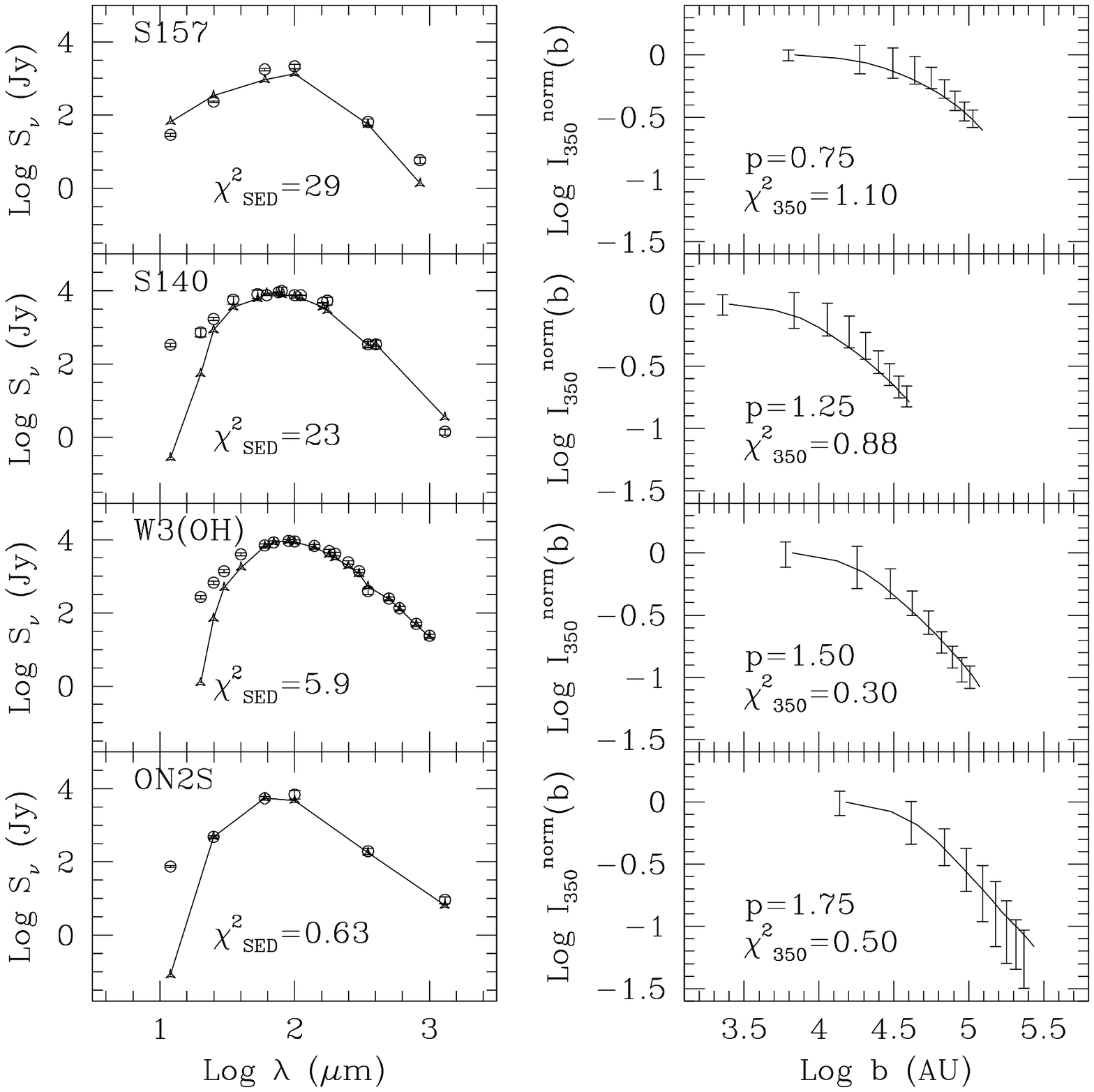}
\vskip 4.0in
\figcaption{\label{modfig}
Observed SEDs (circles) and normalized radial profiles (error bars) for a subsample 
of sources 
with best fit models (solid line) for $p$ = 0.75 to 1.75. The error bars 
in the radial profiles represent the weighted mean
variation in the annulus for which each point of the profile is calculated.
\chisqsed\ is calculated for SED points $>$ 12 \micron.
}
\end{figure}

\clearpage

\begin{figure}
\centering
 \vspace*{7.8cm}
   \leavevmode
   \includegraphics{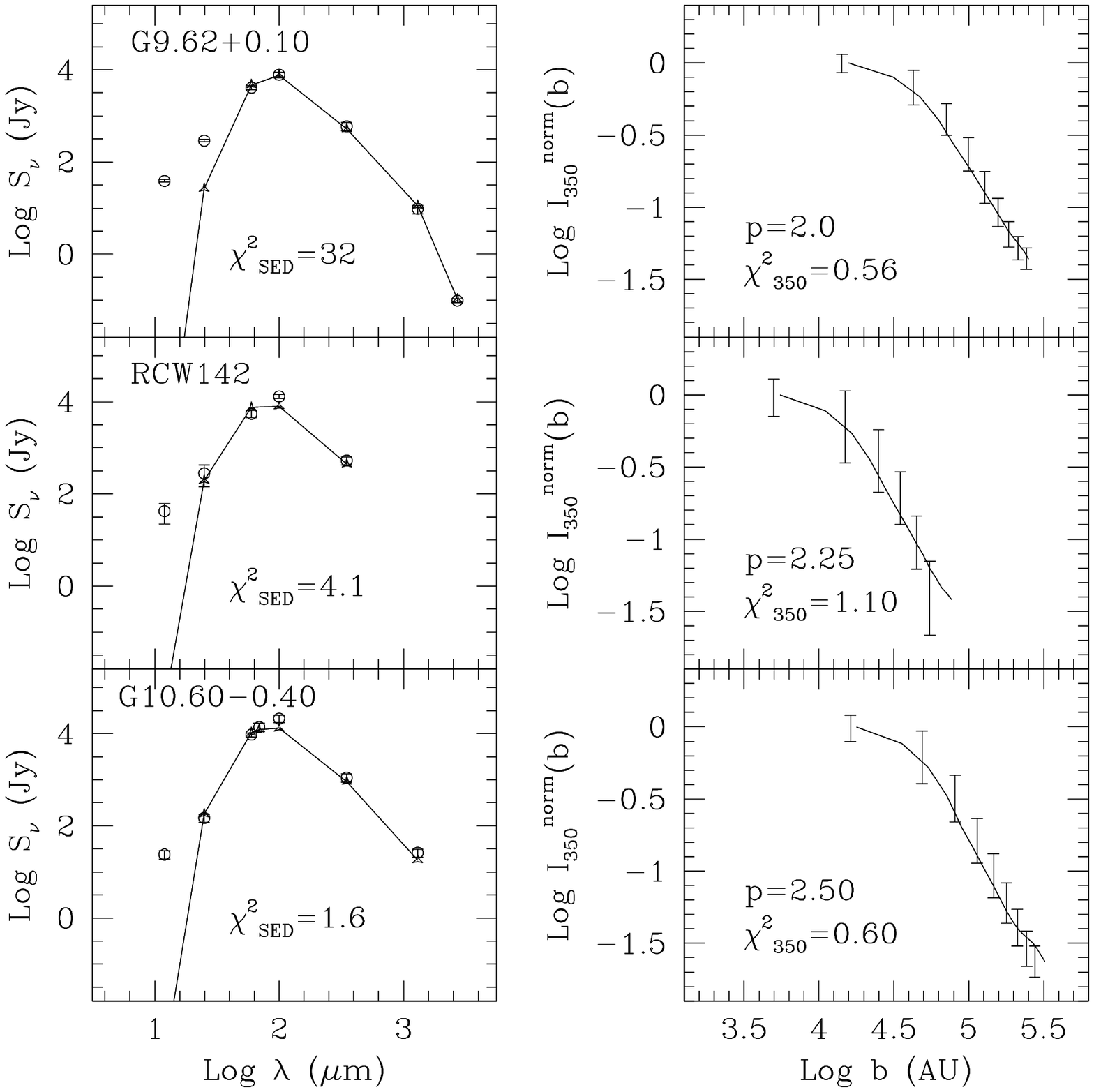}
\vskip 4.0in
\figcaption{\label{modend}
Observed SEDs (circles) and normalized radial profiles (error bars)
 for a subsample of sources
with best fit models (solid line) for $p$ = 2.0 to 2.5. 
The error bars in the radial profiles represent the weighted mean
variation in the annulus for which each point of the profile is calculated.
\chisqsed\ is calculated for SED points $>$ 12 \micron.
}
\end{figure}

\clearpage

\begin{figure}
\plotone{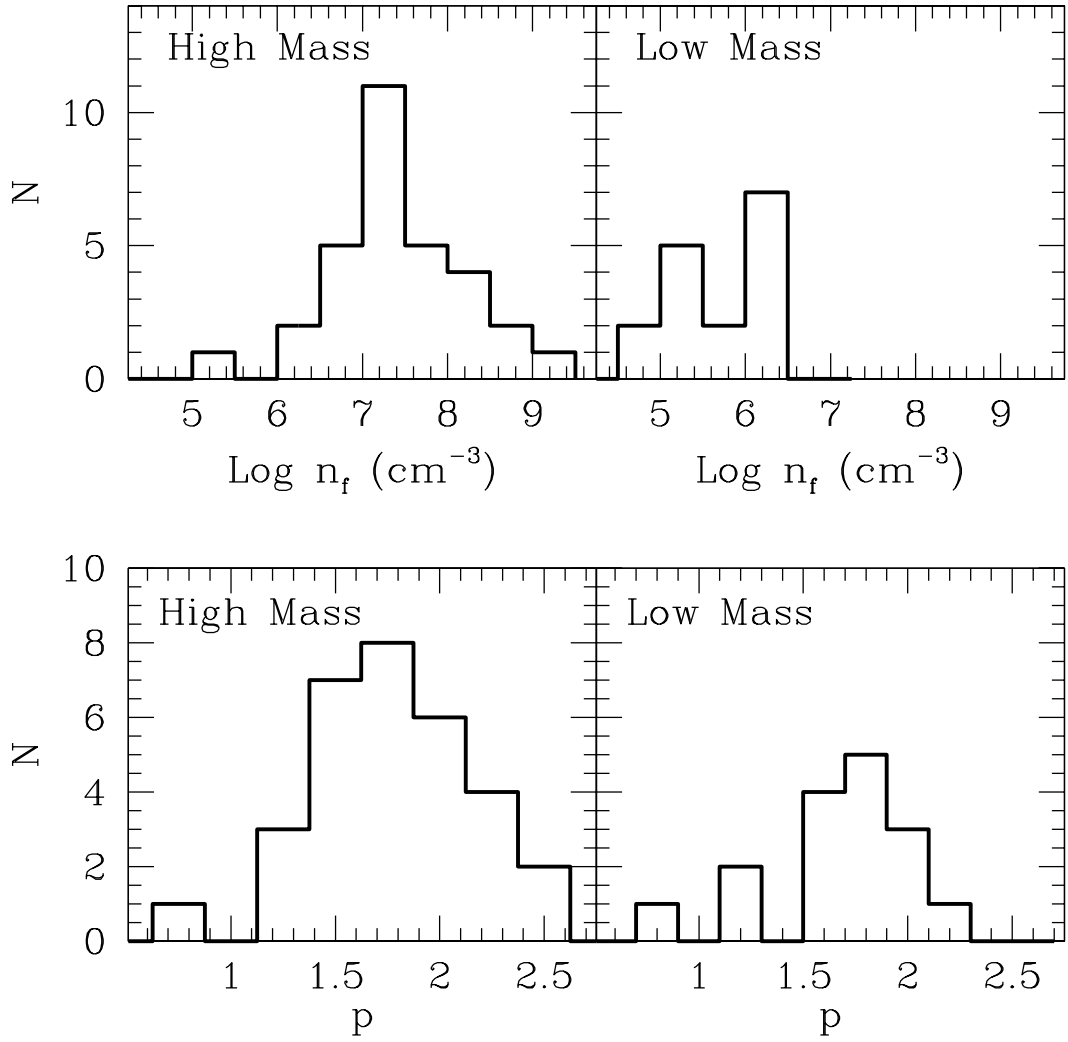}
\figcaption{\label{phist}
Histograms of fiducial densities 
($n_f$) and best fit density power law exponents ($p$) for high (left) 
and low (right) mass 
star forming cores modeled using the same methods. 
For the massive cores, $\langle n_f \rangle = 1.2\ee8$ \cmv\ and $\langle p \rangle 
= 1.8 \pm 0.4$. For the 
low mass cores, $\langle n_f \rangle = 5.2\ee5$ \cmv\ and $\langle p \rangle = 
1.6 \pm 0.4$; $\langle p \rangle = 1.8$ if sources with high aspect ratios
are not included (Young et al.\ 2002).}
\end{figure}

\clearpage

\begin{figure}
\plotone{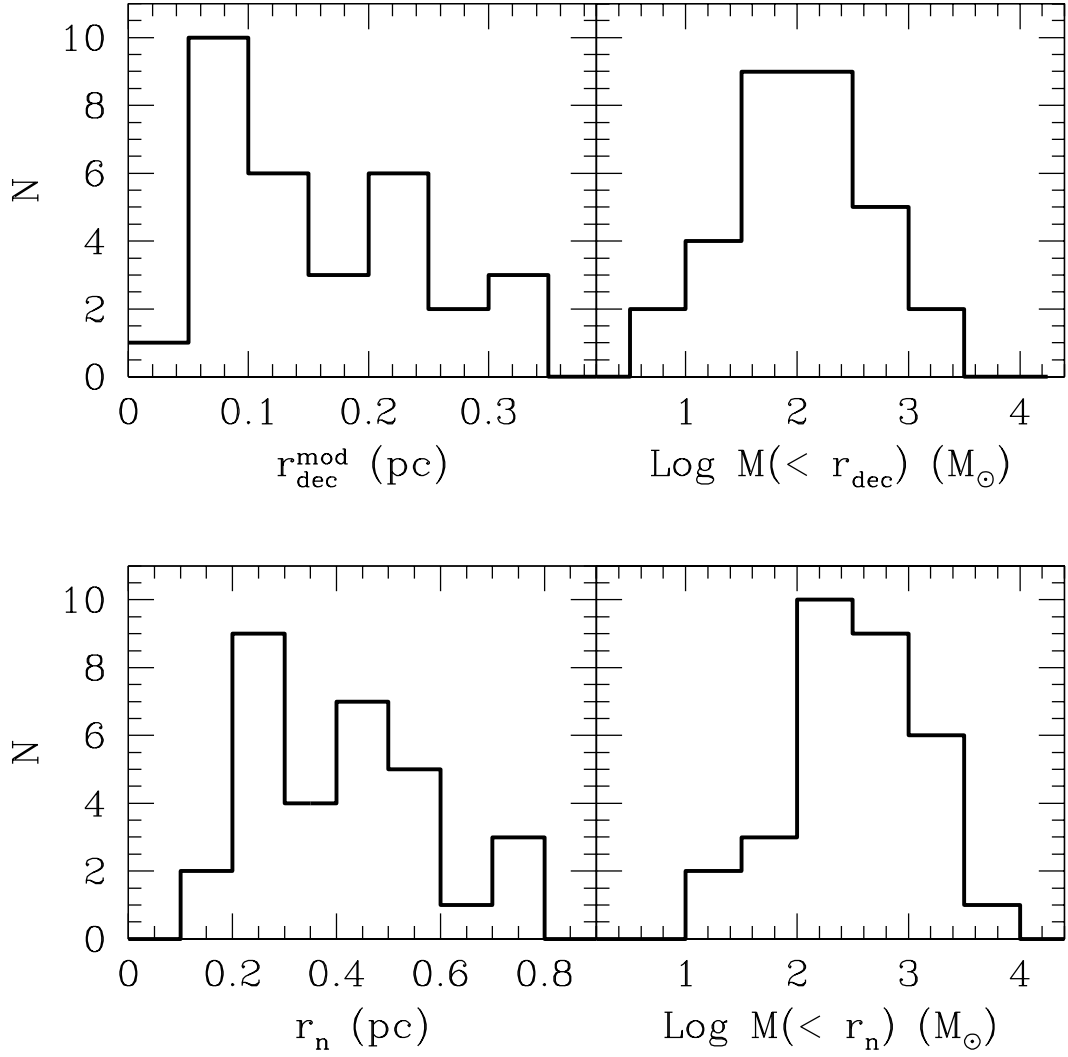}
\figcaption{\label{masshist}
Distributions of radii and integrated masses for the modeled sample. The top
two panels are the histograms of $r_{dec}^{mod}$ and the mass within \rdust. The 
bottom panels show the distributions of $r_n$, the radius at which the density
falls to $10^4$ \cmv, and the mass within that radius.}
\end{figure}

\clearpage

\begin{figure}
\plotone{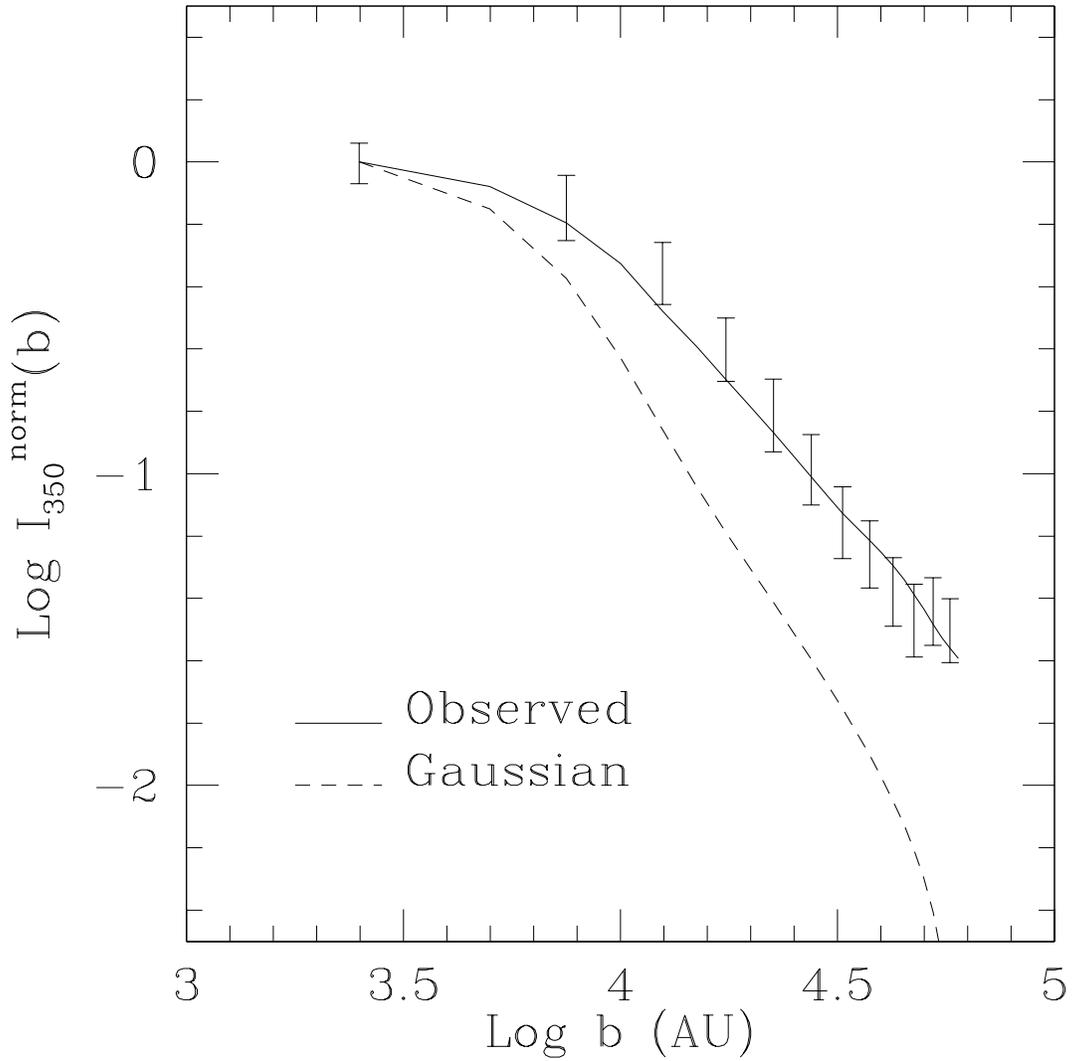}
\figcaption{\label{gaussfig}
Model radial profile with $p = 2.0$ using the observed beam profile
(solid line) and a 10\as\ Gaussian beam (dotted line) as used by van der
Tak et al.\ (2000) plotted with the
observed radial profile of GL2591 (error bars). The error bars in the 
radial profiles represent the weighted mean
variation in the annulus for which each point of the profile is calculated.
The use of a 10\as\ Gaussian beam
decreases the modeled $p$ for GL2591 by 0.75, accounting for the 
discrepancy in the value of $p$ reported by van der Tak et al.\ (2000) 
($\approx$ 1.25) and in this study (2.0).}
\end{figure}

\clearpage

\begin{figure}
\plotone{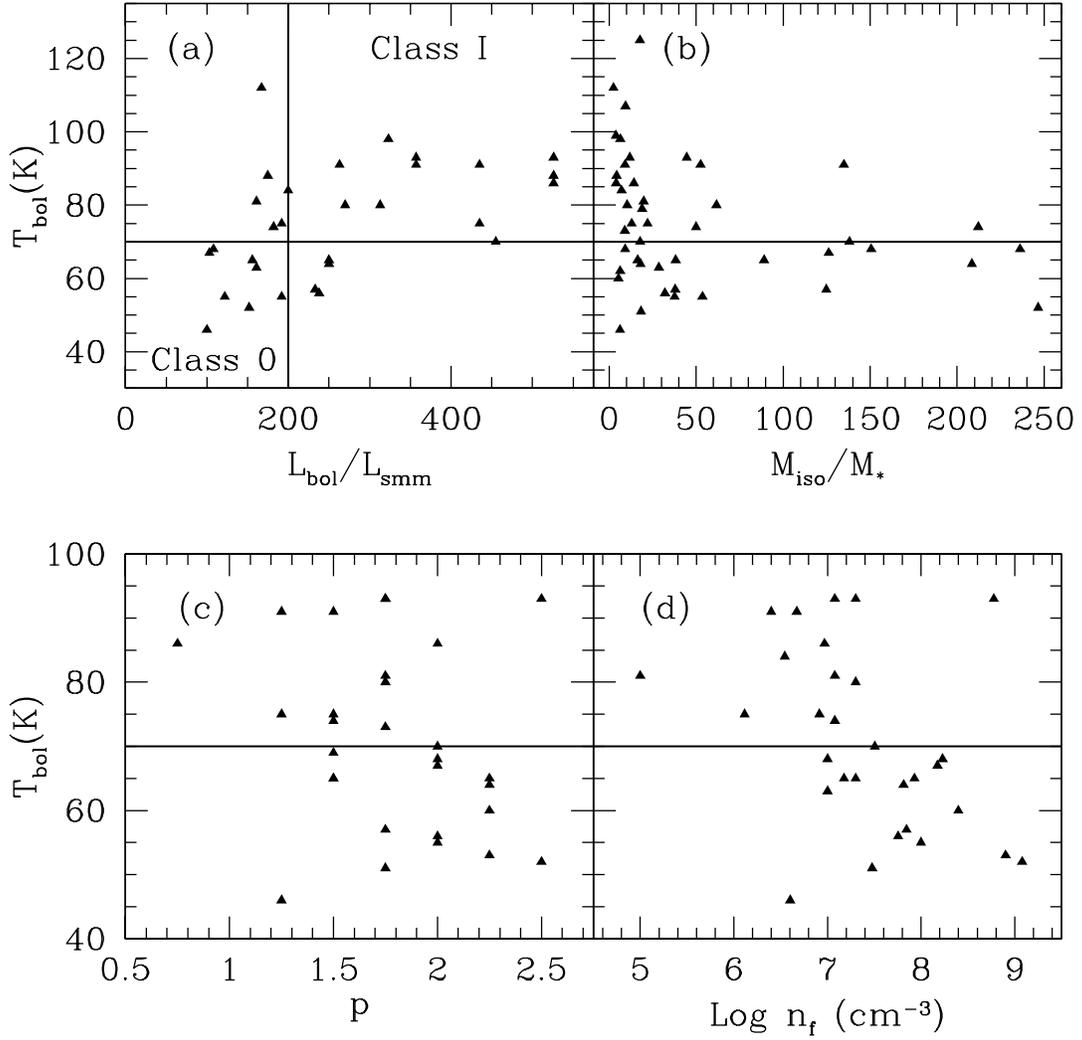}
\figcaption{\label{tbolfig}
Relationships with $T_{bol}$: (a) Relationship between bolometric 
temperature and the ratio of total to submillimeter luminosity, both
considered indicators of evolution in low mass star forming cores. The 
solid lines indicate the divisions between Class 0 and I low mass 
protostars for each indicator. Class I sources have \tbol\ $>$ 70 K and 
$L/\lsmm >$ 200. The linear correlation coefficient, $r$, is 0.52.\
 (b) \tbol\ versus another evolutionary indicator, 
$\miso/M_*$ where $M_* = L^{1/3.5}$ ($r = 0.06$). More evolved sources 
($\miso/M_* < 70$) have higher \tbol\ ($>$ 80). (c) \tbol\ versus the density power 
law index $p$ ($r = -0.25$). (d) \tbol\ versus 
the fiducial density, $n_f$, for the modeled sample ($r = -0.33$).} 
\end{figure} 

\clearpage

\begin{figure}
\plotone{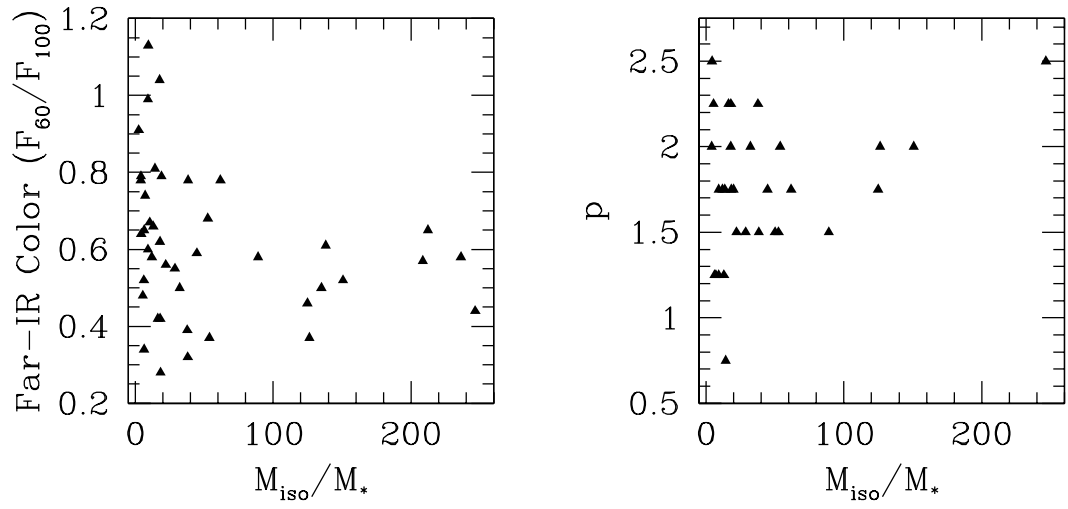}
\figcaption{\label{ratiofig}
Relationships with the envelope to stellar mass ratio. Left: 
the far-infrared color versus \miso/$M_*$ ($r$ = 0.07). 
More evolved sources ($\miso/M_* < 70$) have bluer colors ($>$ 0.7). 
Right: $p$ versus \miso/$M_*$ ($r$ = 0.31).}
\end{figure}

\clearpage

\begin{figure}
\plotone{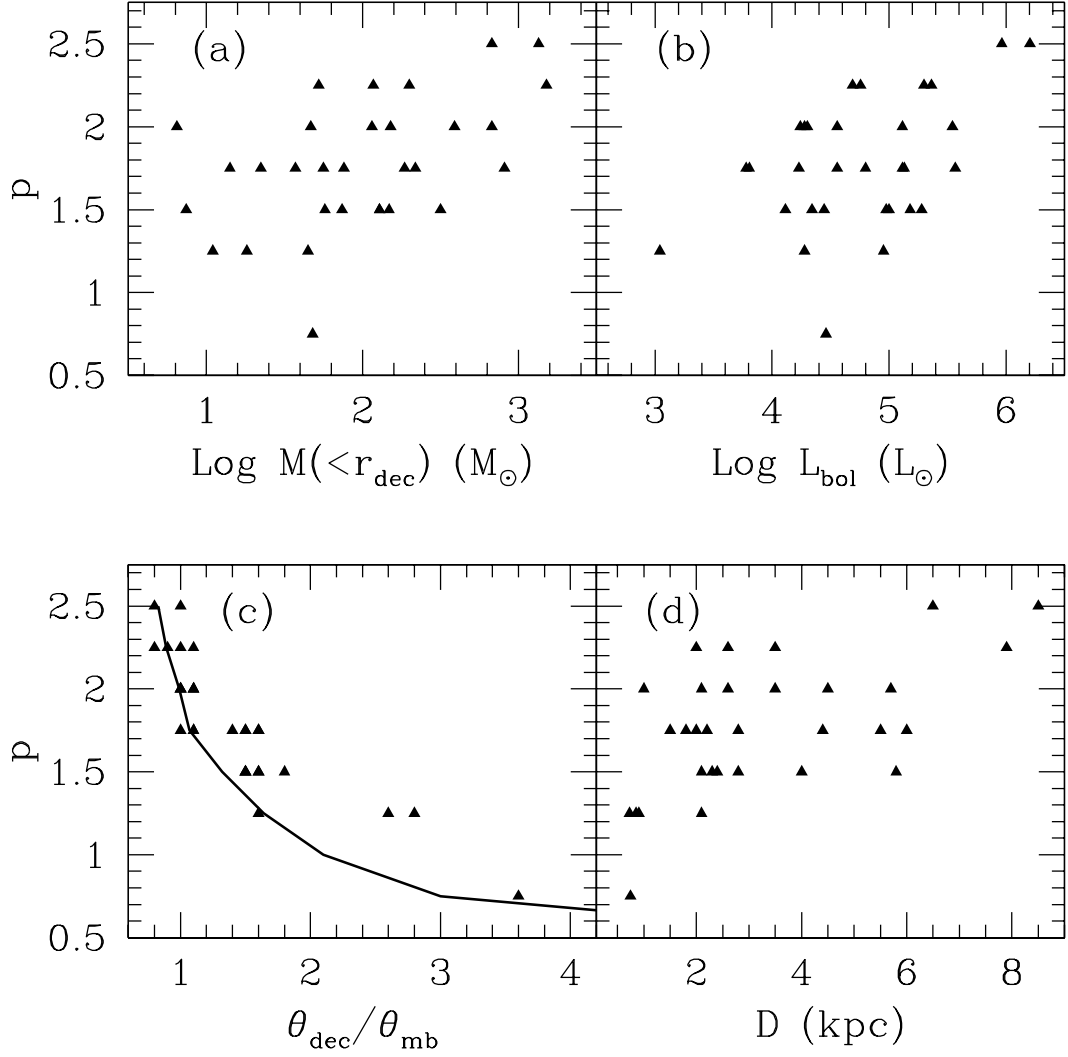}
\figcaption{\label{pfig}
Relationships with $p$: 
(a) the density power law index $p$ 
versus the logarithm of the mass within the 350
\micron\ half-power radius ($r$ = 0.51). 
(b) $p$ versus the logarithm of the modeled internal luminosity 
($r$ = 0.49). 
(c) $p$ versus the ratio of FWHM source size to beam size
 ($r = -$0.86). Only sources with $\theta_{dec}/\theta_{mb} \ge 0.8$ were modeled.
Smaller sources have steeper density profiles. The solid line represents 
models of dust emission with $p$ = 0.5 to 2.5. Sources significantly 
above the line (e.g., S140 and S88B) have extended aspherical
emission. (d) $p$ versus distance 
($r$ = 0.60). 
}
\end{figure}

\clearpage

\begin{figure}
\plotone{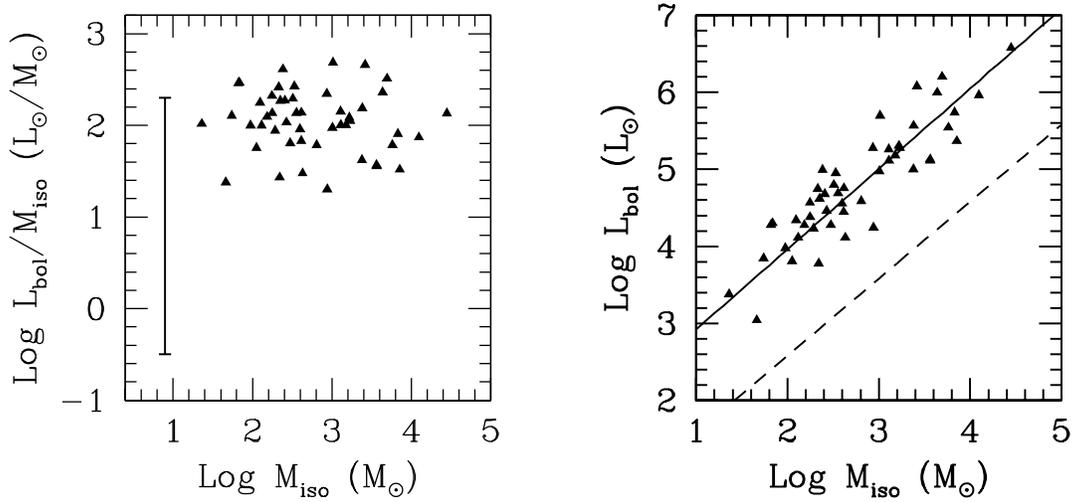}
\figcaption{\label{mlm}
Left: Plot of log($\lobs/\miso$) versus log(\miso). The dispersion in 
log($\lobs/\miso$) is about 1.5 orders of magnitude, significantly less
than in CO observations (2--3 magnitudes). The error bar shows the 
full range from the CO observations (Evans 1991).
Right: Plot of the logarithm
of \lobs\ versus the logarithm of \miso\ ($r = 0.89$). The solid line
is the least squares fit to the data log(\lobs) = 1.9 $+$ log(\miso). 
The dashed line is based on CO masses for molecular clouds containing
\HII\ regions, log(\lobs) = 
0.58 $+$ log(\miso) (Mooney \& Solomon 1988). The luminosity to mass ratio
is much higher in massive star forming cores than in molecular clouds.}
\end{figure}

\clearpage

\begin{figure}
\plotone{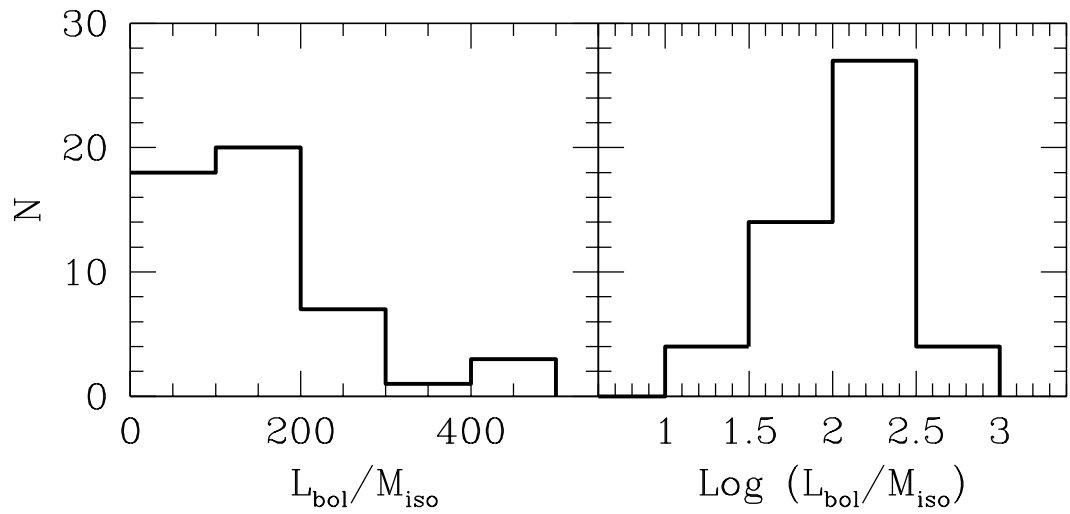}
\figcaption{\label{lmhist}
Distribution of the luminosity to mass ratio, $\lobs/\miso$, (left)
 and log ($\lobs/\miso$) (right). The histogram of $\lobs/\miso$ shows 
a tail of high ratios out to 490 \lsun/\msun.}
\end{figure}

\clearpage


\begin{deluxetable}{lrrrccrc}
\tablecolumns{8}
\tablecaption{Observed Properties \label{obstab}}
\tablewidth{0pt} 
\tablehead{
\colhead{Source}                &
\colhead{$\alpha$ (1950.0)}       &
\colhead{$\delta$ (1950.0)}       &
\colhead{$D$}          &
\colhead{Centroid$^c$}   &
\colhead{350\micron}    &
\colhead{$\theta_{ap}$}              &
\colhead{$D$}           \\
\colhead{}                      &
\colhead{($^h$~~$^m$~~$^s$~)~}    &  
\colhead{($\degree$ ~\arcmin\ ~\arcsec)} &    
\colhead{(kpc)}        &
\colhead{(\as,\as)}       &
\colhead{ $S_{\nu}$$^d$ (Jy)}         &    
\colhead{(\arcsec)}          &          
\colhead{Ref.}  
}

\startdata 
 
 IRAS 00338$+$6312  & 00 33 53.3 & 63 12 31 & 0.85 & ($-$10,$-$5) & 160$\pm$32  & 30 & 10  \\
			&		&  &   &  & 410$\pm$82  & 120 &   \\
 G123.07$-$6.31  & 00 49 29.2 &  56 17 36 & 2.2 & ($-$15,$-$9)& 160$\pm$32  & 30  &  6  \\
		&            &        &   &    & 290$\pm$58  & 120   &  \\  
 W3\tablenotemark{a,b}	& 02 21 53.1 &  61 52 20 & 2.3 & ($-$30,9) & 30$\pm$6  & 30  &  7  \\
		&            &           &  &  & 220$\pm$44  & 120   &  \\ 
 W3(OH)          & 02 23 17.3 &    61 38 58 & 2.4 & (7,$-$2)& 400$\pm$80  & 30  &  6  \\
		&		&	& &   & 1130$\pm$230  & 120 &   \\
 IRAS 02395$+$6244  & 02 39 31.0 &    62 44 16 & 8.1 & ($-$9,12) & 20$\pm$4  & 30 &  5  \\ 
 IRAS 02461$+$6147  & 02 46 11.7 &    61 47 34 & 4.5 & ($-$9,4) & 20$\pm$4 & 30 &  5  \\
 G137.07$-$3.00	& 02 54 11.2 &    56 17 36 & 4.9 & (1,10) & 20$\pm$4 & 30 &  6  \\
 GL490\tablenotemark{a}	& 03 23 38.9 & 58 36 33 & 0.9 & ($-$3,2) & 80$\pm$16 & 30 &  11   \\
		&            &          &  &    & 180$\pm$36  & 120   &   \\
 Ori-IRC2	& 05 32 47.0 & $-$05 24 24 & 0.45 & ($-$4,3) & 1940$\pm$390 & 30 &   5  \\
 		&		&	& &   & 7680$\pm$1540 & 120 &   \\
 S231		& 05 35 51.3 &    35 44 16 & 2.3 & (0,0) & 190$\pm$38  & 30 &  6   \\	
 		&		& &	&   & 522$\pm$100  & 120 &   \\
 S235\tablenotemark{b}	& 05 37 31.8 &  35 40 18 & 1.6 & (1,11) & 50$\pm$10  & 30 &  6  \\
 			&		&	&  &  & 240$\pm$48  & 120 &   \\
 S241    	& 06 00 40.9 &    30 14 54 & 4.7 & (8,9) & 30$\pm$6  & 30 &  6  \\
 			&		&	&  &  & 40$\pm$8  & 120 &   \\
 Mon R2\tablenotemark{b} & 06 05 17.0 & $-$06 22 40 & 0.9 & (15,10) & 150$\pm$30  & 30 & 6 \\
 		&		&	&  &  & 1400$\pm$280  & 120 &   \\
 S252A	    & 06 05 36.5 &    20 39 34 & 1.5 & (2,6)& 130$\pm$26  & 30 &  6  \\
 		&		&	&  &  & 320$\pm$64  & 120 &   \\
 RCW142     & 17 47 04.0 & $-$28 53 42 & 2.0 & ($-$5,8) & 530$\pm$110 & 30 &  6  \\
            &             &             &    & & 670$\pm$130 & 120   &   \\
 W28A2\tablenotemark{a} & 17 57 26.8 & $-$24 03 54 & 2.6 & (0,$-$3) & 950$\pm$190 & 30 & 6  \\
	    &            &           &  &  & 1580$\pm$320  & 120   &    \\ 
 M8E        & 18 01 49.1 & $-$24 26 57 & 1.8 & ($-$8,3) & 210$\pm$42  & 30 & 6  \\
	    &            &           &   &  & 380$\pm$76  & 120   &    \\
 G9.62$+$0.10 & 18 03 16.0 & $-$20 32 01 & 5.7 & (0,3) & 590$\pm$120  & 30 &  4  \\
            &            &           &   &  & 1150$\pm$230  & 120 &      \\
 G8.67$-$0.36 & 18 03 18.6 & $-$21 37 59 & 4.5 & (1,4) & 650$\pm$130  & 30 &  8  \\
	    &            &                &   &   & 1160$\pm$230  & 120 &    \\
 G10.60$-$0.40	& 18 07 30.7 &	$-$19 56 28 & 6.5 & (0,$-$1) & 1110$\pm$220  & 30   & 6  \\
	    &            &                  &  &    & 1900$\pm$380  & 120   &    \\
 G12.42+0.50 & 18 07 56.4 &  $-$17 56 37 & 2.1 & ($-$10,0) & 210$\pm$42  & 30 &   12  \\
	    &            &           &   &  & 440$\pm$88  & 120 &     \\
 G12.89+0.49 & 18 08 56.3 & $-$17 32 16 & 3.5 & (0,0) & 220$\pm$44  & 30   & 8  \\
	    &            &           &  &   & 340$\pm$68  & 120 &     \\
 G12.21$-$0.10 & 18 09 43.7 & $-$18 25 09 & 13.7 & ($-$5,5) & 230$\pm$46  & 30   & 2  \\
 G13.87$+$0.28   & 18 11 41.5 & $-$16 46 34 & 4.4 & (10,$-$8) & 190$\pm$38  & 30   & 1  \\
	    &            &           &   & & 430$\pm$86  & 120 &     \\
 W33A\tablenotemark{a}   & 18 11 44.0 & $-$17 53 09 & 4.0 & (3,4)& 350$\pm$70  & 30 &  5  \\
	    &            &           &   & & 960$\pm$190  & 120 &     \\
 G14.33$-$0.64 & 18 16 00.8 & $-$16 49 06 & 2.6 & (0,$-$4) & 440$\pm$88  & 30 &  8  \\
	    &            &           &  &  & 830$\pm$170  & 120 &     \\
 GL2136\tablenotemark{a}  & 18 19 36.6 & $-$13 31 40 & 2.0 & (8,$-$9) & 240$\pm$48  & 30 &  9  \\
	    &            &           &  &  & 520$\pm$100  & 120 &      \\
 G19.61$-$0.23   & 18 24 50.1 & $-$11 58 22 & 4.0 & (2,$-$6) & 460$\pm$92  & 30 & 6  \\
	    &            &        &   &    & 500$\pm$200  & 120 &     \\
 G23.95$+$0.16\tablenotemark{b}  & 18 31 40.8 & $-$07 57 17 & 5.8 & (7,0) & 100$\pm$20 &30 & 6 \\
	    &            &           &  &  & 320$\pm$64  & 120 &     \\
 G24.49$-$0.04   & 18 33 22.8 & $-$07 33 54 & 3.5 & (7,3) & 190$\pm$37  & 30 &  6  \\
		&            &         &  &    & 190$\pm$37  & 120 &     \\ 
 W43S		& 18 43 26.7 & $-$02 42 40 & 8.5 & (2,4) & 360$\pm$72  & 30 &  6  \\
		&            &        &   &    & 440$\pm$88  & 120 &     \\
 G31.41$+$0.31   & 18 44 59.5 & $-$01 16 07 & 7.9 & ($-$1,0) & 460$\pm$92  & 30 &  1   \\
		&            &        &   &    & 740$\pm$150  & 120 &      \\
 G40.50$+$2.54	& 18 53 45.6 &  07 49 16 & 2.1 & (2,1) & 240$\pm$48  & 30 &  12  \\
		&            &       &    &     & 600$\pm$120  & 120 &     \\
 G35.58$-$0.03   & 18 53 51.4 &    02 16 29 & 3.5 & (2,$-$3) & 110$\pm$22  & 30  &  6 \\
		&            &         &  &    & 120$\pm$24  & 120 &    \\
 G45.07$+$0.13   & 19 11 00.3 &    10 45 42 & 9.7 & ($-$2,$-$2) & 180$\pm$36  & 30  &  6  \\
 G48.61$+$0.02 	& 19 18 13.1 &    13 49 44 & 11.8 & (0,$-$2) & 100$\pm$20  & 30  &  6  \\
		&            &           &  &  & 200$\pm$40  & 120 &    \\
 W51M\tablenotemark{b}	& 19 21 26.2 &    14 24 36 & 7.0 & ($-$29,$-$18) & 280$\pm$56  & 30  &  7  \\
		&            &           &  &  & 3690$\pm$740  & 120   &    \\
 S87\tablenotemark{b}	& 19 44 14.0 &    24 28 10 & 2.3 & (0,$-$3) & 140$\pm$28  & 30  &  6 \\
		&            &           &  &  & 310$\pm$62  & 120  &  \\
 S88B		& 19 44 42.0 &    25 05 30 & 2.0 & (20,$-$10) & 150$\pm$30  & 30 &  6  \\
		&            &           &  &  & 540$\pm$110  & 120 &     \\
 ON 1		& 20 08 09.9 &    31 22 42 & 6.0 & (0,$-$10) & 320$\pm$64  & 30  &  6  \\
		&            &           &  &  & 650$\pm$130  & 120 &    \\  
 ON 2S		& 20 19 48.9 &    37 15 52 & 5.5 & (5,$-$2) & 200$\pm$40  & 30  &  6 \\
		&            &           &   & & 510$\pm$100  & 120 &     \\
 S106\tablenotemark{b}   & 20 25 32.8 &    37 12 54 & 4.1 & (0,$-$5) & 110$\pm$22  & 30   & 7 \\
		&            &           &   &  & 400$\pm$80  & 120 &     \\
 GL2591\tablenotemark{a} & 20 27 35.5 &    40 01 13 & 1.0 & (6,$-$2) & 230$\pm$46  & 30  &  3 \\
		&            &         &   &    & 440$\pm$88  & 120 &    \\
 G97.53$+$3.19  & 21 30 37.0 &    55 40 36 & 8.5 & (0,$-$12) & 50$\pm$10  & 30  &  5 \\
		&            &           &  &   & 90$\pm$18  & 120 &    \\
 BFS 11-B	& 21 41 57.6 &    65 53 17 & 2.0 & (1,$-$1) & 40$\pm$8  & 30  &  6  \\
 		&		& &	&   & 90$\pm$18  & 120 &   \\
 S140\tablenotemark{a}	& 22 17 41.1 &    63 03 42 & 0.90 & ($-$10,$-$4)& 350$\pm$70  & 30  &  12 \\
		&            &           & &   & 1210$\pm$240  & 120 &   \\
 CEP A		& 22 54 19.2 &    61 45 44 & 0.73 & ($-$1,$-$5) & 430$\pm$86 & 30  &  11 \\
 			&		&  &	&   & 1500$\pm$300  & 120 &   \\
 S158      & 23 11 36.1 &    61 10 30 & 2.8 & ($-$3,$-$4) & 250$\pm$50  & 30  &  6  \\
		&		&	&  & & 700$\pm$140  & 120 &   \\
 NGC 7538-1\tablenotemark{a} & 23 11 36.7 &   61 11 51 & 2.8 & ($-$1,1) & 150$\pm$30 & 30  &  6 \\
		&            &           &  &  & 1240$\pm$250 & 120   &  \\
 NGC 7538-9\tablenotemark{a} & 23 11 52.8 &   61 10 59 & 2.8 & (3,0) & 130$\pm$26  & 30  &  6  \\
		&            &              &   &  & 330$\pm$66  & 120   &   \\
 S157           & 23 13 53.1 &    59 45 18 & 2.5 & ($-$9,16) & 70$\pm$14  & 30 &   6 \\
		&		&	& &   & 280$\pm$56  & 120 &   \\

\enddata

\tablewidth{6.5in}
\tablenotetext{a}{Previously studied by van der Tak et al. 2000. $^b$Double or 
multiple peaks in 350 \micron\ map. $^c$350 \micron\ peak offset from the water 
maser position (0,0). $^d$ The 350 \micron\ flux density and uncertainty measured 
in an aperture of diameter $\theta_{ap}$. The uncertainty is 20\%. 
For sources with no reported 120\as\ aperture
flux density, the source was either very weak (IRAS 02395+6244, IRAS 02461+6147,
and G137.07$-$3.00) 
or the source was very centrally peaked and all of the flux is within a 30\as\
aperture (G12.21$-$0.10 and G45.07+0.13).} 
\tablerefs{ 1. Churchwell 1990; 2. Hunter 2000; 3. Mitchell 1992; 4. Olmi 1999; 5. Palagi 1993; 6. Plume 1992; 7. Plume 1997; 8. Val'tts 2000; 9. van der Tak 2000; 10. Yang 1991; 11. Zhou 1996; 12. Zinchenko 1994}

\end{deluxetable}

\clearpage
\begin{deluxetable}{lcccc}
\tablecolumns{5}
\tablecaption{Collected Photometry \label{sedtab}}
\tablewidth{0pt} 
\tablehead{
\colhead{Source}                &
\colhead{$\lambda$}    &
\colhead{\Snu$^b$}    &
\colhead{Beam$^c$}              &
\colhead{Ref.}               \\
\colhead{}                      &
\colhead{($\mu$m)}        &  
\colhead{(Jy)}          &            
\colhead{(\as)}                      &
\colhead{}                                 
}

\startdata 
 
 IRAS 00338$+$6312   & 12$^a$  & 1.8$\pm$0.2 & 300$\times$45  & 1 \\
		  & 25$^a$ & 21$\pm$1  & 300$\times$45 & 1  \\
		  & 60$^a$ & 357$\pm$21  & 90$\times$300 & 1  \\
		  & 100$^a$ & 685$\pm$55  & 180$\times$300 & 1  \\
		 & 143$^a$ & 1615$\pm$323  & 105 & 31  \\
		  & 185$^a$ & 2317$\pm$463  & 102 & 31  \\
		 & 350$^a$ & 160$\pm$32  & 30 &   \\
		 & 350  & 410$\pm$82  & 120 &  \\  
		  & 350 & 111$\pm$5  & 19 & 9  \\
		 & 450$^a$ & 49$\pm$2  & 18 & 9  \\
		 & 450 & 66$\pm$1.5  & 8 & 36  \\
		 & 800$^a$ & 6.2$\pm$0.02  & 16 & 9  \\
		 & 850$^a$ & 6.6$\pm$0.07  & 14.5 & 36  \\
		  & 850 & 17$\pm$3.4  & 18 & 24  \\
		 & 1100$^a$ & 2.3$\pm$0.044  & 18.7 & 9  \\
		 & 1100 & 5.6$\pm$1.1  & 18 & 30  \\
 G123.07$-$6.31  & 12$^a$ & $<$1.8 & 300$\times$45 & 1  \\
                & 25$^a$ & 13$\pm$1  & 300 $\times$ 45 & 1  \\
		 & 60$^a$ & 330$\pm$46 & 90$\times$300 & 1  \\
		  & 100$^a$ & 1166$\pm$117  & 180$\times$300 & 1  \\
		 & 350$^a$ & 160$\pm$32  & 30 &   \\
		  & 350 & 290$\pm$58  & 120 &  \\ 
 W3		 & 40 & 800$\pm$80 & 49 & 23  \\
		  & 350    & 2400$\pm$1200 & 15 & 35  \\	
		  & 350 & 30$\pm$6  & 30 &   \\
		 & 350 & 220$\pm$44  & 120 &   \\ 
	    & 400    & 500$\pm$50  & 49 & 23  \\
	    & 800    & 160$\pm$32  & 19 & 35  \\
	     & 1100   & 20$\pm$3   & 19 & 35  \\
 W3(OH)          & 20$^a$  & 270$\pm$27 & 49 & 23  \\
		 & 25$^a$ & 670$\pm$67  & 49 & 23  \\
		   & 30$^a$ & 1400$\pm$140  & 49 & 23  \\ 
		   & 35 & 25502$\pm$55  & 49 & 23  \\
		  & 40$^a$ & 4000$\pm$400  & 49 & 23  \\
		  & 40 & 4000$\pm$100  & 49 & 43  \\
		 & 50 & 5600$\pm$560  & 49 & 23  \\
		  & 58 & 6500$\pm$480  & 50 & 43  \\
		 & 60$^a$ & 7000$\pm$700  & 49 & 23 \\
		 & 70$^a$ & 8500$\pm$850  & 49 & 23  \\
		 & 80 & 9300$\pm$930  & 49 & 23  \\
		 & 85 & 9500$\pm$440  & 50 & 43  \\
    		& 90$^a$ & 9400$\pm$940  & 49 & 23  \\
		 & 100$^a$ & 9000$\pm$900  & 49 & 23  \\
		  & 120 & 8000$\pm$800  & 49 & 23  \\
		  & 138 & 6900$\pm$450  & 50 & 43  \\
		 & 140$^a$ & 6700$\pm$670  & 49 & 23  \\
		 & 160 & 5800$\pm$580  & 49 & 23  \\
		 & 180$^a$ & 4900$\pm$490  & 49 & 23 \\
		 & 200$^a$ & 4100$\pm$410  & 49 & 23  \\
		 & 250$^a$ & 2400$\pm$240  & 49 & 23  \\
		  & 300$^a$ & 1400$\pm$140  & 49 & 23  \\
		  & 350 & 400$\pm$80  & 30 &   \\
		 & 350$^a$  & 1130$\pm$230  & 120 &  \\
		 & 400 & 520$\pm$52  & 49 & 23  \\
		  & 500$^a$ & 250$\pm$25  & 49 & 23  \\
 		  & 600$^a$ & 135$\pm$14  & 49 & 23  \\
		& 800$^a$ & 51$\pm$5.1  & 49 & 23  \\
		 & 1000$^a$ & 24$\pm$2.4  & 49 & 23  \\
 IRAS 02395$+$6244 & 12  & 11$\pm$0.4  & 300$\times$45 & 1 \\
		 & 25  & 92$\pm$3.7 & 300$\times$45 & 1 \\
		 & 60  & 255$\pm$23  & 90$\times$300 & 1 \\
		 & 100 & 226$\pm$38  & 180$\times$300 & 1 \\  
		 & 350 & 20$\pm$4  & 30 &   \\
 IRAS 02461$+$6147 & 12  & 10$\pm$0.8  & 300$\times$45 & 1 \\
		 & 25  & 84$\pm$5 & 300$\times$45 & 1 \\
		 & 60  & 291$\pm$26  & 90$\times$300 & 1 \\
		 & 100 & 373$\pm$52  & 180$\times$300 & 1 \\   
		 & 350 & 20$\pm$4  & 30 &   \\
 G137.07$-$3.00	 & 12  & 0.35$\pm$0.03  & 300$\times$45 & 1 \\
		 & 25  & $<$0.25   & 300$\times$45 & 1 \\
		 & 60  & $<$0.40   & 90$\times$300 & 1 \\
		 & 100 & $<$12    & 180$\times$300 & 1 \\  
		 & 350 & 20$\pm$4 & 30 &   \\
 GL490		 & 12  & 82$\pm$2.5  & 300$\times$45 & 1 \\
		 & 25  & 278$\pm$8.3 & 300$\times$45 & 1 \\
		 & 60  & 717$\pm$29  & 90$\times$300 & 1 \\
		 & 100 & 785$\pm$55  & 180$\times$300 & 1 \\  
		 & 350 & 80$\pm$16  & 30 &   \\
		 & 350 & 180$\pm$36  & 120 &   \\
		 & 870 & 12$\pm$1.2 & 14  & 6 \\
		 & 1300 & 3.6$\pm$0.2 & 14  & 6  \\
 Ori-IRC2	 & 12  &  120$\pm$14  & 2.3 & 49 \\
		 & 20  &  240$\pm$29 & 2.3 & 49 \\
		 & 57  & 113000$\pm$11000  & 180 & 50 \\
		 & 138  & 219000$\pm$22000 & 180 & 50 \\
		 & 205  & 26000$\pm$2600 & 180 & 50 \\
		 & 350 & 1940$\pm$390 & 30 &   \\
		 & 350  & 7680$\pm$1540  & 120 &  \\
 S231		 & 12$^a$ & 5.6$\pm$0.2 & 300$\times$45 & 1 \\
		 & 25$^a$ & 75$\pm$3.7   & 300$\times$45 & 1  \\
		 & 60$^a$ & 722$\pm$72   & 90$\times$300 & 1  \\
		 & 100$^a$ & 1310$\pm$131 & 180$\times$300 & 1  \\
		 & 350$^a$ & 190$\pm$38  & 12 &   \\
		 & 350  & 522$\pm$100  & 120 &  \\	
 S235		 & 10$^a$ & 37$\pm$7 & 60 & 14  \\
		 & 12 & 28$\pm$1.7  & 300$\times$45 & 1  \\
		 & 20$^a$ & 340$\pm$70   & 60 & 14  \\
		 & 25 & 226$\pm$13.6   & 300$\times$45 & 1  \\
		 & 50$^a$ & 695$\pm$140   & 37 & 14   \\
		 & 60 & 179$\pm$171   & 90$\times$300 & 1  \\
		 & 100 & 1635$\pm$164   & 180$\times$300 & 1  \\
                & 100$^a$ & 740$\pm$150   & 37 & 14  \\
		 & 350$^a$ & 50$\pm$10  & 30 & \\
		 & 350  & 240$\pm$48  & 120 &  \\
 S241    	& 12 & 1.9$\pm$0.3 & 300$\times$45 & 1  \\
		 & 25 & 11.8$\pm$0.7   & 300$\times$45 & 1  \\
		 & 60 & $<$189    &90$\times$300 & 1  \\
		 & 100 & 552$\pm$77   & 180$\times$300 & 1  \\
		 & 350 & 30$\pm$6  & 30 &   \\
		 & 350  & 40$\pm$8  & 120 &  \\
Mon R2	    & 12     & 470$\pm$14   & 300$\times$45 & 1  \\
            & 20     & 500$\pm$250  &     & 25  \\
            & 20     & 2275      &     & 25  \\
            & 25     & 4095$\pm$164  & 300$\times$45 & 1  \\
            & 27     & 5866       &     & 25  \\
             & 40     & 12976      &     & 25  \\
             & 60     & 13070$\pm$1961 & 90$\times$300 & 1  \\
             & 93     & 18825      &     & 25  \\
             & 100    & 20200$\pm$3030 & 180$\times$300 & 1  \\
	      & 140    & 7200       &     & 25  \\
             & 200    & 3300$\pm$830 & 60 & 43  \\
	    & 350 & 150$\pm$30  & 30 &   \\
		 & 350  & 1400$\pm$280  & 120 &  \\
            & 390    & 660        &     & 25  \\
	    & 850 & 21$\pm$4  & 18 & 28  \\
            & 1000   & 58         &     & 25  \\
            & 1300   & 16.64$\pm$.76 & 30 & 45  \\
 S252A	    & 12$^a$ & 16$\pm$0.6 & 300$\times$45 & 1  \\
                & 25$^a$ & 77$\pm$3   & 300$\times$45 & 1  \\
		 & 60$^a$ & 10321$\pm$34   & 90$\times$300 & 1  \\
		 & 100$^a$ & 1715$\pm$189  & 180$\times$300 & 1  \\
		 & 350$^a$ & 130$\pm$26  & 30 &   \\
		 & 350  & 320$\pm$64  & 120 &  \\
 RCW142     & 12$^a$  & $<$42    & 300$\times$45 & 1  \\
	    & 25$^a$     & $<$281    & 300$\times$45 & 1  \\
            & 60$^a$     & 5476$\pm$986 & 90$\times$300 & 1  \\
            & 100$^a$    & 13129$\pm$1313 & 180$\times$300 & 1  \\
	     & 350$^a$   & 530$\pm$110 & 30 &   \\
	     & 350$^a$    & 670$\pm$130 & 120 &   \\
 W28A2      & 12$^a$    & 199$\pm$12  & 300$\times$45 & 1  \\
	    & 25$^a$     & 2190$\pm$131  & 300$\times$45 & 1  \\
            & 60$^a$     & 12790$\pm$3198 & 90$\times$300 & 1  \\
             & 100$^a$    & 26780$\pm$6695 & 180$\times$300 & 1  \\
             & 350    & 830$\pm$166 & 11 & 21 \\
	     & 350$^a$ & 950$\pm$190 & 30 &   \\
	    & 350$^a$ & 1580$\pm$320  & 120 &   \\ 
 M8E        & 10     & 87     &   4 & 39  \\
	    & 12     & 119$\pm$7 & 300$\times$45 & 1  \\
            & 20     & 178      &   4 & 39  \\
            & 25$^a$   & 289$\pm$17 & 300$\times$45 & 1  \\
            & 60$^a$     & 1611$\pm$226 & 90$\times$300 & 1  \\
 	    & 64     & $\leq 3600$   & 210 & 42 \\           
 	    & 69$^a$     & 2600      &  54 & 47  \\
            & 100$^a$    & 2783$\pm$696 & 180$\times$300 & 1  \\
            & 110    & 10000$\pm$3000 & 210 & 42  \\
            & 160    & 5200$\pm$1600 & 210 & 42 \\
	    & 350$^a$ & 210$\pm$42  & 30 &   \\
	     & 350$^a$ & 380$\pm$76  & 120 &   \\
	    & 450$^a$    & 42$\pm$15.8 & 19 & 44  \\
             & 850$^a$    & 9$\pm$1.01 & 25 & 44  \\
 G9.62$+$0.10  & 12$^a$ & 39$\pm$2.3 & 300$\times$45 & 1  \\
            & 25$^a$     & 292$\pm$18 & 300$\times$45 & 1  \\
             & 60$^a$     & 4106$\pm$411 & 90$\times$300 & 1  \\
            & 100$^a$    & 7844$\pm$1098 & 180$\times$300 & 1  \\
            & 350$^a$ & 590$\pm$120  & 30 &   \\
            & 350$^a$ & 1150$\pm$230  & 120 &   \\
	    & 1300$^a$ & 9.6    & 90 & 3  \\
	    & 2700$^a$ & 0.0098    & 3 & 20  \\
 G8.67$-$0.36  & 12  & 19$\pm$1  & 300$\times$45  & 1  \\
             & 25$^a$     & 254$\pm$8  & 300$\times$45 & 1  \\
	    & 60$^a$     & 1895$\pm$303 & 90$\times$300 & 1  \\	
	     & 100$^a$    & 5125$\pm$1128 & 180$\times$300 & 1  \\
	    & 350$^a$ & 650$\pm$130  & 30 &   \\
	     & 350$^a$ & 1160$\pm$230  & 120 &   \\
	    & 450$^a$    & 390$\pm$98  & 18 & 24  \\
	    & 850$^a$ & 49$\pm$10  & 18 & 24  \\
	     & 1300$^a$   & 7.1     & 90 & 3   \\
 G10.60$-$0.40	& 12$^a$ & $<$23   & 300$\times$45 & 1  \\
	    & 18  & 3.1$\pm$0.3  & 29  & 7  \\
             & 25$^a$     & $<$148  & 300$\times$45 & 1  \\
           & 60$^a$     & 9479$\pm$948 & 90$\times$300 & 1  \\
            & 69$^a$    & 14000   & 90 & 15  \\
            & 100$^a$    & 21375$\pm$3847 & 180$\times$300 & 1  \\
	    & 350$^a$ & 1110$\pm$220  & 30 &   \\
	     & 350$^a$ & 1900$\pm$380  & 120 &   \\
             & 1300$^a$   & 26   & 90 & 4  \\
 G12.42+0.50  & 12   & 10.7$\pm$1.1  & 300$\times$45 & 1  \\
	     & 20$^a$    & 100$\pm$10  & 49 & 22  \\
              & 25     & 253$\pm$15.2 & 300$\times$45 & 1  \\
	      & 40$^a$     & 760$\pm$76  & 49 & 22  \\
	     & 59$^a$     & 1490$\pm$149 & 49 & 22  \\
	     & 60     & 1418$\pm$255 & 90$\times$300 & 1  \\
             & 100    & 2380$\pm$595 & 180$\times$300 & 1  \\
             & 101$^a$    & 24202$\pm$42  & 49 & 22  \\
	     & 135$^a$    & 2270$\pm$227  & 49 & 22  \\
	     & 180$^a$    & 2100$\pm$210  & 49 & 22  \\
	     & 350$^a$ & 210$\pm$42  & 30 &   \\
	     & 350$^a$ & 440$\pm$88  & 120 &   \\
	    & 400$^a$    & 160$\pm$16   & 49 & 22  \\
 G12.89+0.49 & 12     & $<$6.1    & 300$\times$45 & 1  \\
	     & 20$^a$     & 30$\pm$3 & 49 & 22 \\
	     & 25     & 45$\pm$4.5  & 300$\times$45 & 1  \\
	     & 40$^a$     & 360$\pm$36 & 49 & 22  \\
             & 59$^a$     & 1100$\pm$110 & 49 & 22  \\
             & 60     & 1248$\pm$225 & 90$\times$300 & 1  \\
	     & 100    & 3148$\pm$787 & 180$\times$300 & 1  \\
             & 101$^a$    & 2200$\pm$220 & 49 & 22  \\
             & 135$^a$    & 2370$\pm$237 & 49 & 22  \\
              & 180$^a$    & 2370$\pm$237 & 49 & 22  \\
	     & 350$^a$ & 220$\pm$44  & 30 &   \\
	     & 350$^a$ & 340$\pm$68  & 120 &   \\
             & 400$^a$    & 210$\pm$21 & 49 & 22  \\
            & 450$^a$    & 200$\pm$50 & 18 & 25  \\ 
 G12.21$-$0.10 &  20$^a$  & 40$\pm$4  & 49 & 23 \\
	     & 40$^a$     & 420$\pm$42  & 49 & 23  \\
	     & 59$^a$     & 810$\pm$81 & 49 & 23  \\	
             & 101$^a$    & 2500$\pm$250  & 49 & 23   \\
	     & 165$^a$    & 1550$\pm$155  & 49 & 23  \\
             & 350    &  560$\pm$112 & 11 & 21  \\
	     & 350$^a$ & 230$\pm$46  & 30 &   \\
	     & 400$^a$    & 80$\pm$8   & 49 & 23  \\
	     & 450$^a$    & 94$\pm$19   & 9 & 18  \\
 	     & 850$^a$    & 14$\pm$0.7   & 15 & 18 \\
	     & 1350$^a$   & 3.2$\pm$0.6   & 22 & 18  \\
	     & 2000$^a$    & 1.1$\pm$0.2   & 34 & 18  \\
 G13.87$+$0.28    & 12  & 75$\pm$4.5 & 300$\times$45 & 1  \\
             & 25$^a$     & 478$\pm$29 & 300$\times$45 & 1  \\
             & 60$^a$     & 3632$\pm$509 & 90$\times$300 & 1  \\
             & 100    & $<$6141   & 180$\times$300 & 1  \\
	     & 350$^a$ & 190$\pm$38  & 30 &   \\
	     & 350$^a$ & 430$\pm$86  & 120 &   \\
	     & 450$^a$    & 24$\pm$5  & 9 & 18  \\
 	     & 850$^a$    & 5.2$\pm$0.3   & 15 & 18  \\
	     & 1300 & 7.3    & 90 & 4  \\
	     & 1350$^a$   & 2.2$\pm$0.4   & 22 & 18  \\
	     & 2000$^a$    & 1.50$\pm$.2  & 34 & 18  \\ 
 W33A        & 12.5   & 22$\pm$3  & 9 & 12  \\
	     & 20     & 50$\pm$20  & 9 & 12  \\
	     & 20$^a$     & 113$\pm$11  & 49 & 23  \\
             & 20     & 113$\pm$5  & 6.8 & 10  \\
            & 25$^a$     & 371$\pm$21  & 6.8 & 10  \\
	    & 25     & 268$\pm$21  & 300$\times$45 & 1  \\
	     & 33$^a$     & 539$\pm$36  & 6.8 & 10  \\
	    & 40$^a$     & 1000$\pm$100  & 49 & 23 \\
            & 42$^a$     & 1300$\pm$130  & 60 & 41  \\
	     & 59$^a$     & 2350$\pm$235 & 49 & 23  \\	
             & 60     & 2206$\pm$530 & 90$\times$300 & 1  \\
            & 73$^a$     & 3400$\pm$340 & 60 & 41  \\
	     & 77$^a$     & 4100$\pm$410 & 60 & 41  \\
            & 101$^a$    & 4050$\pm$405  & 49 & 23  \\
             & 100    & 6183$\pm$1422 & 180$\times$300 & 1  \\
	     & 135$^a$    & 3900$\pm$390  & 49 & 23  \\
             & 135    & 4000$\pm$400  & 60 & 1  \\
	     & 180$^a$    & 2750$\pm$275  & 49 & 23  \\
	    & 350$^a$ & 350$\pm$70  & 30 &   \\
	    & 350$^a$ & 960$\pm$190  & 120 &   \\
	    & 400$^a$    & 300$\pm$30   & 49 & 31  \\
  	    & 450$^a$    & 240$\pm$60   & 18 & 17  \\
	    & 850$^a$    & 45$\pm$9    & 18 & 17  \\
            & 1000$^a$   & 41$\pm$8    & 65   & 2  \\
            & 1300$^a$   & 11       & 90   & 3  \\
	    & 1300   & 7.3        & 90 & 4  \\
            & 1300   & 3.0       & 30 & 17  \\
 G14.33$-$0.64  & 12  & $<$6.4     & 300$\times$45 & 1 \\
	     & 20     & 40$\pm$4  & 49 & 22  \\
	    & 25     & 56$\pm$5.6  & 300$\times$45 & 1  \\
	     & 40$^a$     & 400$\pm$40  & 49 & 22  \\
 	    & 59$^a$     & 740$\pm$74  & 49 & 22   \\	
             & 60     & 994$\pm$249  & 90$\times$300 & 1  \\
             & 100    & 2819$\pm$705 & 180$\times$300 & 1  \\
	     & 101$^a$    & 2000$\pm$200  & 49 & 22  \\
	     & 135$^a$    & 2000$\pm$200  & 49 & 22  \\
	     & 180$^a$    & 2300$\pm$230  & 49 & 22  \\
	     & 350$^a$ & 440$\pm$88  & 30 &   \\
	     & 350$^a$ & 830$\pm$170  & 120 &   \\
	     & 400$^a$    & 120$\pm$12   & 49 & 22   \\
 GL2136	    & 11$^a$ & 40$\pm$8 & 180 & 29   \\
            & 350$^a$ & 240$\pm$48  & 30 &   \\
	     & 350$^a$ & 520$\pm$100  & 120 &   \\
	    & 450$^a$ & 72$\pm$6  & 19 & 26  \\
	     & 800$^a$ & 7.1$\pm$0.2  & 17 & 26  \\
	    & 800 & 6.7$\pm$0.2  & 17 & 26  \\
            & 1100$^a$ & 2.4$\pm$0.04  & 19 & 26  \\
	    & 1100 & 2.4$\pm$0.09  & 19 & 26  \\
	    & 1300$^a$ & 1.7$\pm$0.05  & 19 & 26  \\
	    & 2000$^a$ & 0.5$\pm$0.12  & 19 & 26  \\
 G19.61$-$0.23 & 12 & 48$\pm$4.3 & 300$\times$45 & 1  \\
            & 18 & 103$\pm$10 & 29 & 7  \\
            & 25     & 407$\pm$32.6 & 300$\times$45 & 1  \\
            & 60     & 4635$\pm$417 & 90$\times$300 & 1  \\
            & 100    & 7093$\pm$922 & 180$\times$300 & 1  \\
	    & 350 & 460$\pm$92  & 30 &  \\
	    & 350 & 500$\pm$200  & 120 &   \\
 G23.95$+$0.16    & 12$^a$ & 66.3$\pm$1.2 & 300$\times$45 & 1  \\
             & 25$^a$     & 395$\pm$6.2 & 300$\times$45 & 1  \\
            & 60$^a$     & 2285$\pm$708 & 90$\times$300 & 1  \\
             & 100$^a$    & 3339$\pm$902 & 180$\times$300 & 1  \\
             & 350    & 105   & 22 & 19  \\
	     & 350$^a$ & 100$\pm$20  & 30 &   \\
	    & 350$^a$ & 320$\pm$64  & 120 &   \\
             & 800$^a$    & 14.5   & 50 & 19  \\
 		 & 1300$^a$ & 4.6     & 90 & 3  \\ 
 G24.49$-$0.04    & 12$^a$ & 15.5$\pm$2.2 & 300$\times$45 & 1  \\
            & 25$^a$     & 81$\pm$8.1 & 300$\times$45 & 1  \\
            & 60$^a$     & 1476$\pm$88 & 90$\times$300 & 1  \\
            & 100$^a$    & 3514$\pm$14 & 180$\times$300 & 1  \\
		 & 350$^a$ & 190$\pm$37  & 30 &   \\
		 & 350$^a$ & 190$\pm$37  & 120 &   \\ 
 W43S		 & 12$^a$ & 218$\pm$20 & 300$\times$45 & 1  \\
                & 12.5$^a$ & 235      & 22 & 40  \\
 		 & 12.6  & 121     & 2 & 40  \\
		 & 19$^a$ & 610     & 12 & 40  \\
             & 25$^a$     & 1697$\pm$136 & 300$\times$45 & 1 \\
            & 60$^a$     & 7501$\pm$525 & 90$\times$300 & 1  \\
             & 100$^a$    & 11669$\pm$3151 & 180$\times$300 & 1  \\
		 & 350$^a$ & 360$\pm$72  & 30 &  \\
		  & 350$^a$ & 440$\pm$88  & 120 &  \\
	     & 1300$^a$ & 8    & 90 & 3  \\
		 & 1300     & 20$\pm$1.9  & 12 & 32  \\
 G31.41$+$0.31   & 12 & 4$\pm$0.4 & 300$\times$45 & 1 \\
		 & 25$^a$ & 52$\pm$5.2   & 300$\times$45 & 1  \\
		  & 60$^a$ & 1093$\pm$197   & 90$\times$300 & 1  \\
		  & 100$^a$ & 2815$\pm$394 & 180$\times$300 & 1  \\
		  & 350 & 997$\pm$200 & 11 & 1  \\
  	        & 350$^a$ & 460$\pm$92  & 30 &   \\
		  & 350$^a$ & 740$\pm$150  & 120 &   \\
		 & 450$^a$    & 84$\pm$17   & 9 & 18  \\
 	        & 850$^a$    & 27$\pm$1.4   & 15 & 18  \\
 	  	 & 1300 & 14    & 90 & 3  \\
	        & 1350$^a$   & 4.9$\pm$1.0   & 22 & 18  \\
	        & 2000$^a$    & 2.9$\pm$0.6   & 34 & 18  \\
 G40.50$+$2.54	 & 12$^a$ & 31.8$\pm$4.5 & 300$\times$45 & 1  \\
 		    & 25$^a$ & 242$\pm$24   & 300$\times$45 & 1  \\
		 & 60$^a$ & 2351$\pm$423   & 90$\times$300 & 1 \\
		 & 100$^a$ & 4218$\pm$840   & 180$\times$300 & 1  \\
		 & 350$^a$ & 240$\pm$48  & 30 &   \\
		 & 350$^a$ & 600$\pm$120  & 120 &   \\
		 & 450$^a$ & 215$\pm$54  & 18 & 24  \\
	         & 850$^a$ & 33$\pm$7  & 18 & 24  \\
 G35.58$-$0.03   & 12  & $<$6.0   & 300$\times$45  & 1  \\
		 & 25  & 77$\pm$3.9  & 300$\times$45  & 1  \\
		 & 60  & 1507$\pm$196  & 90$\times$300  & 1 \\
		 & 100 & 2594$\pm$571  & 180$\times$300  & 1 \\
		 & 350 & 110$\pm$22  & 30 &   \\
		 & 350 & 120$\pm$24  & 120 &   \\
 G45.07$+$0.13   & 12  & 58$\pm$3.5 & 300$\times$45 & 1  \\
		 & 25  & 494$\pm$30 & 300$\times$45 & 1  \\ 
		 & 60  & $<$5913   & 90$\times$300 & 1 \\
		 & 100 & $<$7497   & 180$\times$300 & 1 \\
		 & 350 & 180$\pm$36  & 30 &   \\
 G48.61$+$0.02 	  & 12$^a$ & $<$25   & 300$\times$45 & 1   \\
                & 25$^a$ & 175$\pm$10   & 300$\times$45 & 1  \\
		 & 60$^a$ & 3195$\pm$320   & 90$\times$300 & 1  \\
		 & 100$^a$ & 5227$\pm$523   & 180$\times$300 & 1  \\
		 & 350$^a$ & 100$\pm$20  & 30 &   \\
		 & 350$^a$ & 200$\pm$40  & 120 &   \\
 W51M		& 12 & 424$\pm$42  & 300$\times$45 & 1 \\
		& 25 & 4344$\pm$430   & 300$\times$45 & 1 \\
		& 100 & $<$ 26760    & 180$\times$300 & 1 \\ 
		& 350 & 280$\pm$56  & 30 &   \\
		  & 350 & 3690$\pm$740  & 120 &   \\
		 & 1300 & 27$\pm$2.7 & 25 & 37  \\
 S87		& 12 & 47$\pm$3.8 & 300$\times$45 & 1 \\
		 & 25 & 425$\pm$26 & 300$\times$45 & 1  \\
                & 60 & 3446$\pm$310 & 90$\times$300 & 1  \\
                & 100 & 5158$\pm$464 & 180$\times$300 & 1  \\
		 & 350 & 140$\pm$28  & 30 &   \\
		 & 350 & 310$\pm$62  & 120 &   \\
                 & 450 & 110$\pm$28  & 18   & 24  \\
		 & 850 & 17$\pm$3  & 18 & 24  \\
                & 1300 & 3.6    & 90  & 3  \\
 S88B		 & 10.2 & 0.04$\pm$0.01  &  17 & 34 \\
		 & 11.1 & 0.12$\pm$0.024 & 17 & 34   \\
		 & 12$^a$& 93$\pm$6.5 & 300$\times$45 & 1   \\
		& 12.6$^a$ & 22$\pm$4.4 & 17 & 34  \\
		  & 17$^a$ & 40$\pm$8.0   & 17 & 34  \\
	         & 25$^a$ & 1185$\pm$71.1   & 300$\times$45 & 1  \\
		 & 60$^a$ & 8686$\pm$1129   & 90$\times$300 & 1  \\
                & 100$^a$ & 13214$\pm$1057   & 180$\times$300 & 1  \\
	 	 & 350$^a$ & 150$\pm$30  & 30 &   \\
		 & 350$^a$ & 540$\pm$110  & 120 &   \\
		 & 1300$^a$ & 3.4    & 90 & 3  \\
 ON 1		 & 12$^a$  & 1.1$\pm$0.1  & 300$\times$45 & 1 \\
		   & 25$^a$ & 58.8$\pm$4.7   & 300$\times$45 & 1  \\
		  & 60$^a$ & 1431$\pm$115   & 90$\times$300 & 1  \\
		 & 100$^a$ & 3119$\pm$312  & 180$\times$300 & 1  \\
		 & 350$^a$ & 320$\pm$64  & 30 &   \\
		 & 350$^a$ & 650$\pm$130  & 120 &   \\  
 ON 2S		 & 12 & 330  &  & 8  \\
		 & 12$^a$ & 74$\pm$4.5   & 300$\times$45 & 1  \\
		 & 25$^a$ & 481$\pm$29   & 300$\times$45 & 1  \\
		  & 60 & 4200     &  & 8 \\
		 & 60$^a$ & 5446$\pm$545   & 90$\times$300 & 1  \\
		 & 100 & 13000     &  & 8  \\
		  & 100$^a$ &$<$6985     & 180$\times$300 & 1  \\
		  & 350$^a$ & 200$\pm$40  & 30 &   \\
		  & 350$^a$ & 510$\pm$100  & 120 &   \\
		  & 1300$^a$ & 9     & 90 & 4  \\
 S106		 & 12 & 204$\pm$12 & 300$\times$45 & 1  \\
		 & 25 & 2510$\pm$176   & 300$\times$45 & 1  \\
		  & 60 & 10100$\pm$1111  & 90$\times$300 & 1  \\
		 & 100 & 13100$\pm$1179   & 180$\times$300 & 1  \\
		 & 350 & 110$\pm$22  & 30 &   \\
		  & 350 & 400$\pm$80  & 120 &   \\
 GL2591	 & 12 & 439$\pm$26.3 & 300$\times$45 & 1 \\
		  & 12.6 & 680    & 6.6 & 28  \\
		  & 19.5$^a$ & 630    & 6.6 & 28  \\
		 & 23$^a$ & 920   & 6.6 & 28  \\
		  & 25 & 1112$\pm$67   & 300$\times$45 & 1  \\
		  & 60$^a$ & 4600$\pm$920 & 49 & 28  \\
		  & 60 & 5314$\pm$425   & 90$\times$300 & 1  \\
                   & 95$^a$ & 5800$\pm$300 & 49 & 28  \\
		  & 100 & 5721$\pm$858 & 180$\times$300 & 1  \\
                 & 110$^a$ & 5500$\pm$1100 & 49 & 28  \\
                    & 160$^a$ & 34006$\pm$80 & 49 & 28  \\
		    & 350$^a$ & 230$\pm$46  & 30 &   \\
		   & 350$^a$ & 440$\pm$88  & 120 &   \\
		   & 450$^a$ & 170$\pm$43  & 18 & 24  \\
		   & 850$^a$ & 19$\pm$4  & 18 & 24  \\
                    & 1300$^a$ & 5.7$\pm$1.57 & 30 & 45  \\
		    & 1300 & 2.49  & 30 & 17  \\
                    & 3300$^a$ & 0.115$\pm$0.034 & 75 & 38  \\
 G97.53$+$3.19     & 350 & 50$\pm$10  & 30 & \\
		 & 350 & 90$\pm$18  & 120 & \\
 BFS 11-B	 & 25 & 79$\pm$5  & 300$\times$45  & 1  \\
                & 60 & 688$\pm$76 & 90$\times$300 & 1  \\
                 & 100 & 1215$\pm$97 & 180$\times$300 & 1 \\
                & 175 & 410$\pm$110  &    & 25  \\
		  & 350 & 40$\pm$8  & 30 &   \\
		 & 350  & 90$\pm$18  & 120 &  \\
                & 450 & 60$\pm$15  &    & 25  \\
                & 800 & 4.2$\pm$8  &    & 25  \\
                & 1100 & 1.2     &   &  25  \\
 S140		 & 10 & 150$\pm$38 & 3.5 & 48  \\
		 & 12$^a$ & 332$\pm$40  & 30 & 48  \\
		  & 20$^a$ & 740$\pm$185  & 3.5 & 48  \\
		 & 25$^a$ & 1694$\pm$170  & 30 & 48  \\
		  & 35$^a$ & 5700$\pm$1425  & 34 & 48  \\
		 & 53$^a$ & 8200$\pm$2050  & 17 & 48  \\
		  & 60 & 11374$\pm$1200   & 60 & 48  \\
		  & 62$^a$ & 7600$\pm$130  & 49 & 48  \\
		  & 76$^a$ & 9200$\pm$150  & 49 & 48  \\
		   & 80$^a$ & 9900$\pm$2475  & 37 & 48  \\
		 & 100 & 8600$\pm$2150  & 38 & 48  \\
		 & 100 & 13000$\pm$1300  & 120 & 48  \\
		  & 101$^a$ & 7700$\pm$150  & 49 & 48  \\
                 & 111$^a$ & 7500$\pm$150  & 49 & 48  \\
		 & 162$^a$ & 4700$\pm$120  & 49 & 48  \\
 		 & 175$^a$ & 54001$\pm$350  & 45 & 48  \\
		  & 350 & 333$\pm$50  & 30 & 17  \\
		 & 350$^a$ & 350$\pm$70  & 30 &   \\
		 & 350$^a$ & 1210$\pm$240  & 120 &   \\
		 & 400$^a$ & 3508$\pm$8  & 49 & 48  \\
		  & 1300$^a$ & 1.4$\pm$0.25 & 30 & 17  \\
		 & 1300 & 15.9   & 30 & 45  \\
 CEP A		 & 12$^a$ & 170$\pm$60 & 300$\times$45 & 1  \\
		 & 20 & 416$\pm$.2  & 4 & 11  \\
		 & 25$^a$ & 860$\pm$215  & 300$\times$45 & 1  \\
		  & 30 & 27$\pm$5.65  & 30 & 45  \\
		 & 50$^a$ & 10600$\pm$2650  & 20 & 11  \\
		 & 60$^a$ & 17000$\pm$3400  & 90$\times$300 & 1  \\
 		  & 85  & 46500$\pm$12900  & 270 & 27  \\
		  & 100$^a$ & 230004$\pm$600  & 180$\times$300 & 1  \\
		  & 100$^a$ & 20200$\pm$5050  & 30 & 11  \\
                  & 125$^a$ & 33100$\pm$9170  & 50 & 13  \\
		  & 150 & 23400$\pm$6800  & 270 & 27  \\
		  & 350$^a$ & 430$\pm$86 & 30 &   \\
		 & 350$^a$  & 1500$\pm$300  & 120 &  \\
 	 	 & 400$^a$ & 2570$\pm$741 & 50 & 13  \\
		 & 450$^a$ & 737$\pm$140  & 20 & 33  \\
                 & 550 & 27900$\pm$6300 & 50 & 13  \\
                & 800$^a$ & 86$\pm$10  & 20 & 33  \\
		 & 865 & 84$\pm$18  & 8 & 16  \\
		  & 1300$^a$ & 26$\pm$8  & 40 & 16  \\
                & 1300 & 27$\pm$1.6 & 30 & 45  \\
 S158            & 12 & 243$\pm$7.3 & 300$\times$45 & 1  \\
		 & 25 & 1780$\pm$71   & 300$\times$45 & 1  \\
		 & 60 & 7073$\pm$495   & 90$\times$300 & 1  \\
		 & 100 & 14138$\pm$1131   & 180$\times$300 & 1  \\
 		  & 350 & 249$\pm$50 & 30 & 5  \\
		 & 350 & 250$\pm$50  & 30 &   \\
		 & 350  & 700$\pm$140  & 120 &  \\
		 & 1300 & 15   & 30 & 5  \\
 NGC 7538-1	 & 12.5$^a$ & 149$\pm$21 & 7.5 & 1 \\
		  & 20$^a$ & 250$\pm$50  & 6 & 46  \\
		 & 25$^a$ & 640$\pm$130  & 6 & 46  \\
   		 & 30$^a$ & 2300$\pm$700  & 40 & 46  \\
		 & 50$^a$ & 6700$\pm$2010  & 40 & 46  \\
		 & 100$^a$ & 11000$\pm$3300  & 55 & 46 \\
		  & 350$^a$ & 150$\pm$30 & 30 &    \\
		 & 350$^a$ & 1240$\pm$250 & 120 &    \\
		  & 1000$^a$ & 30$\pm$9  & 55 & 46  \\
 NGC 7538-9	 & 12.5$^a$ & 74$\pm$13 & 9 & 46  \\
		 & 20$^a$ & 124$\pm$30  & 6 & 46  \\
		 & 25$^a$ & 260$\pm$50  & 6 & 46  \\
		 & 30$^a$ & 500$\pm$150  & 40 & 46  \\
		 & 50$^a$ & 1300$\pm$390  & 40 & 46  \\
		 & 100$^a$ & 2700$\pm$810  & 55 & 46  \\
		& 350$^a$ & 130$\pm$26  & 30 &   \\
		 & 350$^a$ & 330$\pm$66  & 120 &   \\
		 & 1000$^a$ & 51$\pm$5  & 55 & 46  \\ 
 S157           & 12$^a$ & 29$\pm$3  & 300$\times$45 & 1  \\
		 & 25$^a$ & 233$\pm$12 & 300$\times$45 & 1   \\
                & 60$^a$ & 1759$\pm$123 & 90$\times$300 & 1  \\
                 & 100$^a$ & 264$\pm$303 & 180$\times$300 & 1  \\
		 & 350$^a$ & 70$\pm$14  & 30 &   \\
		& 350  & 280$\pm$56  & 120 &  \\
                & 850$^a$ & 5.9$\pm$1.2  & 18  & 24 \\

\enddata

\tablenotetext{a}{Flux value used in model. $^b$Observed flux density and 
uncertainty. 350 \micron\ flux 
densities with Beam = 30\as\ or 120\as\ are from this work. $^c$FWHM beam size 
for observed flux density.}
\tablewidth{6.5in}
\tablerefs{1. {\it IRAS} PSC, 2. Cheung 1980, 3. Chini 1986a, 4. Chini 1986b, 5. Chini 1986c, 6. Chini 1991, 7. De Buizer 2000, 8. Dent 1988, 9. Dent 1998, 10. Dyck 1977, 11. Ellis 1990, 12. Evans 1979, 13. Evans 1981a, 14. Evans 1981b, 15. Fazio 1978, 16. Gordon 1990, 17. Guertler 1991, 18. Hatchell 2000, 19. Hoare 1991, 20. Hofner 1996, 21. Hunter 2000, 22. Jaffe 1984, 23. Jaffe 2001 (priv.\ comm.), 24. Jenness 1995, 25. Jenness 1996, 26. Kastner 94, 27. Koppenaal 1979, 28. Lada 1984, 29. Lebofsky 1976, 30. McCutcheon 1995, 31. Mooherjea 1999, 32. Moony 1995, 33. Moriarity-Schieven 1991, 34. Pipher 1977, 35. Richardson 1989, 36. Sandell 2001, 37. Schoerb 1987, 38. Schwartz 1977, 39. Simon 1985, 40. Soifer 1975, 41. Stier 1984, 42. Thronson 1979, 43. Thronson 1980, 44. Tothill 1999, 45. Walker 1990, 46. Werner 1979, 47. Wright 1977, 48. Zhou 1994, 49. Gezari 1998, 50. Mookerjea 2000}

\end{deluxetable}

\clearpage
\begin{deluxetable}{lccccc}
\tablecolumns{6}
\tablecaption{Modeled and Observed Source Properties \label{modtab}}
\tablewidth{0pt} 
\tablehead{
\colhead{Source}                &
\colhead{$p^a$}       &
\colhead{$n_f$$^b$}         &
\colhead{\lobs$^c$}       &
\colhead{\tbol$^d$}     &
\colhead{$\theta_{dec}/\theta_{mb}$$^e$}             \\
\colhead{}                      &
\colhead{}    &     
\colhead{($10^7$ \cmv)}        &
\colhead{($10^4$ \lsun)} & 
\colhead{(K)}                    &
\colhead{}
}

\startdata 

IRAS 00338$+$6312 & 1.25  & 0.4 & 0.11  & 46 & 1.6 \\
G123.07$-$6.31    & 1.75  & 3.0 & 0.60   & 51 & 1.1 \\
W3              &  --    & -- & 2.4 & 77 & 1.0 \\
W3(OH)		& 1.50  & 1.5 & 9.5    & 65 & 1.6 \\
IRAS 02395+6244 &  --   & -- & 5.6  & 107 & 0.8 \\
IRAS 02461+6147 &  --   & -- & 1.9 & 99 & 1.1 \\
GL490           &  --   & -- & 0.24  & 112 & 1.5 \\
Ori-IRC2        &  --   & -- & 9.9 & 73 & 1.5 \\
S231            & 1.50  & 1.0& 1.3   & 63 & 1.5 \\
S235            &  --    & -- & 0.95  & 125 & 2.2 \\
S241            &  --    & -- & 1.3 & 62 & 1.0 \\
MonR2           &  --    & -- & 3.7 & 98 & -- \\
S252A		& 1.75   & 1.0 & 0.64   & 68 & 1.0 \\
RCW142          & 2.25   & 6.5 & 5.7   & 64 & 1.0 \\
W28A2           & 2.25  & 25 & 20   & 60 & 0.9 \\
M8E 		& 1.75	 & 1.2 & 1.7  & 93  & 1.6 \\
G9.62$+$0.10 	& 2.00	 & 17 & 35   & 68  & 1.1 \\
G8.67$-$0.36 	& 2.00	& 15 & 13    & 67  & 1.0 \\
G10.60$-$0.40 	& 2.50	& 120 & 92    & 52  & 1.0 \\
G12.42$+$0.50 	& 2.00	 & 3.2 & 1.9   & 70 & 1.1 \\
G12.89$+$0.49 	& 2.00   & 5.7 & 3.9   & 56 & 1.0 \\
G12.21$-$0.10   &  --   & -- & 55  & 57 & 0.5 \\
G13.87$+$0.28	& 1.75	 & 2.0 & 13   & 93 & 1.6 \\
W33A		& 1.50	 & 2.0 & 10  & 65 & 1.5 \\
G14.33$-$0.64	& 2.00   & 1.8 & 10   & 55 & 1.0 \\
GL2136		& 1.75  & 6.3 & 1.0     & 173 & 1.1 \\
G19.61$-$0.23   &  --   & -- & 18  & 74 & 1.1 \\
G23.95$+$0.16	& 1.50	 & 0.5 & 19  & 91 & 1.5 \\
G24.49$-$0.04   & 2.25   & 8.5 & 4.9    & 65 & 1.1 \\
W43S            & 2.50  & 60 & 160   & 93 & 0.8 \\
G31.41$+$0.31   & 2.25   & 80 & 23  & 55 & 0.8 \\
G40.50$+$2.54	& 1.50	 & 0.8 & 2.8   & 75 & 1.6 \\
G35.58$-$0.03   &  --   & -- & 4.2  & 66 & 1.3 \\
G45.07$+$0.13   &  --    & -- & 120 & 79 & 1.3 \\
G48.61$+$0.02   &  --   & -- & 100  & 70 & 0.5 \\
W51M		&  --   & -- & 380  & 88 & 1.0 \\
S87             &  --   & -- & 4.8  & 80 & 1.3 \\
S88B		& 1.25	& 0.1  & 9.0   & 75 & 2.8 \\
ON1		& 1.75   & 7.0 & 15   & 57 & 1.4 \\
ON2S		& 1.75	 & 2.0 & 37  & 80 & 1.5 \\
S106            &  --   & -- & 50   & 94 & 0.5 \\
GL2591		& 2.00	& 0.9 & 2.0    & 86 & 1.1 \\
BFS 11-B        &  --    & -- & 0.70  & 64 & -- \\
S140		& 1.25 	 & 0.3 & 1.9   & 91 & 2.6 \\
Cep A		& 1.50	 & 0.5 & 2.2   & 84 & 1.8 \\
S158            &  --   & -- & 19  & 91 & 1.0 \\
NGC 7538-1	& 1.50	 & 1.2 & 15  & 74 & 1.8 \\
NGC 7538-9 	& 1.75	 & 1.2 & 3.6   & 81 & 1.5 \\
S157		& 0.75  & 0.01 & 2.9    & 86 & 3.6 \\

\enddata

\tablewidth{6in}
\tablenotetext{a}{The best fit model power law density distribution exponent. 
$^b$The model density at 1000 AU. $^c$The bolometric luminosity calculated from 
the SED. $^d$The temperature of a blackbody
with the same mean frequency of the SED. $^e$The ratio of the deconvolved source 
size to the beam size. Sources with $\theta_{dec}/\theta_{mb} < 0.8$ were not 
modeled.}

\end{deluxetable}

\clearpage
\begin{deluxetable}{lcccccc}
\tablecolumns{9}
\tablecaption{Masses and Sizes\label{masstab}}
\tablewidth{0pt} 
\tablehead{
\colhead{Source}                &
\colhead{\rdust$^a$}    &
\colhead{\md$^b$}    & 
\colhead{$r_n$$^c$}    &
\colhead{\mn$^d$}    &  
\colhead{\miso$^e$}    &  
\colhead{$\lobs/\miso$}     \\
\colhead{}         &  
\colhead{(\eten4\ AU)}          &
\colhead{(\msun)} &
\colhead{(\eten4\ AU)}          &
\colhead{(\msun)} &
\colhead{(\msun)}        &
\colhead{(\lsun / \msun)}               
}

\startdata 

IRAS 00338$+$6312 & 1.0 & 10 & 12 & 810  & 50 & 24 \\
G123.07$-$6.31    & 1.9 & 80 & 9.4 & 570  & 220 & 27 \\
W3                 & 1.6 & -- & -- & --   & 180 & 140 \\
W3(OH)		   & 2.9 & 130 & 13 & 1210  & 1010 & 94 \\
IRAS 02395$+$6244  & 4.6 & -- & -- & --   & 210 & 260 \\
IRAS 02461$+$6147  & 3.8 & -- & -- & --  & 70 & 290 \\
G137.07$-$3.00    & 1.9 & -- & -- & --   & 80 & -- \\
GL490              & 1.0 & -- & -- & --  & 20 & 100 \\
Ori-IRC2	    & 0.5 & -- & -- & -- & 240 & 410 \\
S231               & 2.7 & 70 & 9.9 & 530  & 430 & 30 \\
S235                & 2.7 & -- & -- & --  & 100 & 100 \\
S241                & 3.4 & --& -- & --  & 130 & 100 \\
MonR2              & --  & --& -- & --  & 180 & 210 \\
S252A		    & 1.2 & 10 & 5.0 & 90   & 110 & 57 \\
RCW142             & 1.4 & 50 & 4.8 & 130   & 410 & 140 \\
W28A2     	   & 1.4 & 200 & 9.0 & 790 & 1660 & 120 \\
M8E 		    & 2.2 & 40 & 5.6 & 120   & 190 & 88 \\
G9.62$+$0.10 	   & 4.8 & 670 & 12 & 1660  & 5780 & 61 \\
G8.67$-$0.36 	   & 3.2 & 390 & 12 & 1470  & 3650 & 36 \\
G10.60$-$0.40 	    & 4.6 & 1330 & 10 & 1960 & 12500 & 74 \\
G12.42$+$0.50 	    & 1.8 & 50 & 5.4 & 140  & 300 & 64 \\
G12.89$+$0.49 	    & 2.5 & 120 & 7.5 & 350  & 640 & 61 \\
G12.21$-$0.10      & 5.3 & -- & -- & --  & 6770 & 81 \\
G13.87$+$0.28	  & 5.3 & 190 & 7.7 & 300  & 1290  & 100 \\
W33A		  & 4.4 & 320 & 15 & 1990   & 2390 & 42 \\
G14.33$-$0.64	    & 1.9 & 150 & 9.8 & 800  & 880 & 20 \\
GL2136		   & 1.7 & 20 & 5.2 & 90    & 320 & 200 \\
G19.61$-$0.23     & 3.4 & -- & -- & --   & 1290 & 140 \\
G23.95$+$0.16	 & 6.4 & 130 & 5.9 & 120   & 1690 & 110 \\
G24.49$-$0.04     & 3.0 & 120 & 5.6 & 190   & 360 & 140 \\
W43S              & 4.8 & 680 & 8.3 & 890  & 4940 & 320 \\
G31.41$+$0.31     & 4.5 & 1500 & 15 & 3830  & 7150 & 33 \\
G40.50$+$2.54	   & 2.5 & 60 & 8.5 & 350  & 410 & 68 \\
G35.58$-$0.03      & 3.4 & --& -- & --   & 230 & 190 \\
G45.07$+$0.13     & 9.5 & --& -- & --   & 2620 & 460 \\
G48.61$+$0.02      & 4.6 & -- & -- & --  & 4350 & 230 \\
W51M              & 5.0 & --& -- & --   & 28100 & 140 \\
S87                 & 2.2 & -- & -- & --  & 260 & 190 \\
S88B		  & 4.4 & 50 & 4.6 & 50    & 260 & 270 \\
ON1		   & 6.3 & 810 & 15 & 2390  & 3650 & 37 \\
ON2S		  & 6.0 & 220 & 7.2 & 270   & 2410 & 150 \\
S106               & 0.2 & -- & -- & --  & 1030& 490 \\
GL2591  	  & 0.8 & 10 & 2.9 & 20   & 68  & 290 \\
G97.53$+$3.19      & 9.3 & -- & -- & -- & 1030 & -- \\
BFS 11-B            & -- & -- & -- & --  & 60 & 130 \\
S140		  & 1.8 & 20 & 8.0 & 250   & 150 & 120 \\
Cep A		  & 1.0 & 10 & 11 & 280   & 120 & 180 \\
S158               & 2.0 & -- & -- & --   & 860 & 220 \\
NGC 7538-1	  & 3.7 & 150 & 11 & 750   & 1510 & 100 \\
NGC 7538-9	   & 3.1 & 60 & 5.6 & 120   & 400 & 180 \\
S157		   & 6.8 & 50 & 3.9 & 10   & 270 & 110 \\

\enddata

\tablewidth{6.5in}
\tablenotetext{a}{The radius of the deconvolved FWHM size, $\theta_{dec}$, of
the 350 \micron\ emission. $^b$The integrated mass within $r_{dec}$. $^c$The 
density equals $10^4$ \cmv\ at this radius. $^d$The integrated mass within 
$r_{n}$. $^e$Calculated with the isothermal temperature from Eq. (6) for
the modeled sample and $\langle T_{iso} \rangle$ = 29 K for the sources that 
were not modeled.}

\end{deluxetable}


\begin{references}

\reference{} Allers, K. N., Jaffe, D. T., Evans, N. J. II, Plume, R., \&
    van Dishoeck, E. F. 2002, in prep.
\reference{} Andr\'{e}, P., Ward-Thompson, D., \& Barsony, M. 1993, \apj, 406, 122
\reference{} Benford, D. J., Cox, P., Omont, A., Phillips, T. G., \& McMahon, R. G. 1999, \apj, 518, L65
\reference{} Beuther, H., Schilke, P., Menten, K. M., Motte, F., Sridharan, T. K., \& Wyrowski, R. 2002, \apj, 566, 945
\reference{} Blain, A. W., Smail, I., Ivison, R. J., Kneib, J. -P., Frayer, D. T. 2002, Phys. Reports, in press (astro-ph/0202228)
\reference{} Blitz, L., Fich, M., \& Stark, A. A., 1982, ApJS, 49, 183
\reference{} Bronfman, L., Casassus, S., May, J., Nyman, L.-$\mathring{A}$. 2000, A\&A, 358, 521
\reference{} Bronfman, L., Nyman, L.-$\mathring{A}$., May, J. 1996, A\&AS, 115, 81 
\reference{} Calzetti, D., Armus, L., Bohlin, R. C., Kinney, A. L., Koornneef, J., \& Storchi-Bergmann, T. 2000, \apj, 533, 682
\reference{} Carpenter, J. M., 2000, \aj, 120, 3139
\reference{} Cesaroni, R., Palagi, F., Felli, M, Catarzi, M., Comoretto, G., Di Franco, S., Giovanardi, G., \& Palla, F. 1988, A\&AS, 76, 445
\reference{} Chen, H., Myers, P. C., Ladd, E. F., \& Wood, D. O. S. 1995, ApJ, 445, 377
\reference{} Chini, R., Henning, Th., Pfau, W. 1991, A\&A, 247, 157
\reference{} Chini, R., Kreysa, E., Mezger, P. G., \& Gemuend, H. -P. 1986a, A\&A, 154, L8
\reference{} Chini, R., Kreysa, E., Mezger, P. G., \& Gemuend, H. -P. 1986b, A\&A, 157, L1
\reference{} Chini, R., Kruegel, E., \& Kreysa, E. 1986c, A\&A, 167, 315
\reference{} Cheung, L. H., Frogel, J. A., Gezari, D. Y., \& Hanser, M. G. 1980, ApJ, 240, 74
\reference{} Churchwell, E., Wolfire, M. G., \& Wood, D. O. S. 1990, \apj, 354, 247
\reference{} De Buizer, J. M., 2000, PhD Thesis
\reference{} Dent, W. R. F., MacDonald, G. H., \& Anderson, M. 1988, MNRAS, 235, 1379
\reference{} Dent, W. R. F., Matthews, H. E., \& Ward-Thompson, D. 1998, MNRAS, 301, 1049
\reference{} Draine, B. T. \& Lee, H. M. 1984, 285, 89
\reference{} Dunne, L. \& Eales, S. A. 2001, MNRAS, 327, 697
\reference{} Dyck, H. M. \& Simon, T. 1977, ApJ, 211, 421
\reference{} Egan, M. P., Leung, C. M., \& Spagna, G. R., 1988, Comput. Phys. Comm., 48, 857
\reference{} Elitzur, M., Hollenbach, D. J., \& McKee, C. F. 1989, \apj, 346, 983
\reference{} Ellis Jr., H. B., Lester, D. F., Harvey, P. M., Joy, M., Telesco, C. M., Decher, R., \& Werner, M. W. 1990, ApJ, 365, 287
\reference{} Emerson, D. T., Klein, U. \& Haslam, C. G. T., 1979, A\&A, 76, 92
\reference{} Evans, N. J. II, 1991, in Frontiers of Stellar Evolution, ed. D. L.
Lambert, ASP Conf. Ser., 20, 45
\reference{} Evans, N. J. II, 1999, ARA\&A, 37, 311
\reference{} Evans, N. J. II, et al.\ 1981a, \apj, 244, 115
\reference{} Evans, N. J. II, Beckwith, S., Brown, R. L., \& Gilmore, W. 1979, \apj, 227, 450
\reference{} Evans, N. J. II, Beichman, C., Gatley, I., Harvey, P., Nadeau, D., \& Sellgren, K. 1981b, ApJ, 246, 409
\reference{} Evans, N. J. II, Rawlings, J. M. C., Shirley, Y. L., \& Mundy, L. G. 2001, \apj, 557, 193
\reference{} Evans, N. J. II, Shirley, Y. L.,  Mueller, K. E., \& Knez, C., 
2002, in Hot Star Workshop III: The Earliest Phases of Massive Star Birth, 
ed. P. A. Crowther, ASP Conf. Ser. 267, in press.
\reference{} Fazio, G. G., Lada, C. J., Kleinmann, D. E., Wright, E. L., Ho, P. T. P., \& Low, F. J. 1978, ApJ, 221, L77
\reference{} Garay, G. \& Lizano, S. 1999, PASP, 111, 1049
\reference{} Gezari, D. Y., Backman, D. E., Werner, M. W. 1998, ApJ, 509, 283
\reference{} Gordon, M. A. 1990, \apj, 352, 636
\reference{} Guertler, J., Henning, Th., Kruegel, E., \& Chini, R. 1991, A\&A, 252, 801
\reference{} Hatchell, J., Fuller, G. A., Millar, T. J., Thompson, M. A., \& MacDonald, G. H. 2000, A\&A, 357, 637
\reference{} Hoare, M. G., Roche, P. F., \& Glencross, W. M. 1991, MNRAS, 251, 584 
\reference{} Hofner, P., Kurtz, S., Churchwell, E., Walmsley, C. M., \& Cesaroni, R. 1996, \apj, 460, 359
\reference{} Hunter, T. R. 1997, Ph.D. Thesis, Caltech
\reference{} Hunter, T. R., Benford, D. J., Serabyn, E. 1996, \pasp, 108, 1042 
\reference{} Hunter, T. R., Churchwell, E., Watson, C., Cox, P., Benford, D. J.,
\& Roelfsema, P. R. 2000, \apj , 119, 2711
\reference{} Infrared Astronomical Satellite Catalogs and Atlases, vol. 1, Explanatory Supplement. 1988, ed. C. Beichman, et al.\ (NASA RP-1190) (Washington:GPO) ({\it IRAS} PSC)
\reference{} Ivezi$\acute{c}$, Z. \& Elitzur, M. 1997, MNRAS, 287, 799
\reference{} Jaffe, D. T., Hildebrand, R. H., Keene, J., Harper, D. A., Loewenstein, R. F., Moran, J. M. 1984, \apj, 281, 225
\reference{} Jenness, T., Scott, P. F., Padman, R. 1995, MNRAS, 276, 1024
\reference{} Jenness, T. 1996, PhD Thesis
\reference{} Kastner, J. H., Weintraub, D. A., Snell, R. L., Sandell, G., Aspin, C., Hughes, D. H., \& Bass, F. 1994, ApJ, 425, 695
\reference{} Kennicutt Jr., R. C., 1998, ARA\&A, 36, 189 
\reference{} Koppenaal, K., van Duinen, R. J., Aalders, J. W. G., Sargent, A. I., \& Nordh 1979, AA, 75, L1
\reference{} Kurtz, S., Churchwell, E., Wood, D. O. S. 1994, ApJS, 91, 659
\reference{} Lada, C. J. \& Wilking, B. A. 1984, \apj, 287, 610
\reference{} Lada, C. J., Thronson Jr., H. A., Smith, H. A., Schwartz, P. R., \& Glaccum, W.  1984, ApJ, 286, 302
\reference{} Lada, C. J. 1987, in IAU Symp. 115, Star Formation Regions, ed. M. Peimbert \& J. Jugaku (Dordrecht:Reidel), 1 
\reference{} Lebofsky, M. J., Kleinmann, S. G., Reike, G. H., Low, F. J. 1976, ApJ, 206, L157
\reference{} McCutcheon, W. H., Sato, T., Purton, C. R., Matthews, H. E., \& Dewdney 1995, AJ, 110, 1762
\reference{} McKee C. F. \& Tan, J. C. 2002, Nature, 416, 59
\reference{} McLaughlin, D. E. \& Pudritz, R. E. 1997, ApJ, 476, 750
\reference{} Mead, K. N., Kutner, M. L., \& Evans, N. J. II 1990, 
  \apj, 354, 492
\reference{} Mitchell, G. F., Hasegawa, T. I., \& Schella, J. 1992, \apj, 386, 694 
\reference{} Mookerjea, B., Ghosh, S. K., Karnik, A. D., Rengarajan, T. N., Tandon, S. N., \& Verna, R. P. 1999, ApJ, 522, 285
\reference{} Mookerjea, B., Ghosh, S. K., Rengarajan, T. N., Tandon, S. N., \& Verna, R. P. 2000, AJ, 120, 1954
\reference{} Mooney, T. J., \& Solomon, P. M. 1988, \apj, 334, L51
\reference{} Mooney, T. J., Shivers, A., Mezger, P. G., Solomon, P. M., Krey, S. E., Haslam, C. G. T. \& Lemke, R. 1995, A\&A, 299, 869
\reference{} Moriarty-Schieven, G. H., Snell, R. L., \& Hughes, V. A. 1991, ApJ, 374, 169 
\reference{} Myers, P. C., \& Ladd, E. F. 1993, \apj, 413, L47
\reference{} Olmi, L. \& Cesaroni, R. 1999, A\&A, 352, 266
\reference{} Omont, A., Cox, P., Bertoldi, F., McMahon, R. G., Carilli, C., \& Isaak, K. G. 2002, A\&A, in press (astro-ph/0107005)
\reference{} Osorio, M., Lizano, S., \& D'Alessio, P. 1999, \apj, 525, 808
\reference{} Palagi, F., Cesaroni, R., Comoretto, G., Felli, M., \& Natale, V. 1993, A\&AS, 101, 153
\reference{} Peeters, E., et al.\ 2002, A\&A, 381, 571
\reference{} Pipher, J. L., et al.\ 1977, A\&A, 59, 215
\reference{} Plume R., Jaffe, D. T., \& Evans, N. J. II,  1992, ApJS, 78, 505
\reference{} Plume R., Jaffe, D. T., \& Evans, N. J. II, Martin-Pintado, J., 
Gomez-Gonzalez, J. 1997, \apj, 476, 730
\reference{} Richardson, K. J., White, G. J., Sandell, G., Duncan, W. D., \& Krisciunas, K. 1989, A\&A, 221, 95
\reference{} Sandell, G. 1994, MNRAS, 271, 75
\reference{} Sandell, G. \& Weintraub, D. A. 2001, ApJS, 134, 115
\reference{} Scalo, J. M. 1986, Fund. Cos. Phys., 11, 1
\reference{} Schloerb, F. P., Snell, R. L., \& Schwartz, P. R. 1987, ApJ, 319, 426
\reference{} Shu, F. H. 1977, ApJ, 214, 488
\reference{} Schwartz, P. R. \& Spencer, J. H. 1977, MNRAS, 180, 29
\reference{} Schweitzer, F. 1998, in Saas-Fee Advanced Course 26, Galaxies: Interactions and Induced Star Formation, ed. D. Friedli, L. Martinet, \& D. Pfenniger (Berlin: Springer), 105
\reference{} Scott, P. F. 1978, MNRAS, 183, 435
\reference{} Shirley, Y. L., Evans, N. J. II, \& Rawlings, J. M. C. Gregerson, E. M., 2000, ApJS, 131, 249 
\reference{} Shirley, Y. L., Evans, N. J. II, \& Rawlings, J. M. C. 2002a, ApJ, in press (astro-ph/0204024)
\reference{} Shirley, Y. L., Evans, N. J. II, Mueller, K. E., Knez, C., Jaffe, D. T., 2002b, in prep.
\reference{} Simon, M., Peterson, D. M., Longmore, A. J., Storey, J. M. V., \& Tokumaga, A. T. 1985, ApJ, 298, 328
\reference{} Soifer, B. T. \& Pipher, J. L. 1975, ApJ, 199, 663
\reference{} Sridharan, T. K., Beuther, H., Schilke, P., Menten, K. M., \& Wyrowski, F. 2002, \apj, 566, 931
\reference{} Stier, M. T. 1984, ApJ, 283, 573
\reference{} Tan, J. C., \& McKee, C. F. 2002, in Hot Star Workshop III: The Earliest Phases of Massive Star Birth, ed. P. A. Crowther, ASP Conf. Ser. 267, in press.
\reference{} Thompson, R. 1984, ApJ, 283, 165 
\reference{} Thronson Jr., H. A., Lowenstein, R. F., \& Stokes, G. M. 1979, AJ, 84, 1328
\reference{} Thronson Jr., H. A., Gatley, I., Harvey, P. M., Sellgren, K., \& Werner, M. W. 1980, ApJ, 237, 66
\reference{} Tofani, G., Felli, M., Talor, G. B., Hunter, T. R. 1995, A\&AS, 112, 299
\reference{} Tothill, N.F.H. 1999, PhD Thesis
\reference{} Val'tts, I. E., Ellingsen, S. P., Slysh, V. I., Kalenskii, S. V., Otrupcek, R., \& Larionov, G. M. 2000, MNRAS, 317, 315
\reference{} van der Tak, F. F. S., van Dishoeck, E. F., Evans, N. J. II, Bakker, E. J., \& Blake, G. A. 1999 \apj, 522, 991
\reference{} van der Tak, F. F. S., van Dishoeck, E. F., Evans, N. J. II, \& Blake, G. A. 2000 \apj, 537, 283
\reference{} Walker, C. K., Adams, F. C., \& Lada, C. J. 1990, \apj, 349, 515
\reference{} Walsh, A. J., Bertoldi, F., Burton, M. G., Nikola, T. 2001, MNRAS, 326, 36 
\reference{} Walsh, A. J., Burton, M. G., Hyland, A. R., Robinson, G. 1998, MNRAS, 301, 640
\reference{} Werner, M. W., Becklin, E. E., Gatley, I., Matthews, K., Neugebauer, G., \& Wynn-Williams, C. G. 1979, MNRAS, 188, 463 
\reference{} Wilner, D. J., Welch, W. J., Forster, J. R. 1995, ApJ, 449, L73
\reference{} Wood, D. O. S, \& Churchwell, E. 1989, ApJ, 340, 265
\reference{} Wright, E. L., Lada, C. J., Fazio, G. G., Low, F. J., \& Kleinman, D. E. 1977, AJ, 82, 132
\reference{} Yang, J., Umemoto, T., Iwata, T., \& Fukui, Y. 1991, \apj, 373, 137
\reference{} Young, C. H., Shirley, Y. L., Evans, N. J. II, \& Rawlings, J. M. C. 2002, \apj, submitted
\reference{} Zhou, S., Butner, H. M., Evans, N. J. II, Guesten, R., Kutner, M. L. \& Mundy, L. G. 1994, \apj, 428, 219
\reference{} Zhou, S., Evans II, N. J., \& Wang, Y. 1996, \apj, 466, 296 
\reference{} Zinchenko, I., Forsstroem, V., Lapinov, A., \& Mattila, K. 1994, A\&A, 288, 601


\end{references}
\end{document}